\documentclass[preprint,review,authoryear,12pt,3p]{elsarticle}

\usepackage{amsmath}
\usepackage{graphicx,amssymb,psfrag,subfigure,color}
\usepackage{epsfig}
\newcommand{\beq}{\begin{equation}}
\newcommand{\eeq}{\end{equation}}
\newcommand{\bpm}{\begin{pmatrix}}
\newcommand{\epm}{\end{pmatrix}}
\newcommand{\beqa}{\begin{eqnarray}}
\newcommand{\eeqa}{\end{eqnarray}}
\newcommand{\beqas}{\begin{eqnarray*}}
\newcommand{\eeqas}{\end{eqnarray*}}

\renewcommand{\d}{\mathrm{d}}
\newcommand{\mR}{\mathbb{R}}

\newcommand{\CL}{\mathcal{L}}

\newcommand{\pdhfrac}[2]{\mathchoice{\frac{#1}{#2}}{#1/#2}{#1/#2}{#1/#2}}

\newcommand{\pd}[2]{\pdhfrac{{\partial}#1}{{\partial}#2}}

\newcommand{\vb}{\mathbf{b}}

\newcommand{\vt}{\mathbf{t}}
\newcommand{\vr}{\mathbf{r}}
\newcommand{\vn}{\mathbf{n}}
\newcommand{\vm}{\mathbf{m}}
\newcommand{\vl}{\boldsymbol{l}}
\newcommand{\vk}{\mathbf{k}}

\newcommand{\vv}{\mathbf{v}}
\newcommand{\vu}{\mathbf{u}}

\newcommand{\vsi}{\boldsymbol{\sigma}}

\newcommand{\ka}{\kappa}

\newcommand{\eps}{\epsilon}

\newcommand{\tvr}{\tilde{\vr}}

\def\XXint#1#2#3{{\setbox0=\hbox{$#1{#2#3}{\int}$ }
\vcenter{\hbox{$#2#3$ }}\kern-.6\wd0}}

\begin{document}

\begin{frontmatter}

\title{A continuum model for dislocation dynamics in three dimensions using the dislocation density potential functions and its application to micro-pillars}
\author[HKUST]{Yichao~Zhu}
\author[HKUST]{Yang~Xiang\corref{cor1}}
\ead{maxiang@ust.hk}

\cortext[cor1]{Corresponding author}

\address[HKUST]{Department of Mathematics, The Hong Kong University of Science and Technology, Clear Water Bay, Kowloon, Hong Kong}


\begin{abstract}
In this paper, we present a dislocation-density-based three-dimensional continuum model, where the dislocation substructures are represented by pairs of dislocation density potential functions (DDPFs), denoted by $\phi$ and $\psi$. The slip plane distribution is characterized by the contour surfaces of $\psi$, while the distribution of dislocation curves on each slip plane is identified by the contour curves of $\phi$ which represents the plastic slip on the slip plane. By using DDPFs, we can explicitly write down an evolution equation system, which is shown consistent with the underlying discrete dislocation dynamics. The system includes i) A constitutive stress rule, which describes how the total stress field is determined in the presence of dislocation networks and applied loads; ii) A plastic flow rule, which describes how dislocation ensembles evolve. The proposed continuum model is validated through comparisons with discrete dislocation dynamics simulation results and experimental data. As an application of the proposed model, the ``smaller-being-stronger'' size effect observed in single-crystal micro-pillars is studied. A scaling law for the pillar flow stress $\sigma_{\text{flow}}$ against its (non-dimensionalized) size $D$ is derived to be $\sigma_{\text{flow}}\sim\log(D)/D$.

\end{abstract}

\begin{keyword}
Dislocation density  \sep Crystal plasticity  \sep  Continuum model \sep Size effect \sep Micro-pillars \sep Finite elements


\end{keyword}

\end{frontmatter}

\section{Introduction \label{Sec_introduction}}

It is widely agreed that plasticity theories that properly integrate the accumulated knowledge in small-scale physics can facilitate the design of high-end materials. The continuum crystal plasticity (CCP) theories \citep[e.g.][]{Rice1971, Peirce1983, Hutchinson1993, Gao1998, Gurtin2002} have shown their values in understanding the elasto-plastic behavior of crystals, but they are still phenomenological.
 On the other hand, the (three-dimensional) discrete dislocation dynamical (DDD) models take the dislocation microstructural evolution into account based on the fact that plastic deformation of crystals is carried out by the motion of a large number of dislocations \citep[e.g.][]{Kubin1992,Zbib1998,Fivel1998,Ghoniem2000, Gumbsch2001, Needleman2002,Xiang2003_Acta,Needleman2004, Xiang_thin_film2006, Cai2007,Rao2007,ElAwady2008, Tang2008, Weygand2008,ElAwady2009, Rickman2010,Xiang2012_local_expansion, Zhou2012,Ryu2013,Zhu_JMPS2013,Zhu_MSEA2014,WangJ2014}. In DDD models, dislocations are treated as line singularities embedded into an elastic medium. The kinematics of individual dislocations is governed by a collection of laws for dislocation gliding, climb, multiplication, annihilation, reaction, etc., and the microstructural changes within crystals are then captured by the evolution of dislocation curves. DDD models have been well applied to provide insights in understanding many plastic deformation processes observed in micro- or nano-crystalline structures, such as in thin films and interfaces \citep[e.g.][]{Gumbsch2001, Needleman2002, Xiang_thin_film2006,Zhou2012,WangJ2014}  and in micro-pillars \citep[e.g.][]{Rao2007,ElAwady2008, Tang2008,Weygand2008,ElAwady2009,Ryu2013}. However, three-dimensional DDD models become too computationally intensive when the specimen size exceeds the order of several microns.

Therefore, a successful dislocation-density-based theory of plasticity (DDBTP) whose associated length scale lies between CCP's and DDD's is still highly expected. The development of DDBTP dates back to the works of \citet{Nye1953}, where a dislocation network is represented by a continuously distributed second-order tensor, known as the Nye dislocation tensor. Nowadays with more knowledge in physics taking place on smaller scales, a successful DDBTP should be constituted by laws that are consistent with the underlying discrete dislocation dynamics from the following two aspects: i) A constitutive stress rule to determine the stress field in the presence of a continuous dislocation density distribution and applied loads; ii) A plastic flow rule to capture the motion of dislocation ensembles (in response to the calculated stress field), which results in plastic flows in crystals.

As the simplest dislocation configuration, systems of straight and mutually parallel dislocations have been analyzed relatively well at the continuum level~\citep[e.g.][]{Groma2003,Berdichevsky2006,Voskoboinikov2007JMPS,Kochmann_IJP2008, Cameron2011,Oztop2013,Geers2013,Le_IJP2014,Schulz2014,Zhu_2Ddipoles2014,Le2015}. In this case, each dislocation can be treated as a point singularity in a plane that is perpendicular to all dislocations. As a result, the  Nye dislocation density tensor is reduced to several scalar dislocation density functions. The geometric complexity of the dislocation networks is dramatically reduced in this case. However, the development of three-dimensional DDBTP is still far from satisfactory despite a number of valuable works~\citep[e.g.][]{Nye1953,Kroener1963,Kosevich1979,Nelson1981,Mura1987,Head1993, ElAzab2000,Acharya2001,Svendsen2002,Arsenlis2002,Sedlacek2003, Alankar2011, Sandfeld2011,Engels2012, Hochrainer2014,Zbib2014,Ngan2014}.
One of the main barriers in establishing a successful three-dimensional theory is due to the fact that the complex networks of curved
dislocation substructures make the upscaling of discrete dislocation dynamics extremely difficult.

To overcome such difficulties, \citet{Xiang2009_JMPS} introduced the idea of a coarse-grained disregistry function (CGDF), which is defined to approximate the exact disregistry function (plastic slip) used in the Peierls-Nabarro models~\citep{Peierls1940,Nabarro1947,Xiang2008_Acta}, by a smoothly varying profile without resolving details of dislocation cores. By this way, the density distribution of a discrete curved dislocation network in a single slip plane (after local homogenization) can be simply represented by the scalar CGDF (more precisely, the  spatial derivatives of CGDF), and  dislocation dynamics on the slip plane at the continuum level is explicitly formulated in terms of the evolution of the CGDF~\citep{Xiang2010_PhilMag}. Using this representation of CGDF for dislocation density distribution, connectivity condition of dislocations is automatically satisfied.
The underlying topological changes of dislocations are automatically handled by the evolution equation of the CGDF, and no law for dislocation annihilation needs to be further imposed. The dynamics of dislocations in the continuum model is derived from the DDD model, and the dislocation velocity in the continuum model depends on a continuum version of the Peach-Koehler force on dislocations.
It has been rigorously shown by~\citet{Xiang2009_JMPS}  that in the continuum model, the Peach-Koehler force due to the resolved shear stress of a family of curved dislocations can be decomposed into a long-range dislocation  interaction force and a short-range self line tension force, and they can both be expressed in terms of the spatial derivatives of CGDFs.
The Frank-Read  source, which is one of the major mechanisms  for dislocation multiplication, is also well incorporated into this continuum framework \citep{Zhu2014_IJP}. As one application of this continuum model using CGDFs, a two-dimensional Hall-Petch law, which relates the flow stress of a polycrystal not only to the physical dimension of its constituent grains, but also to the grain aspect ratio, is derived without any adjustable parameters~\citep{Zhu2014_IJP}.

In this paper, we generalize our previous single-slip-plane model to that for dislocation ensembles in three-dimensions, where the density distribution of dislocations is locally co-determined by an in-plane dislocation density distribution and a slip plane distribution. To take into account the spatial variation from these two aspects, we define a pair of {\bf dislocation density potential functions} (DDPFs) for each active slip system. One DDPF $\psi$ is employed to describe the slip plane distribution (after local homogenization) by its contour surfaces, and the other DDPF $\phi$ is defined such that $\phi$ restricted on each slip plane describes the plastic slip across the slip plane and identifies the density distribution of dislocation curves (after local homogenization) on that plane.
 Here we name $\phi$ and $\psi$ by density potential functions, because the Nye dislocation density tensor is represented in terms of the spatial derivatives of these two functions. As our previous continuum model in a single slip plane~\citep{Xiang2009_JMPS}, the major advantage of this three-dimensional continuum model lies in its simple representation of distributions of curved dislocations (after local homogenization) using two scalar DDPFs, which automatically satisfies the connectivity condition of dislocations.

To derive the constitutive stress rule in the continuum framework with the DDPFs, we sequentially express the Nye dislocation density tensor, the plastic distortion and the elastic strain tensor in terms of the DDPFs. As in our previous continuum model in a single slip plane~\citep{Xiang2009_JMPS}, the continuum Peach-Koehler force due to the resolved shear stress consists of a long-range dislocation  interaction force and a short-range self line tension force. The long-range stress field is determined by the derived constitutive stress rule and the equilibrium equations along with  boundary conditions, and a finite element (FE) formulation is proposed to compute this long-range stress field.
The local self line tension effect can be explicitly formulated in terms of the spatial derivatives of $\phi$ and $\psi$.
The plastic flow rule is described by evolution equations of the DDPFs $\phi$ and $\psi$.
For face-centered-cubic (fcc) crystals considered in this paper, the motion of dislocations is limited to their respective slip planes, and the plastic flow rule is given by an evolution equation of $\phi$.
Frank-Read sources in our single-slip-plane continuum model~\citep{Zhu2014_IJP} is generalized to three-dimensional case with continuous distributions of a number of sources.
The derived equations form a closed system evolving in time as summarized in Eqs.~\eqref{eqn_stress_tensor1} to \eqref{eqn_mobility_law} in Sec.~\ref{Sec_summary_constitutive_eqns}.

With this continuum model, we investigate the size effect on crystalline strength widely observed in the uniaxial compression tests of monocrystalline micro-pillars \citep[e.g.][]{Uchic2004, Uchic_review2009, Gao_Nanopillars2013}. Practically, an empirical power law is adopted to relate the pillar flow stress $\sigma_{\text{flow}}$ to the pillar size $D$ by $
\sigma_{\text{flow}} \sim D^{-m}$, where $m$ is found to be from 0 to 1, varying by study~\citep{Uchic_review2009}. Typically there are two classes of models proposed to rationalize this size effect. The first type falls into the family of the ``dislocation starvation'' models \citep{Greer2005,Greer2006}. They argued that a crystal small in size does not provide enough space for dislocation multiplication and the flow strength gets increased as a result. The second category of models attribute the observed size effect to the stochastics of dislocation source lengths in the small-size specimens~\citep{Parthasarathy2007}.  There are also models using statistical approaches to reproduce the power law expression \citep[e.g.][]{Ngan2013_JMPS}. Analysis of heterogeneous deformation  in single crystal micropillars under compression has been performed by using a hybrid elasto-viscoplastic simulation model which couples DDD model, and a scaling law $\sigma_{\text{flow}} \sim D^{-n}\log D+ \alpha D^{-1}$ with two parameters $\alpha$ and $ 0<n<1$ was proposed \citep{Akarapu2010}. In the last part of this paper, we apply the proposed continuum model to study the size effect on the strength of micro-pillars, by following the trace of the stochastic source length models which have been employed to rationalize the size effect observed in DDD simulations~\citep[e.g.][]{ElAwady2008, Weygand2008, ElAwady2009, Zbib_size_effect2014}. We find that the flow stress scales with the sample size by
\beq \label{strength_scale_with_log}
\sigma_{\text{flow}} \sim \frac{b}{D}\log\left(\frac{D}{b}\right).
\eeq
This relation is validated through comparison with experimental data conducted in several fcc crystals for pillar size ranging from submicrons to tens of microns.

The rest of this paper is organized as follows. In Sec.~2, the (three-dimensional) DDPFs are introduced, and this is followed by the derivation of the constitutive stress rule and the plastic flow rule needed at the continuum level. In Sec.~3, numerical schemes for the derived equation system are presented. In Sec.~4, the derived continuum model is validated through  comparisons with DDD simulation results. In Sec.~5, the continuum model is applied to study the size effect arising in the uniaxial compression tests of single crystalline micro-pillars.

To better illustrate the derivation of the continuum model, following notations are used throughout the article unless specified. The Cartesian coordinates are denoted by $\vr = (x,y,z)$. The $i$-th entry of a vector, for example $\vr$, is denoted by $r_i$, and the $ij$-th entry of a second-order tensor, for example $\vsi$, is denoted by $\sigma_{ij}$. Unless specified, the following notations are used: the vector gradient $(\nabla \vu)_{ij} = \partial u_i/\partial r_j$; the cross product $(\vm\times\vn)_i = \sum_{j,k=1}^3\eps_{ijk}m_jn_k$ with $\eps_{ijk}$ the permutation tensor; the inner product of two vectors $\vm\cdot\vn = \sum_{i=1}^3 m_in_i$; the inner product of two second-order tensors $\boldsymbol{\alpha}:\boldsymbol{\beta} = \sum_{i,j=1}^3\alpha_{ij}\beta_{ij}$; the inner product of a fourth-order tensor and a second-order tensor $\boldsymbol{\CL}:\boldsymbol{\beta}=\sum_{k,l=1}^3\CL_{ijkl}\beta_{kl}$; the magnitude of a vector $|\vu| = \sqrt{\vu\cdot\vu}$; the symmetric part of a second-order tensor $\text{sym}(\boldsymbol{\alpha}) = (\boldsymbol{\alpha}+\boldsymbol{\alpha}^{\text{T}})/2$; the outer product of two vectors $(\mathbf{a}\otimes\vb)_{ij} = a_ib_j$; and the row ``curl'' of a second-order tensor $(\nabla\times\boldsymbol{\alpha})_{ij} = \sum_{k,l=1}^3\eps_{jkl}\alpha_{il,k}$.

\section{Continuum plasticity model based on dislocation density potential functions}

In this section, we first review the continuum model for dislocation dynamics in one slip plane using a two-dimensional coarse-grained disregistry function~\citep{Xiang2009_JMPS,Xiang2010_PhilMag,Zhu2014_IJP}.
Then by using the DDPFs, we  build the three-dimensional continuum model for dislocation dynamics and plasticity.

\subsection{Review of the continuum model in a single slip plane\label{Sec_CGDF_2d}}
\begin{figure}[!ht]
\centering
\subfigure[]{\includegraphics[width = .53\textwidth]{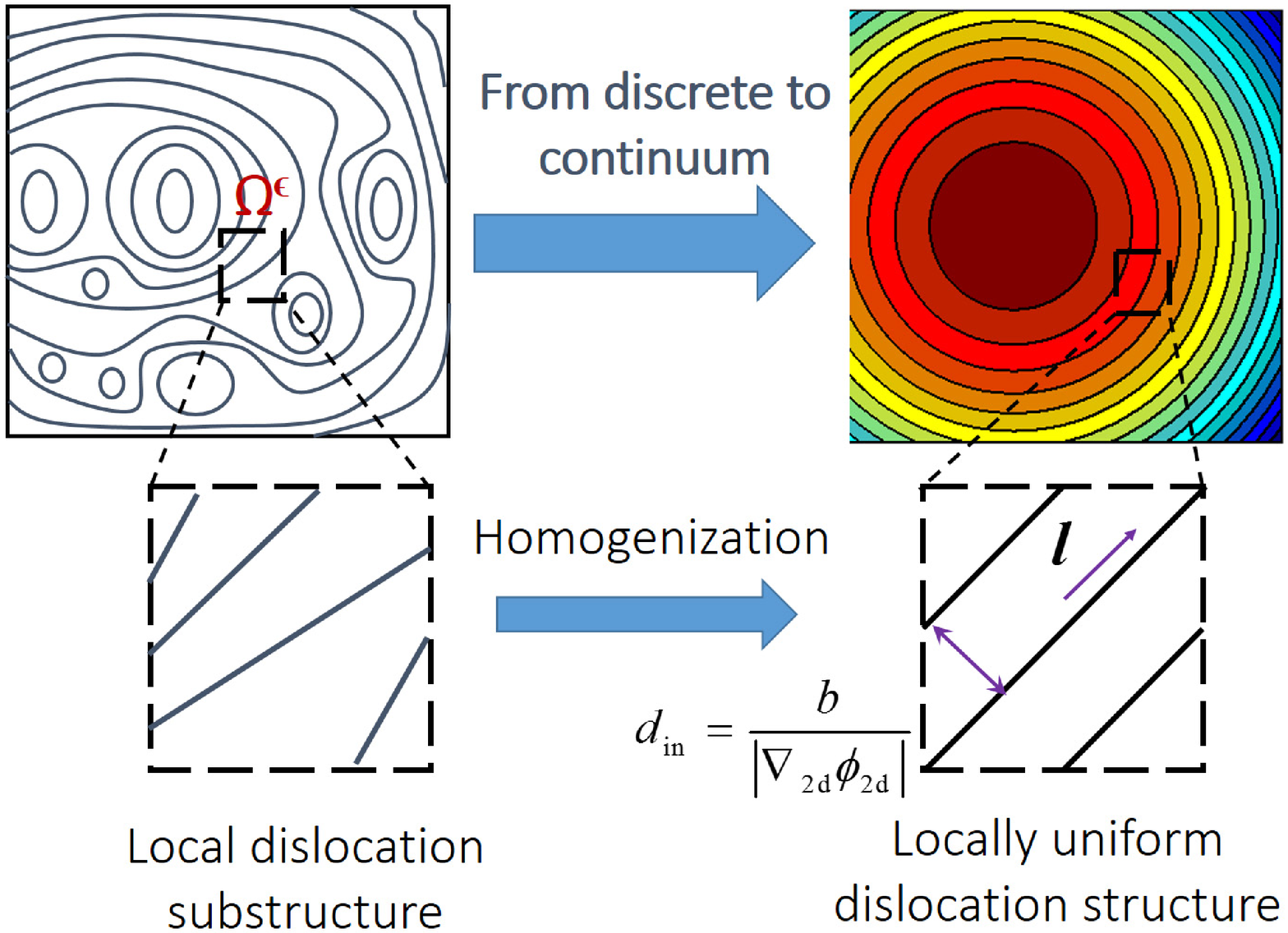}}
\subfigure[]{\includegraphics[width = .45\textwidth]{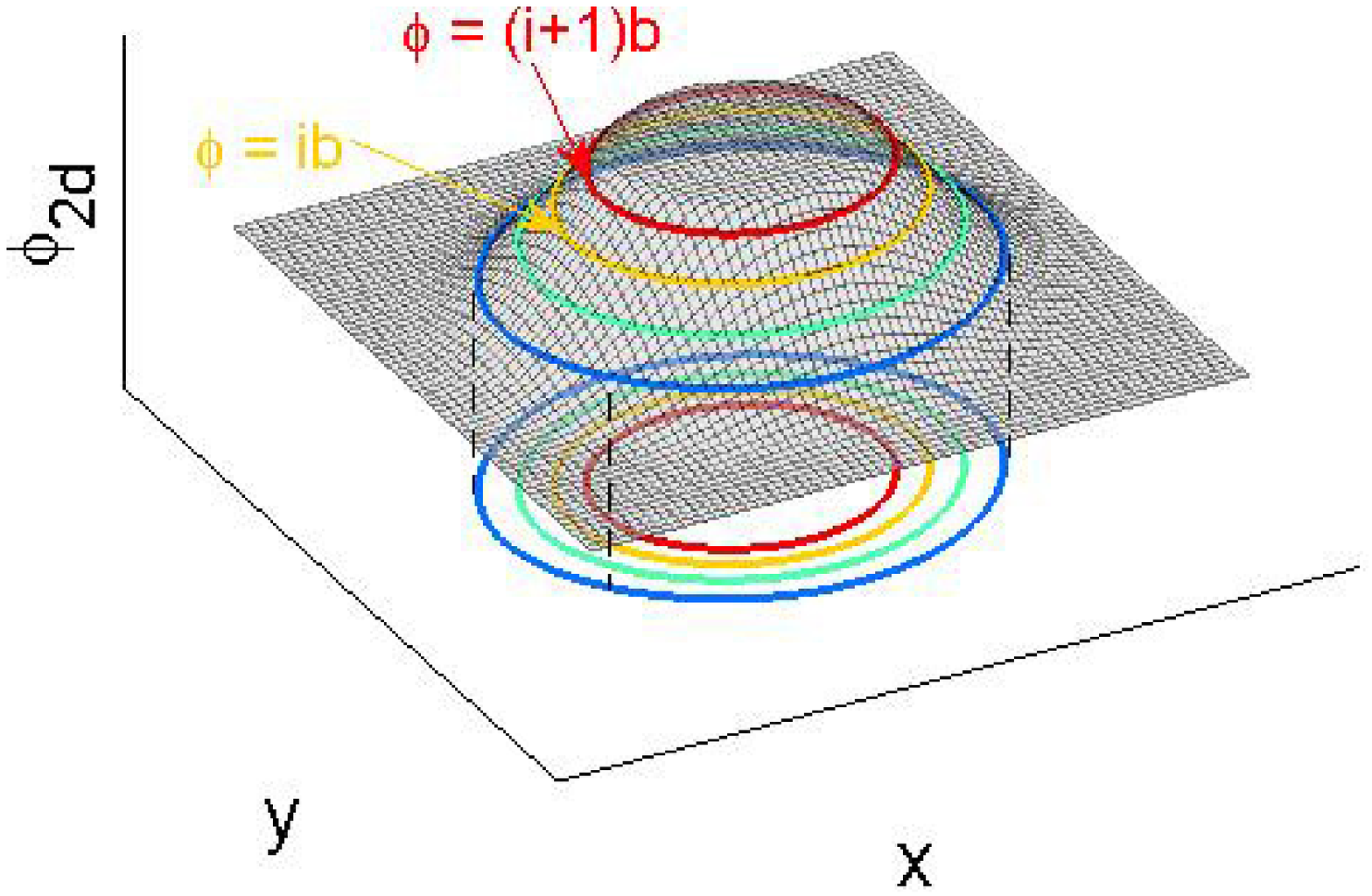}}
\caption{(a) Turning a discrete dislocation network in a single slip plane into a dislocation continuum. The discrete dislocation network with Burgers vector $\vb$ is approximated by a family of smoothly varying dislocation curves with local line direction $\vl$ and dislocation spacing $d_{\text{in}}$, averaged locally over
 a representative area $\Omega^{\eps}$ of the discrete dislocation network, see Eq.~\eqref{vl_rho_def_2d}.   (b) Representation of the dislocation continuum in (a) (on the top right) using the CGDF $\phi_{\text{2d}}$. The $i$-th dislocation curve  corresponds to the contour of $\phi_{\text{2d}}$ with height $\phi_{\text{2d}}=ib$. \label{Fig_illu_CGDF}}
\end{figure}

We consider describing a given discrete dislocation network in a single slip plane and with the same Burgers vector $\vb$ (e.g. the configuration on the top left of Fig.~\ref{Fig_illu_CGDF}(a)) by a dislocation continuum.
For this purpose, we use  two field quantities at the continuum level: the average dislocation line direction $\vl$ and the average dislocation spacing  $d_{\text{in}}$, which equals the reciprocal of the net dislocation length per area $\rho_{\text{g}}^{\text{2d}}$.
The values of the two field quantities $\vl$ and $d_{\text{in}}$ at each point in the continuum model come from the averaging over  a representative area $\Omega^{\eps}$ of size $\eps$ in the discrete model, where
$d_{\text{in}}<<\eps<<D$ with $D$ being the sample size in the continuum model. Under this assumption,
all dislocations inside $\Omega^{\eps}$ can be treated as line segments, as schematically shown in Fig.~\ref{Fig_illu_CGDF}(a).
By superpositioning all dislocation segments inside $\Omega^{\eps}$, we can obtain a super line segment denoted by $\mathbf{L}$, i.e. $\mathbf{L}=\sum_i \mathbf{s}_i$ for line segment $\mathbf{s}_i\in\Omega^{\eps}$.
The quantities of interest at the continuum level can then be defined by
\beq \label{vl_rho_def_2d}
\vl = \lim_{\eps/D\rightarrow0}\frac{\mathbf{L}}{|\mathbf{L}|}, \qquad\qquad
d_{\text{in}} = \frac1{\rho_{\text{g}}^{\text{2d}}} = \lim_{\eps/D\rightarrow0}\frac{|\Omega^{\eps}|}{|\mathbf{L}|},
\eeq
where $|\Omega^{\eps}|$ is the area of $\Omega^{\eps}$.

By performing such averaging process everywhere, the original discrete dislocation network is turned into a dislocation continuum as schematically shown in Fig.~\ref{Fig_illu_CGDF}(a) (on the top right)
 and we can then use field variables to express it.
  \citet{Xiang2009_JMPS} introduced a coarse-grained disregistry function $\phi_{\text{2d}}$, such that
the $i$-th dislocation curve in the averaged dislocation continuum corresponds to the contour of $\phi_{\text{2d}}$ with height $\phi_{\text{2d}}=ib$, see Fig.~\ref{Fig_illu_CGDF}(b). With the function $\phi_{\text{2d}}$, the local dislocation line direction and inter-dislocation spacing can be given by
\beq \label{in_plane_tangent_2D}
\vl = \frac1{ |\nabla_{\text{2d}}\phi_{\text{2d}}|}\left(\frac{\partial \phi_{\text{2d}}}{\partial y},\,-\frac{\partial \phi_{\text{2d}}}{\partial x} \right)
\eeq
and\
\beq \label{average_spacing_2d}
d_{\text{in}} = \frac1{\rho_{\text{g}}^{\text{2d}}} = \frac{b}{|\nabla_{\text{2d}}\phi_{\text{2d}}|},
\eeq
respectively, where $\nabla_{\text{2d}} = \left( \frac{\partial}{\partial x},\,\frac{\partial}{\partial y} \right)$.  A fact used to derive these formulas is that the normal vector $\vn$ of the $i$-th dislocation curve in the averaged dislocation continuum
\beq \label{in_plane_normal_2d}
\vn = \frac{\nabla_{\text{2d}}\phi_{\text{2d}}}{ |\nabla_{\text{2d}}\phi_{\text{2d}}|},
\eeq
which is the unit vector in the slip plane and normal to $\vl$.

With the introduction of $\phi_{\text{2d}}$, we can capture the resolved shear stress including that due to the long-range dislocation interaction by an integral
\beq \label{stress_long_range_2D}
\begin{aligned}
\tau_{\text{long}}^{\text{2d}} &= \frac{\mu}{4\pi}\int_{\mR^2} \frac{(x-\tilde{x})\frac{\partial \phi_{\text{2d}}(\tilde{x},\tilde{y})}{\partial \tilde{x}} + (y-\tilde{y})\frac{\partial \phi_{\text{2d}}(\tilde{x},\tilde{y})}{\partial \tilde{y}} }{((x-\tilde{x})^2+(y-\tilde{y})^2)^{3/2}} \d \tilde{x} \d \tilde{y} \\
& \quad + \frac{\mu\nu}{4\pi(1-\nu)b^2}\int_{\mR^2} \frac{\bigg(b_1\frac{\partial \phi_{\text{2d}}(\tilde{x},\tilde{y})}{\partial \tilde{x}}+b_2\frac{\partial \phi_{\text{2d}}(\tilde{x},\tilde{y})}{\partial \tilde{y}}\bigg)(b_1(x-\tilde{x})+b_2(y-\tilde{y}))}{((x-\tilde{x})^2+(y-\tilde{y})^2)^{3/2}} \d \tilde{x} \d \tilde{y},
\end{aligned}
\eeq
where $\mu$ and $\nu$ are the shear modulus and the Poisson's ratio, respectively, and a contribution due to the local line tension effect
\beq \label{self_stress_2D}
\begin{aligned}
\tau_{\text{self}}^{\text{2d}} &=\frac{\mu b \kappa}{4\pi}\left(\frac{1+\nu}{1-\nu} - \frac{3\nu}{1-\nu}\frac{(b_1\frac{\partial \phi_{\text{2d}}}{\partial x} + b_2\frac{\partial \phi_{\text{2d}}}{\partial y})^2/b^2}{\sqrt{\left(\frac{\partial \phi_{\text{2d}}}{\partial x}\right)^2 + \left(\frac{\partial \phi_{\text{2d}}}{\partial y}\right)^2}}\right) \log\left(\frac{b/r_{\text{c}}}{2\pi \sqrt{\left(\frac{\partial \phi_{\text{2d}}}{\partial x}\right)^2 + \left(\frac{\partial \phi_{\text{2d}}}{\partial y}\right)^2}}+1\right),
\end{aligned}
\eeq
where $r_{\text{c}}$ is a parameter depending on the dislocation core and $\kappa=-\nabla\cdot(\nabla\phi_{\text{2d}}/|\nabla\phi_{\text{2d}}|)$ is the local (average) curvature of the dislocation. These stress formulas in the continuum model were derived rigorously from the discrete dislocation model by asymptotic analysis~\citep{Xiang2009_JMPS}.

In the single-slip-plane continuum model, $\phi_{\text{2d}}$ measures the plastic slip across the slip plane in the direction of the Burgers vector, and the plastic flow is governed by an evolution equation of $\phi_{\text{2d}}$~\citep{Xiang2010_PhilMag,Zhu2014_IJP}:
\beq \label{evolution_CGDF_2D}
\pd{\phi_{\text{2d}}}{t} + v_n\sqrt{\left(\pd{\phi_{\text{2d}}}{x}\right)^2 + \left(\pd{\phi_{\text{2d}}}{y}\right)^2} = s_{\text{2d}}.
\eeq
Here $v_n$ is the dislocation moving speed along its normal direction~\citep{Xiang2010_PhilMag}
\begin{equation}
v_n=m_g(\tau_{\text{long}}^{\text{2d}}+\tau_{\text{self}}^{\text{2d}})b,
\end{equation}
where $m_g$ is the dislocation glide mobility and the applied stress  can also be included in the above equation.
 The right-hand term $s_{\text{2d}}$ in Eq.~(\ref{evolution_CGDF_2D}) formulates the effect due to the dislocation multiplication by Frank-Read sources \citep{Zhu2014_IJP}, which will be reviewed in detail in Sec.~\ref{Sec_FR_source_2D}.

\subsection{Dislocation substructures in three dimensions represented by  dislocation density potential functions\label{Sec_DDPFs}}

From now on, we present our continuum model for the dynamics of dislocation structures in three dimensions. In this subsection, we introduce the representation of dislocation substructures in three dimensions using DDPFs.

\begin{figure}[htbp]
\centering
\subfigure[]{\includegraphics[width = .58\textwidth]{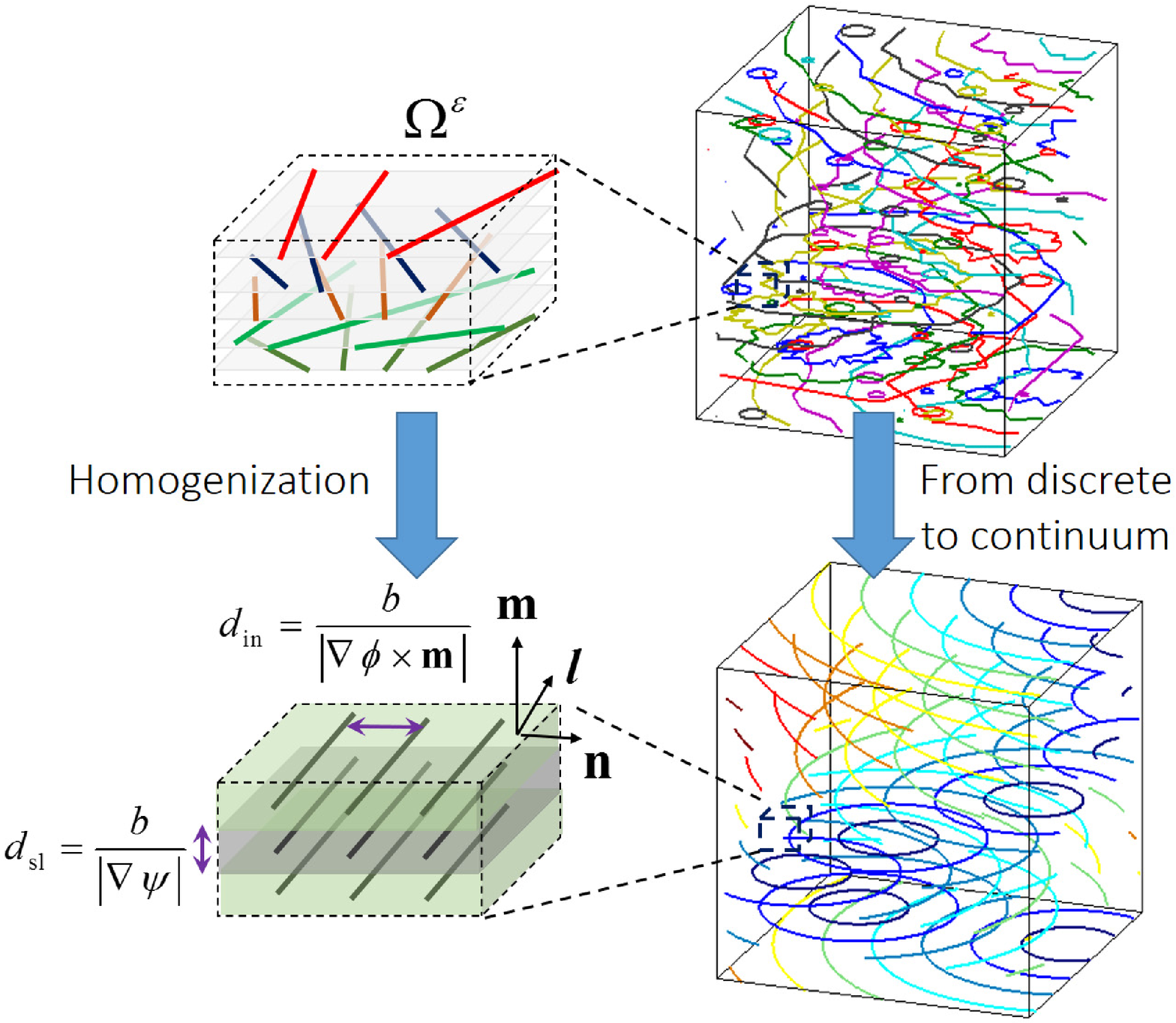}}
\subfigure[]{\includegraphics[width = .4\textwidth]{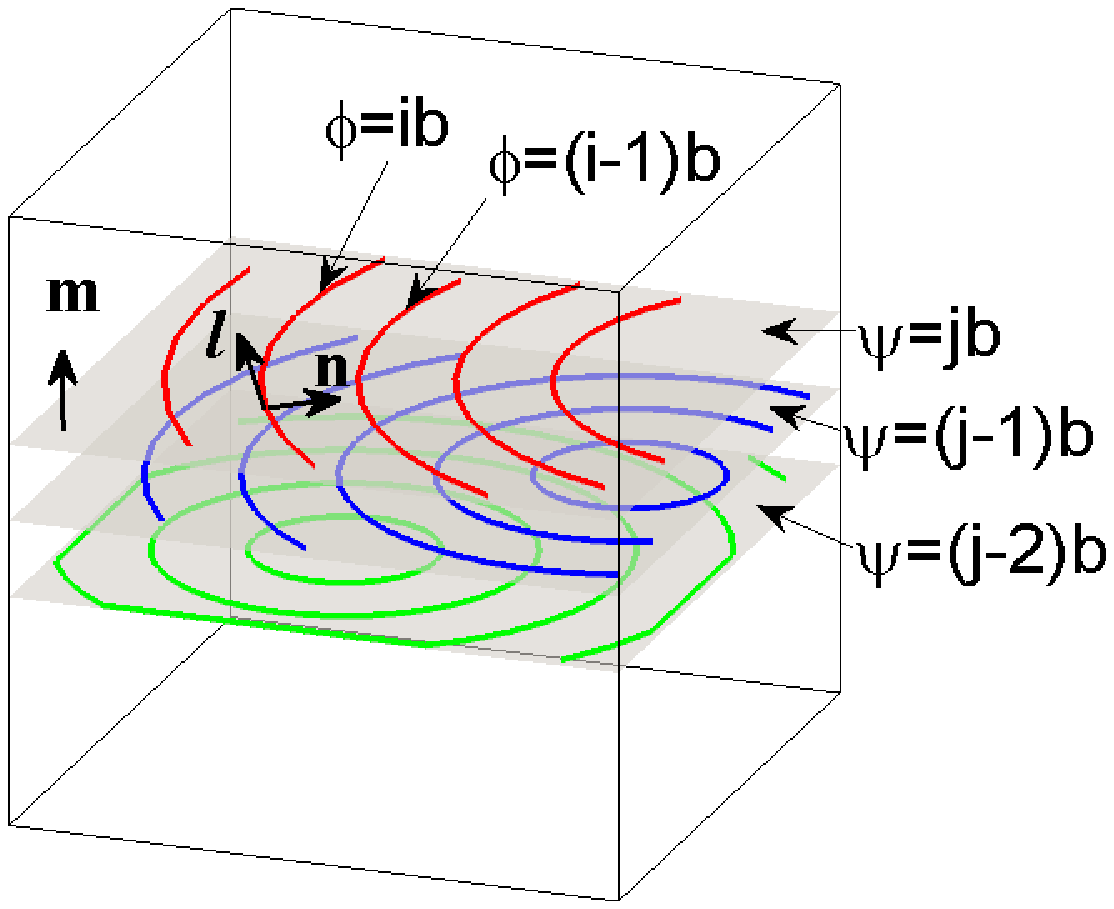}}
\caption{(a)  A discrete dislocation network in three dimensions  is approximated by a dislocation continuum, after local homogenization of dislocation ensembles within some representative volume $\Omega^{\eps}$. These dislocations in the network are associated with the same slip system with Burgers vector $\vb$ and slip plane normal direction $\vm$. The dislocation continuum is characterized by the dislocation line direction $\vl$, the slip plane spacing $d_{\text{sl}}$ and the in-plane dislocation spacing $d_{\text{in}}$ averaged over the representative volume $\Omega^{\eps}$, see Eqs.~(\ref{vl_rho_def}) and (\ref{rho_num_3d}). (b) Representation of the dislocation continuum in (a) using a pair of DDPFs $\phi$ and $\psi$.  The DDPF $\psi$ is employed such that the $j$-th slip plane  is the contour plane of $\psi$ of height $\psi=jb$.  The DDPF $\phi$ is defined such that $\phi$ restricted on each slip plane describes the dislocation distribution on that plane, i.e., the $i$-th dislocation on the slip plane is  the contour line $\phi=ib$. \label{Fig_illu_psi_phi}}
\end{figure}

We first focus on  a given discrete dislocation network in a single slip system with Burgers vector $\vb$ and slip plane normal direction $\vm$. We take a representative cuboid volume $\Omega^{\eps}$ with size $\eps$, the Nye dislocation density tensor over $\Omega^{\eps}$ is~\citep{Nye1953,Kroener1963,Kosevich1979}
$\boldsymbol{\alpha} = \sum_i\vb\otimes\mathbf{s}_i/|\Omega^{\eps}|$ for dislocation line segment $\mathbf{s}_i\in\Omega^{\eps}$, where $|\Omega^{\eps}|$ is the volume of $\Omega^{\eps}$, see Fig.~\ref{Fig_illu_psi_phi}(a). Denoting $\mathbf{L}=\sum_i \mathbf{s}_i$ for all dislocation line segments $\mathbf{s}_i\in\Omega^{\eps}$ which is the superposition of all dislocation segments inside $\Omega^{\eps}$, and assuming that $\eps<<D$ where $D$ is the domain size of the continuum model,
we write the Nye dislocation density tensor over $\Omega^{\eps}$  as
\beq \label{Nye_density_tensor_def_single_slip_system}
\boldsymbol{\alpha} = \rho_{\text{g}}\vb\otimes\vl,
\eeq
where $\vl$ is the average dislocation line direction over $\Omega^{\eps}$ and $\rho_{\text{g}}$ is the dislocation density over $\Omega^{\eps}$ (net length per unit volume):
\beq \label{vl_rho_def}
\vl = \lim_{\eps/D\rightarrow0}\frac{\mathbf{L}}{|\mathbf{L}|},\qquad\qquad
\rho_{\text{g}} = \lim_{\eps/D\rightarrow0}\frac{|\mathbf{L}|}{|\Omega^{\eps}|}.
\eeq
 See Fig.~\ref{Fig_illu_psi_phi}(a) for an illustration of this average. We characterize $\rho_{\text{g}}$ by two variables
\beq \label{rho_num_3d}
\rho_{\text{g}} = \frac1{d_{\text{sl}}d_{\text{in}}},
\eeq
where $d_{\text{sl}}$ is the slip plane spacing and $d_{\text{in}}$ is the in-plane dislocation spacing, see Fig.~\ref{Fig_illu_psi_phi}(a).

By performing such averaging process everywhere, the original three-dimensional discrete dislocation network is turned into a dislocation continuum in three dimensions as schematically shown in Fig.~\ref{Fig_illu_psi_phi}(a).
We introduce a pair of {\bf dislocation density potential functions} (DDPFs) $\phi$ and $\psi$
to represent this resulting dislocation continuum consisting of dislocations in the same slip system. The DDPF $\psi$ is employed to describe the distribution of the slip planes in the averaged dislocation continuum, such that the $j$-th slip plane  is the contour plane of $\psi$ with $\psi=jb$.
In this paper, we focus on fcc crystals, in which dislocations stay in their respective slip planes. We further assume a uniform distribution of the slip planes with prescribed slip plane spacing $d_{\text{sl}}$. Under these conditions, the DDPF $\psi$ for the slip system with Burgers vector $\vb$ and slip normal direction $\vm$ can be written as
\begin{equation} \label{psi_def_no_cross_slip}
\psi(\vr) = \frac{b\vm \cdot (\vr-\vr^0)}{d_{\text{sl}}},
\end{equation}
where $\vr^0$ is some point on the $0$-th slip plane, see Fig.~\ref{Fig_illu_psi_phi}(b).

With slip planes in the averaged dislocation continuum described by $\psi$, we introduce another DDPF $\phi$ to describe the distribution of dislocations on each slip plane, such that $\phi$ restricted on each slip plane is the two-dimensional CGDF $\phi_{\text{2d}}$ in the single-slip plane model reviewed in the previous subsection.
That is, the $i$-th dislocation curve on the $j$-th slip plane in the averaged dislocation continuum can be represented by
\beq \label{dislocation_loop_3D}
\{\vr | \phi(\vr) = ib,\,\psi(\vr) = jb, \, {\rm for \ integers}\ i, j\},
\eeq
see Fig.~\ref{Fig_illu_psi_phi}(b). As $\phi_{\text{2d}}$ in the single-slip plane model, here $\phi$ restricted on each slip plane describes the plastic slip across the slip plane in the direction of the Burgers vector. Local geometric quantities of dislocations and the Nye dislocation density tensor in the averaged dislocation continuum are simply expressed in terms of these DDPFs $\phi$ and $\psi$, see the following subsections.

The major advantage of this three-dimensional continuum model characterized by DDPFs lies in its simple representation of distributions of curved dislocations (after local homogenization). This representation also automatically satisfies the connectivity condition of dislocations, see Sec.~\ref{Sec_density_tensor}. In addition, dislocation annihilation within the same slip plane is automatically handled as in the previous single-slip-plane model~\citep{Xiang2009_JMPS,Xiang2010_PhilMag}.
When there are multiple slip systems activated, one can assign a pair of DDPFs to each active slip system.

In body-centered-cubic (bcc) crystals or fcc crystals at high temperatures where dislocations are able to move out of their slip planes by climb, dislocations no longer stay on planar slip planes.  Dislocation networks in these crystals can still be represented under the continuum framework characterized by the DDPFs $\phi$ and $\psi$. In these cases,  the contour surface $\psi(\vr)=jb$ may describe a curved surface rather than a plane, and the local slip plane normal direction in the averaged dislocation continuum is determined by
\beq \label{slip_normal_psi}
\vm = \frac{\nabla \psi}{|\nabla\psi|}.
\eeq

\subsection{Geometrical structures of dislocation continuum described by DDPFs\label{Sec_dislocation_geometry}}

Now we present expressions of the local geometric quantities of dislocations in the averaged dislocation continuum  in terms of the DDPFs $\phi$ and $\psi$. The slip plane normal direction  is  given by Eq.~(\ref{slip_normal_psi}).
The local dislocation line direction $\vl$ is
\beq \label{tangent_def}
\vl = \frac{\nabla \phi \times \nabla \psi}{|\nabla \phi \times \nabla \psi|}.
\eeq
This is because the dislocation is contained in both the contour surfaces of $\phi=ib$ and $\psi=jb$ (see Eq.~(\ref{dislocation_loop_3D})), thus it is perpendicular to both $\nabla \phi$ and  $\nabla \psi$.
The local dislocation normal $\vn$ (which is in the slip plane) is also calculated in terms of DDPFs by
\beq \label{in_plane_normal_def}
\vn = \vm \times \vl = \frac{\nabla \psi \times (\nabla \phi \times \nabla \psi)}{|\nabla \phi \times \nabla \psi||\nabla \psi|}.
\eeq
Moreover, by the Frenet-Serret formulas, we have
$\ka\vn = \left( \vl\cdot\nabla l_1,\, \vl\cdot\nabla l_2, \, \vl\cdot\nabla l_3 \right)$,
where $\kappa$ is the curvature of the local dislocation. Thus with the expressions for $\vl$ and $\vn$ in Eqs.~\eqref{tangent_def} and \eqref{in_plane_normal_def},  $\kappa$ is also represented by the DDPFs as
\beq \label{curvature}
\ka = \vn\cdot (\ka\vn) = \sum_{i=1}^3 n_i\vl\cdot\nabla l_i.
\eeq

Finally, the in-plane dislocation spacing and the slip plane spacing are respectively
\beq\label{din-dsl}
d_{\text{in}}=\frac{b}{|\nabla\phi\times\vm|}, \ \ d_{\text{sl}}=\frac{b}{|\nabla\psi|}.
\eeq
Here we have used the definition in Eq.~(\ref{dislocation_loop_3D}) and the fact that
the gradient of $\phi$ over the slip plane is $
\nabla_{\text{in-plane}} \phi = \nabla \phi - (\vm\cdot\nabla \phi)\vm=\nabla\phi\times\vm$. Note that the above formulas hold for a general DDPF $\psi$ for the definition of dislocations in Eq.~(\ref{dislocation_loop_3D}), including but not limited to the particular expression in Eq.~(\ref{psi_def_no_cross_slip}).

\subsection{Representation of the Nye dislocation density tensor by DDPFs\label{Sec_density_tensor}}

After the local homogenization process described in Sec.~\ref{Sec_DDPFs}, the Nye dislocation density tensor $\boldsymbol{\alpha}$ for a single slip system is calculated by Eq.~(\ref{Nye_density_tensor_def_single_slip_system}) in terms of the averaged dislocation line direction $\vl$ and the net length per volume dislocation density $\rho_{\text{g}}$ over some representative volume, and $\rho_{\text{g}}$ is given by  Eq.~(\ref{rho_num_3d}) using the slip plane spacing  $d_{\text{sl}}$ and the in-plane dislocation spacing $d_{\text{in}}$.
Further using the formulas in Eqs.~(\ref{slip_normal_psi}), (\ref{tangent_def}) and (\ref{din-dsl}) with the DDPFs $\phi$ and $\psi$, we have
\beq \label{num_density_formula}
\rho_{\text{g}} = \frac{|\nabla\phi\times\vm|\cdot|\nabla\psi|}{b^2} = \frac{|\nabla \phi\times \nabla \psi|}{b^2}.
\eeq

Therefore, incorporating Eqs.~\eqref{tangent_def} and \eqref{num_density_formula} into Eq.~(\ref{Nye_density_tensor_def_single_slip_system}) gives the following expression for the Nye dislocation density tensor based on the DDPFs
\beq \label{Nye_density_tensor_single_slip_system}
\begin{aligned}
\boldsymbol{\alpha} &= \frac{\vb}{b^2}\otimes (\nabla \phi\times\nabla \psi)\\
& = \frac{1}{b^2}\begin{pmatrix} b_1\left(\frac{\partial \psi}{\partial z}\frac{\partial \phi}{\partial y} - \frac{\partial \psi}{\partial y}\frac{\partial \phi}{\partial z}\right) & b_1\left(\frac{\partial \psi}{\partial x}\frac{\partial \phi^{\alpha
}}{\partial z} - \frac{\partial \psi}{\partial z}\frac{\partial \phi}{\partial x}\right) & b_1\left(\frac{\partial \psi}{\partial y}\frac{\partial \phi}{\partial x} - \frac{\partial \psi}{\partial x}\frac{\partial \phi}{\partial y}\right) \\ b_2\left(\frac{\partial \psi}{\partial z}\frac{\partial \phi}{\partial y} - \frac{\partial \psi}{\partial y}\frac{\partial \phi}{\partial z}\right) & b_2\left(\frac{\partial \psi}{\partial x}\frac{\partial \phi}{\partial z} - \frac{\partial \psi}{\partial z}\frac{\partial \phi}{\partial x}\right) & b_2\left(\frac{\partial \psi}{\partial y}\frac{\partial \phi}{\partial x} - \frac{\partial \psi}{\partial x}\frac{\partial \phi}{\partial y}\right) \\
b_3\left(\frac{\partial \psi}{\partial z}\frac{\partial \phi}{\partial y} - \frac{\partial \psi}{\partial y}\frac{\partial \phi}{\partial z}\right) & b_3\left(\frac{\partial \psi}{\partial x}\frac{\partial \phi}{\partial z} - \frac{\partial \psi}{\partial z}\frac{\partial \phi}{\partial x}\right) & b_3\left(\frac{\partial \psi}{\partial y}\frac{\partial \phi}{\partial x} - \frac{\partial \psi}{\partial x}\frac{\partial \phi}{\partial y}\right) \end{pmatrix}.
\end{aligned}
\eeq

For the case with multiple slip systems, we use one pair of DDPFs $\phi^{\lambda}$ and $\psi^{\lambda}$ for the $\lambda$-th slip system, and the Nye dislocation density tensor is expressed by
\beq \label{Nye_density_tensor_formula}
\boldsymbol{\alpha} = \sum_{\lambda} \frac{\vb^{\lambda}}{(b^{\lambda})^2}\otimes (\nabla \phi^{\lambda}\times\nabla \psi^{\lambda}).
\eeq

It is easy to check that $\boldsymbol{\alpha}$ given by Eq.~\eqref{Nye_density_tensor_formula} satisfies
\begin{equation}
\nabla\cdot\boldsymbol{\alpha}=\textbf{0},
\end{equation}
which means $\sum_{j=1}^3\frac{\partial \alpha_{ij}}{\partial r_j}=0$ for all $i=1$, $2$ and $3$. This is the connectivity condition of dislocations, which is due to the fact that dislocation lines can never begin or end inside the sample.

\subsection{Constitutive stress rule\label{Sec_constitutive_stress}}

In this subsection, we present the constitutive stress rule in our continuum plasticity model using DDPFs. The derivation is based on the constitutive relations commonly used in classical dislocation-density-based models, including (when the specimen experiences small deformations):
\begin{itemize}
\item The total distortion, which is the gradient of the displacement field $\vu$, can be decomposed into an elastic distortion  $\boldsymbol{\beta}^{\text{e}}$ and a plastic distortion $\boldsymbol{\beta}^{\text{p}}$
\beq \label{total_distortion}
\nabla\vu = \boldsymbol{\beta}^{\text{e}} + \boldsymbol{\beta}^{\text{p}}.
\eeq
\item The Nye dislocation density tensor equals the curl gradient of the plastic distortion
\beq \label{eqn_plastic_distortion}
\nabla \times \boldsymbol{\beta}^{\text{p}} = -\boldsymbol{\alpha} = - \sum_{\lambda} \frac{\vb^{\lambda}}{(b^{\lambda})^2}\otimes (\nabla \phi^{\lambda}\times\nabla \psi^{\lambda}).
\eeq
\item The stress field $\vsi$ satisfies the Hooke's law (e.g. the isotropic case):
\beq \label{Hooke_law}
\vsi = 2\mu \boldsymbol{\epsilon}^{\text{e}} + \frac{2\mu \nu}{1-2\nu} \text{tr}(\boldsymbol{\epsilon}^{\text{e}}) \textbf{I},
\eeq
where $\boldsymbol{\epsilon}^{\text{e}}$ is the elastic strain tensor related to the elastic distortion by $
\boldsymbol{\epsilon^{\text{e}}} = \text{sym}\left(\boldsymbol{\beta}^{\text{e}}\right)$,
$\text{tr}(\boldsymbol{\epsilon}^{\text{e}})$ is the trace of $\boldsymbol{\epsilon}^{\text{e}}$,
and $\textbf{I}$ is the $3\times3$ identity matrix.
\item
The equilibrium condition in absence of body forces is
\beq \label{eqn_force_balance}
\nabla\cdot\vsi = \boldsymbol{0}.
\eeq
\end{itemize}

By generalizing the expression for $\boldsymbol{\beta}^{\text{p}}$ for the case of discrete dislocations \citep[e.g., Sec.~1.6 of][]{Mura1987}, and using the fact that $\phi$
describes the plastic slip across the slip plane in the direction of the Burgers vector and $\psi$ describes the distribution of the slip planes, in the continuum model we have
\beq \label{plastic_distortion_exp}
\boldsymbol{\beta}^{\text{p}} = -\sum_{\lambda}\frac{\phi^{\lambda}}{(b^{\lambda})^2}(\vb^{\lambda}\otimes \nabla\psi^{\lambda}).
\eeq
 It can be checked that $\boldsymbol{\beta}^{\text{p}}$ given by Eq.~\eqref{plastic_distortion_exp} satisfies Eq.~\eqref{eqn_plastic_distortion}.
A comparison between Eq.~\eqref{plastic_distortion_exp} and its counterpart in continuum crystal plasticity theories  suggests that the scalar
\begin{equation}\label{shear_slip_exp0}
\gamma^{\lambda} = \frac{\phi^{\lambda}|\nabla\psi^{\lambda}|}{b^{\lambda}}
\end{equation}
measures the magnitude of the plastic shearing in the direction of Burgers vector $\vb^{\lambda}$ in the $\lambda$-th slip system.

By using Eqs.~\eqref{total_distortion} and \eqref{plastic_distortion_exp}, we can express the elastic strain tensor by
\beq \label{elastic_strain_exp}
\boldsymbol{\epsilon}^{\text{e}} =  \text{sym}(\nabla \vu) + \sum_{\lambda}\frac{\phi^{\lambda}}{(b^{\lambda})^2}\text{sym}(\vb^{\lambda}\otimes \nabla\psi^{\lambda}).
\eeq
Incorporating Eq.~\eqref{elastic_strain_exp} with the Hooke's law \eqref{Hooke_law} and using the fact that $\vb^{\lambda}\cdot\nabla\psi^{\lambda}=0$ (the Burgers vector is always perpendicular to the normal direction of the slip plane), we obtain
\beq \label{eqn_constitutional_law}
\vsi = \boldsymbol{\CL}:\nabla\vu + 2\mu\sum_{\lambda} \frac{\phi^{\lambda}}{(b^{\lambda})^2} \text{sym}(\vb^{\lambda}\otimes\nabla\psi^{\lambda}),
\eeq
where the symmetric fourth-order tensor $\boldsymbol{\CL}$ is defined such that
\beq \label{L_operator_def}
\boldsymbol{\CL}:\nabla\vu = 2\mu \left(\text{sym}\left(\nabla \vu\right) + \frac{\nu}{1-2\nu} (\nabla \cdot \vu) \textbf{I}\right).
\eeq

When a solid body is purely elastic (dislocation free), $\phi^{\lambda}=0$ and Eq.~\eqref{eqn_constitutional_law} becomes $\vsi = \boldsymbol{\CL}:\nabla\vu$,
which is exactly the constitutive relation in the linear elasticity theory.

In summary, with given dislocation distributions described by DDPF pairs $\phi^{\lambda}$ and $\psi^{\lambda}$ for $\lambda$ varies over all the slip systems, the stress field $\vsi$ is obtained by solving Eq.~(\ref{eqn_constitutional_law}) and the equilibrium condition in Eq.~(\ref{eqn_force_balance}) over the sample $\Omega$, subject to appropriate boundary conditions on the boundary $\partial \Omega$.  One boundary condition is the displacement boundary condition imposed on part of the boundary denoted by $\partial \Omega_{\text{d}}$:
\beq \label{BC_Dirichlet}
\vu|_{\partial \Omega_{\text{d}}} = \vu^{\text{b}},
\eeq
another boundary condition is the traction boundary conditions imposed on the rest of the boundary denoted by $\partial \Omega_{\text{t}}$:
\beq \label{BC_Noemann}
\vsi|_{\partial \Omega_t}\cdot\boldsymbol{k} = \textbf{t}^{\text{b}},
\eeq
with $\boldsymbol{k}$ the outer unit normal to the surface $\partial \Omega_{\text{t}}$. Here $\partial\Omega = \partial\Omega_{\text{d}}\cup\partial\Omega_{\text{t}}$.

\subsection{Short-range dislocation line tension effect}

It is well-known that there are various short-range dislocation interactions that play important roles in the plastic deformation processes, in addition to the long-range effect of dislocations. In this subsection, we  first show that the continuum stress formulation presented in the previous subsection is for the long-range dislocation interaction. We then  present the formulation for the short-range line tension effect in the DDPFs based three dimensional continuum model.

When the medium is the whole three dimensional space $\mR^3$, using the Nye dislocation density tensor formula in Eq.~(\ref{Nye_density_tensor_formula}) in our continuum model,
the stress formulas given by the DDD model and the continuum model are as follows~\citep{Hirth,Mura1987}.
In the DDD model, the stress field is
\beq \label{PK_stress}
\begin{aligned}
\vsi_{\text{dd}} (\vr)&= \frac{\mu}{2\pi}\sum_{\lambda} \sum_{i=1}^{N_\lambda} \int_{\gamma_i^\lambda} \text{sym}\left(\frac{\vb^{\lambda}\times (\vr-\tvr)}{|\vr-\tvr|^3} \otimes \vl \d s \right) \\
& \quad + \frac{\mu}{4\pi(1-\nu)} \sum_{\lambda} \sum_{i=1}^{N_\lambda} \int_{\gamma_i^\lambda} (\vl \cdot(\vb^{\lambda}\times\nabla))(\nabla\otimes\nabla - \textbf{I}\nabla^2) |\vr-\tvr| \d s,
\end{aligned}
\eeq
where $\gamma_i^\lambda$ denotes the $i$-th dislocation in the $\lambda$-th slip system, $N_\lambda$ is the number of dislocations of the $\lambda$-th slip system,  $\tvr$ goes over all points on $\gamma_i^\lambda$,
$s$ is the arclength of  $\gamma_i^\lambda$, and $\nabla$ denotes the gradient with respect to $\vr$.
In the continuum framework characterized by the DDPFs, the stress field is
\beq \label{stress_int_3D}
\begin{aligned}
\vsi(\vr) &= \sum_{\lambda}\frac{\mu}{2\pi (b^\lambda)^2}  \int_{\mR^3}\text{sym}\left(\frac{\vb^\lambda\times(\vr-\tvr)}{|\vr-\tvr|^3}\otimes(\nabla \phi^\lambda(\tvr)\times\nabla \psi^\lambda(\tvr))\right) \d \tilde{V} \\
& \quad + \sum_{\lambda}  \frac{\mu}{4\pi(b^\lambda)^2 (1-\nu)} \int_{\mR^3}(\vb^\lambda\cdot\nabla\phi^\lambda(\tvr))(\nabla\psi^\lambda(\tvr)\cdot\nabla) (\nabla\otimes\nabla - \textbf{I}\nabla^2) |\vr-\tvr| \d \tilde{V},
\end{aligned}
\eeq
where $\d \tilde{V}$ is an infinitesimal volume associated with $\tvr$.
Since the Nye dislocation density of the DDD model is averaged into the Nye dislocation density of the continuum model in the coarse-graining process as described in Secs. \ref{Sec_DDPFs}  and \ref{Sec_density_tensor}, from Eqs. (\ref{PK_stress}) and (\ref{stress_int_3D}),
we have
\begin{equation}
\vsi_{\text{dd}} \longrightarrow \vsi
\end{equation}
in the coarse-graining process from the DDD model to the continuum model.

When we consider a material with finite size, an image stress is added in the DDD model to accommodate the boundary conditions~\citep{Needleman1995}. Following the same coarse-graining process, such a DDD stress solution gives a stress solution that satisfies all the equations and boundary conditions of the continuum model, which is the same as that obtained by directly solving the continuum model due to the uniqueness of the solution. Therefore in this case, the stress field solved from the continuum model is also the leading order approximation of the stress field of the DDD model.

The stress field $\vsi$ solved from the elasticity system in the continuum model describes the long-range elastic interaction of dislocations. There are also various short-range dislocation interactions that play important roles in the plastic processes. How to capture these short-range effects at the continuum level is an important issue in the development of dislocation based plasticity theories.

One way to incorporate the short-range dislocation interactions in the continuum model is to add complementary  resolved shear stresses that describes these effects, because
  the dynamics of dislocations depends on the resolved component of the stress field in the slip plane~\citep{Hirth}. For the $\lambda$-th slip system, the resolved shear stress due to the long-range dislocation interaction in the continuum model is
 \beq \label{long_range_stress_res}
\tau^{\lambda}_{\text{long}} = \frac{\vb^{\lambda}}{b^{\lambda}}\cdot \vsi \cdot \frac{\nabla\psi^{\lambda}}{|\nabla\psi^{\lambda}|}.
\eeq
Recall that $\vm^{\lambda}=\nabla\psi^{\lambda}/|\nabla\psi^{\lambda}|$ is the normal direction of the slip plane.
The principle to include additional shear stresses in the continuum model is to obtain a better approximation to the shear stress in the DDD model
\beq \label{stress_approximate_exp}
\tau_{\text{dd}}^{\lambda}\longrightarrow\tau_{\text{total}}^{\lambda} = \tau_{\text{long}}^{\lambda} + \tau_{\text{short}}^{\lambda}
\eeq
in the coarse-graining process from the DDD model to the continuum model, where $\tau_{\text{short}}^{\lambda}$ includes contributions to the resolved shear stress due to all the important short-range effects, and
$\tau_{\text{dd}}^{\lambda}=\frac{\vb^{\lambda}}{b^{\lambda}}\cdot \vsi_{\text{dd}} \cdot \vm^{\lambda}$ is the resolved shear stress of the $\lambda$-th slip system using the DDD model.

In this paper, we focus on the dislocation line tension effect which is one of the important short-range dislocation effects, and the total shear stress in the continuum model is
 \beq \label{short_stress}
\tau_{\text{total}}^{\lambda} = \tau_{\text{long}}^{\lambda} + \tau_{\text{self}}^{\lambda},
\eeq
where $\tau_{\text{self}}^{\lambda}$ is the contribution to the resolved stress due to the line tension effect. The dislocation line tension effect  plays crucial roles for example in dislocation multiplication by Frank-Read sources.
For dislocation distributions in a single slip plane, we have used asymptotic analysis to rigorously derive the expression of the line tension effect in the continuum model from the DDD model \citep{Xiang2009_JMPS} as reviewed in Sec.~\ref{Sec_CGDF_2d}.
 Here for dislocation distributions in three dimensions represented by DDPFs, for the $\lambda$-th slip system, $\tau_{\text{self}}^{\lambda}$ is given by
\beq \label{self_stress_res}
\tau_{\text{self}}^{\lambda}= \frac{\mu b^{\lambda}}{4 \pi} \left(\frac{1+\nu}{1-\nu}-\frac{3 \nu}{1-\nu}\cdot \frac{|\nabla\psi^{\lambda}|^2(\vb^{\lambda} \cdot \nabla \phi^{\lambda})^2}{(b^{\lambda})^2 |\nabla\psi^{\lambda} \times \nabla \phi^{\lambda}|^{2}}\right) \kappa \cdot \log \left(\frac{b^{\lambda}|\nabla\psi^{\lambda}|}{2 \pi r_{c} |\nabla\psi^{\lambda} \times \nabla \phi^{\lambda}|} +1\right),
\eeq
where the curvature $\kappa$ of the local dislocation is calculated by Eq.~\eqref{curvature}. In order to derive this formula, we have used the fact that the DDPF $\phi$ restricted onto each slip plane which is a contour surface of the DDPF $\psi$ reduces to our previous single slip plane model as well as the expressions of geometric characterizations of the local dislocation given in Sec.~\ref{Sec_dislocation_geometry}.

There are other short-range effects to be included in $\tau_{\text{short}}$, such as that due to the short-range  interaction between dislocations from different slip systems. These will be discussed elsewhere.

\subsection{Plastic flow rule\label{Sec_plastic_flow_rule}}

In our continuum model characterized by DDPFs, the plastic flow rule is given by
\beq \label{eqn_phi_evolution0}
\dot{\phi}^{\lambda} + v^{\lambda}_{\text{n}} \frac{|\nabla \phi^{\lambda} \times \nabla\psi^{\lambda}|}{|\nabla\psi^{\lambda}|} = s^{\lambda}
\eeq
for the $\lambda$-th slip system, where $\dot{\phi}^{\lambda} = \pd{\phi^{\lambda}}{t}$, $v^{\lambda}_{\text{n}}$ is the speed of the local dislocation in its normal direction in the slip plane, and $s^{\alpha}$ formulates the dislocation generation by Frank-Read sources to be discussed in details in the next subsection. Being the three-dimensional version of Eq.~\eqref{evolution_CGDF_2D}, Eq.~\eqref{eqn_phi_evolution0} is also established based on the conservation of plastic shear slips (\citet{Xiang2010_PhilMag}, see also the level set DDD method in \citet{Xiang2003_Acta}) and the fact that the gradient of $\phi^{\lambda}$ over the slip plane is $\nabla\phi^{\lambda} \times \vm^{\lambda}$, where $\vm^{\lambda}=\nabla \psi^{\lambda}/|\nabla \psi^{\lambda}|$ is the normal direction of the slip planes.
Note that there is no need to assign extra rule for dislocation annihilation on the same slip plane, which is automatically handled by the topological changes in the contours of the DDPF $\phi^{\lambda}$ during its evolution.

The dislocation speed $v^{\lambda}_{\text{n}}$ in Eq.~\eqref{eqn_phi_evolution0} is determined by a mobility law following the DDD models:
\beq \label{mobility_law}
v^{\lambda}_{\text{n}}= m_{\text{g}} b^{\lambda} \tau^{\lambda}_{\text{total}}= m_{\text{g}} b^{\lambda} \left(\tau^{\lambda}_{\text{long}} + \tau^{\lambda}_{\text{self}}\right),
\eeq
where $m_{\text{g}}$ is the dislocation glide mobility, $\tau^{\lambda}_{\text{long}}$ is the component of the long-range stress field $\vsi$ determined by the elasticity problem  in Sec.~\ref{Sec_constitutive_stress} and resolved in the $\lambda$-th slip system given in Eq.~(\ref{long_range_stress_res}),
and  $\tau^{\lambda}_{\text{self}}$ is the  effective stress due to  the line tension effect given in Eq.~(\ref{self_stress_res}) in the previous subsection.

In general, boundary conditions are needed for Eq.~\eqref{eqn_phi_evolution0}.  One extreme case is that dislocations can exit the specimen freely, which can be described by the Neumann boundary condition $\frac{\partial \phi^{\lambda}}{\partial \vk_s}=0$ where $\vk_s$ is the outer normal direction of the intersection of the slip plane and the specimen surface ($\vk_s=\vm\times(\vm\times \vk)/|\vm\times(\vm\times \vk)|$ where $\vk$ is the outer normal direction of the specimen surface).
 Another extreme case is that the dislocations are impenetrable to a specimen surface, where $\phi^{\lambda}$ is fixed on the specimen surface.

The rate of Nye dislocation density tensor $\dot{\boldsymbol{\alpha}}$, the rate of the plastic distortion  $\dot{\boldsymbol{\beta}}^{\text{p}}$ and the rate of the scalar plastic shear $\dot{\gamma}^{\lambda}$ can all be easily calculated using $\dot{\phi}^{\lambda}$ in Eq.~(\ref{eqn_phi_evolution0}) and the expressions of $\boldsymbol{\alpha}$, $\boldsymbol{\beta}^{\text{p}}$ and $\gamma^{\lambda}$ in terms of $\phi^{\lambda}$ and $\psi^{\lambda}$ in Eqs.~(\ref{Nye_density_tensor_formula}), (\ref{plastic_distortion_exp}) and (\ref{shear_slip_exp0}). Note that in this paper, the DDPF $\psi^{\lambda}$ that describes the distribution of the slip planes is fixed.

Many other quantities that are useful in understanding the plastic behavior of crystals can also be expressed in terms of the DDPFs $\phi$ and $\psi$. For example, the total dislocation density (length per unit volume) within the specimen $\Omega$ can be formulated by
\beq \label{total_number_density}
\rho_{\text{tot}} =  \frac1{|\Omega|} \sum_{\lambda}\int_{\Omega} \frac{|\nabla \phi^{\lambda} \times \nabla\psi^{\lambda}|}{(b^{\lambda})^2} \d V+\rho_0,
\eeq
where $\rho_0$ is the dislocation density due to the initial distribution of dislocation segments of the Frank-Read sources (see the next subsection).
 Moreover, the total plastic strain rate, which is conventionally defined to be the rate of area swept by all dislocations multiplied by the length of the respective Burgers vector per volume, can be expressed by
\beq \label{total_plastic_strain}
\dot{\eps}^{\text{p}}_{\text{tot}} = \frac1{|\Omega|}\sum_{\lambda}\int_{\Omega} \dot{\phi}^{\lambda} \frac{|\nabla\psi^{\lambda}|}{b^{\lambda}}\d V.
\eeq

\subsection{Incorporation of Frank-Read sources into the continuum model}

In this subsection, we present the expression for the source term $s^{\lambda}$ in the plastic flow rule in Eq.~\eqref{eqn_phi_evolution0} for a continuous distribution of Frank-Read sources in three dimensions, which is generalized from our previous continuum model for individual sources on a single slip plane.

\subsubsection{Review of Frank-Read sources in the single-slip plane continuum model\label{Sec_FR_source_2D}}

We first review our continuum model for individual sources on a single slip plane \citep{Zhu2014_IJP}.

A Frank-Read source is a dislocation segment pinned at its two ends. When the resolved shear stress $\tau$ acting on it exceeds a critical value, known as the activation stress  $\tau_c$, it will keep injecting dislocation loops to the system~\citep{Hirth}. The time it takes for a Frank-Read source to perform an operating cycle is known as the nucleation time $t_{\text{nuc}}$.

However, if observed at the continuum level, what we see is not the detailed loop-releasing process, but continuous dislocation flux originating from a small source region. In the single-slip plane model \citep{Zhu2014_IJP} reviewed in Sec.~\ref{Sec_CGDF_2d}, the operation of a Frank-Read source is controlled by three parameters all determined by the underlying DDD model: the source activation stress $\tau_c$, the source operating rate which equals to $1/t_{\text{nuc}}$, and the source region $\Omega_{\text{s}}^{\text{2d}}$ which is  the two-dimensional region enclosed by a newly released dislocation loop.

The nucleation stress $\tau_c$ is evaluated by adopting the critical stress formula given by \beq \label{activation_stress_single_source}
\tau_c = \frac{C_{\text{s}}\mu b}{2\pi l} \log\left(\frac{l}{r_{\text{c}}}\right),
\eeq
where $C_{\text{s}}$ depends on the source character and the Poisson's ratio $\nu$ (with $\nu=1/3$, $C_{\text{s}}=1$ for an edge-oriented source and $C_{\text{s}}=1.5$ for an screw-oriented source), $l$ is the length of the Frank-Read source, and $r_{\text{c}}$ is a parameter depending on the dislocation core.

 The nucleation time $t_{\text{nuc}}$ is calculated to be
\beq \label{source_nucleation_time}
t_{\text{nuc}} = \frac{Q_{\text{ch}}l}{m_{\text{g}}b(|\tau| - \tau_c)},
\eeq
where $\tau$ is the resolved shear stress due to the long-range stress field, and $Q_{\text{ch}}$ depends only on the source orientation fitted from the DDD simulation ($Q_{\text{ch}} = 6.1278$ for edge-oriented source, $Q_{\text{ch}} = 3.0413$ for screw-oriented source). For a Frank-Read source of length $l$ parallel to the $x$ axis and centered at $(x_{\text{s}},y_{\text{s}})$, the source region is approximately bounded by an ellipse
\beq \label{source_region_2d0}
\Omega_{\text{s}}^{\text{2d}} = \left\{(x,y)\left|\frac{(x-x_{\text{s}})^2}{(a_1 l)^2} + \frac{(y-y_{\text{s}})^2}{(a_2 l)^2} \le 1\right.\right\},
\eeq
where $a_1$ and $a_2$ are calculated to be $2.4610$ and $2.2488$, respectively.

With $\tau_c$, $t_{\text{nuc}}$ and $\Omega_{\text{s}}^{\text{2d}}$ determined by Eqs.~\eqref{activation_stress_single_source}, \eqref{source_nucleation_time} and \eqref{source_region_2d0}, the Frank-Read source is incorporated into the single-slip plane model in the following sense. When the resolved shear stress $\tau>\tau_c$, a Frank-Read source keeps changing the value of $\phi^{\text{2d}}$ at a speed of $1/t_{\text{nuc}}$  such that   dislocation loops enclosing area $\Omega_{\text{s}}^{\text{2d}}$ are continuously  generated into the system. Mathematically, the source term $s^{\text{2d}}$ in the single-slip plane continuum model in Eq.~\eqref{evolution_CGDF_2D} is given by
\beq \label{source_term_2D}
s^{\text{2d}} = -\frac{m_{\text{g}}b^2(\tau - \text{sign}(\tau)\tau_c)}{ Q_{\text{ch}}l}H(|\tau|-\tau_c)\cdot\chi_{\Omega_{\text{s}}^{\text{2d}}},
\eeq
where  $H(z)$ is the Heaviside function that equals $1$ when $z>0$ and $0$ otherwise, and $\chi_{\Omega_{\text{s}}^{\text{2d}}}$ is the characteristic function in $\Omega_{\text{s}}^{\text{2d}}$, i.e. $\chi_{\Omega_{\text{s}}^{\text{2d}}}$ is $1$ in $\Omega_{\text{s}}^{\text{2d}}$ and vanishes elsewhere.

\subsubsection{Incorporation of Frank-Read sources into the three-dimensional continuum model}

Now we derive the expression of the Frank-Read sources in the three-dimensional continuum model. We first write down a formulation for individual Frank-Read sources in three dimensions by generalizing the single-slip plane model reviewed above, and then derive a formulation based on a source continuum for the three-dimensional continuum model. The formulation is affiliated with a slip system, and we omit the slip system superscript $\lambda$ for simplicity of notations.

We first write down a formulation for individual Frank-Read sources in three dimensions. Following the construction of three-dimensional dislocation continuum described in Sec.~\ref{Sec_DDPFs}, we generalize the source region $\Omega_{\text{s}}^{\text{2d}}$ in the single slip plane model given in Eq.~(\ref{source_region_2d0}) to three dimensions by a
cylinder $\Omega_{\text{s}}^{\text{3d}}=\Omega_{\text{s}}^{\text{2d}}\times [-d_{\text{sl}}/2,d_{\text{sl}}/2]\vm$, i.e. the intersection of the cylinder $\Omega_{\text{s}}^{\text{3d}}$ with a slip plane is $\Omega_{\text{s}}^{\text{2d}}$, and the height of the cylinder is $d_{\text{sl}}$ in the slip plane normal direction $\vm$, where $d_{\text{sl}}$ is recalled to be the averaged slip plane spacing.

Since the physical dimension of the source region $\Omega_{\text{s}}^{\text{3d}}$ is much smaller compared to that of the domain size $D$, we further approximate the source region $\Omega_{\text{s}}^{\text{3d}}$ by a regularized point source with same volume at the continuum level, i.e.
$\chi_{\Omega_{\text{s}}^{\text{3d}}}(\cdot)\approx|\Omega_{\text{s}}^{\text{3d}}|\delta_{\text{reg}}(\cdot)$, where $\delta_{\text{reg}}(\cdot)$ is a regularized Dirac function of the source region and $|\Omega_{\text{s}}^{\text{3d}}|=\pi a_1a_2l^2d_{\text{sl}}$ is the volume of $\Omega_{\text{s}}^{\text{3d}}$.
When there are $S$ Frank-Read sources operating, a preliminary way to incorporate them into the continuum model in Eq.~\eqref{eqn_phi_evolution0} is to  express the source term $s$ as the sum of contributions from all the individual Frank-Read sources located at $\vr_{\text{s}}^k$, $k=1,2,\cdots, S$:
\beq \label{source_term_3d_sum}
s_{\text{ind}} = -m_{\text{g}}\pi a_1a_2b^2d_{\text{sl}}\sum_{k=1}^{S} \frac{l_k(\tau - \text{sign}(\tau)\tau_c^k)}{Q_{\text{ch}}^k}H(|\tau|-\tau_c^k) \delta_{\text{reg}}(\vr - \vr_{\text{s}}^k).
\eeq

However, in the continuum model, we do not want to resolve individual Frank-Read sources. For this purpose, we introduce a source continuum to approximate the collective effect of all the individual Frank-Read sources. This is achieved by the following source term in the continuum model in Eq.~\eqref{eqn_phi_evolution0}:
\beq \label{source_term_con_exp}
s = g(\vr)(\tau - \text{sign}(\tau)\tau_0(\vr))H(|\tau| - \tau_0(\vr)),
\eeq
where at any point $\vr$, $\tau_0(\vr)$ is the  source activation stress at $\vr$, and $(\tau - \tau_0(\vr))g(\vr)$ measures the rate of plastic shear slip initiated at $\vr$ by the source continuum.

Now we determine the functions $\tau_0(\vr)$ and $g(\vr)$  from the discrete source model in Eq.~(\ref{source_term_3d_sum}). When the shear stress $\tau$ exceeds  the critical stresses of all the sources that influence point $\vr$, all the Heaviside functions in
Eqs.~(\ref{source_term_3d_sum}) and (\ref{source_term_con_exp}) can be dropped,
 a comparison between them gives
\beq \label{effective_source_density}
g(\vr) =  -\pi m_{\text{g}} a_1a_2b^2 d_{\text{sl}}\sum_{k} \frac{l_k}{Q^{k}_{\text{s}}} \delta_{\text{reg}}(\vr-\vr_{\text{s}}^{k})
\eeq
and
\beq \label{effective_activation_stress}
\tau_0(\vr) = -\frac{m_{\text{g}}\mu a_1a_2 b^3 d_{\text{sl}}}{2g(\vr)} \cdot \sum_{k} \left(\frac{C^{k}_{\text{s}} \delta_{\text{reg}}(\vr-\vr_{\text{s}}^{k}) }{Q^k_{\text{ch}}} \log\left(\frac{l_k}{r_{\text{c}}}\right)\right), \ \ {\rm where}\ g(\vr)\neq 0.
\eeq
Here we have used the expression of $\tau_c$ in Eq.~\eqref{activation_stress_single_source}. The source continuum formula of $s$ in Eq.~(\ref{source_term_con_exp}) with these expressions of $\tau_0(\vr)$ and $g(\vr)$ still provides a good approximation to $s_{\text{ind}}$ in Eq.~(\ref{source_term_3d_sum}) when the shear stress $\tau$ does not exceed the critical stresses of all the sources that influence point $\vr$, and it becomes exact again when the shear stress $\tau$ falls below all  the critical stresses of  the sources that influence point $\vr$ (In this case, $s(\vr)=s_{\text{ind}}(\vr)=0$).

The functions  $\tau_0(\vr)$ and $g(\vr)$ for the source continuum can be calculated from the arrangement of the discrete Frank-Read sources. In this paper, the locations and parameters of the Frank-Read sources are given initially and do not change in the simulation, and we calculate these two functions only once for the initial distribution of the sources.

\begin{figure}[!ht]
\centering
\subfigure[]{\includegraphics[width=.25\textwidth]{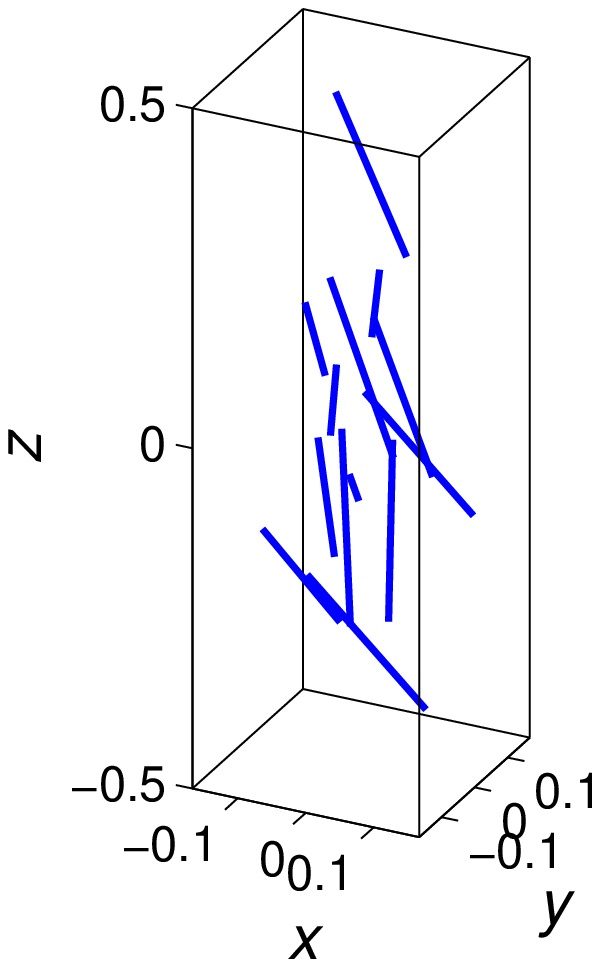}}
\subfigure[]{\includegraphics[width=.28\textwidth]{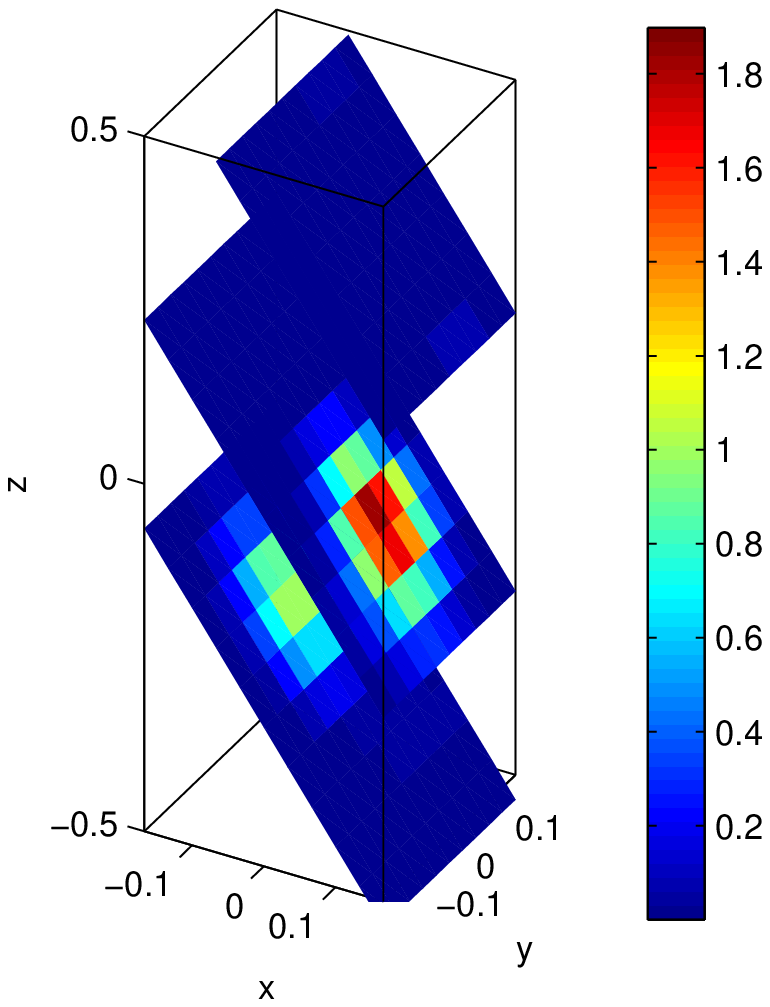}}
\caption{Converting a number of Frank-Read sources to a source continuum in a cuboid sample. (a) The discrete sources. (b) The corresponding function $g(\vr)$ in the source continuum obtained by Eq.~\eqref{effective_source_density} on several slip planes. The unit of the color bar is $m_gb$. \label{Fig_source_dis_con_example}}
\end{figure}

An example of the Frank-Read source continuum is given in
Fig.~\ref{Fig_source_dis_con_example}.
Fig.~\ref{Fig_source_dis_con_example}(a) shows a set of individual Frank-Read sources with their positions and lengths generated randomly following a normal and a uniform distribution, respectively, and the profiles of its corresponding $g(\vr)$ on several selected slip planes are drawn in Fig.~\ref{Fig_source_dis_con_example}(b). It can be observed that $g(\vr)$ attains a relatively high value around the cuboid center, because the number density of individual sources are high in the same place in Fig.~\ref{Fig_source_dis_con_example}(a).

\subsection{Summary of the governing equations in our continuum model\label{Sec_summary_constitutive_eqns}}

 To summarize, the derived continuum model based on the DDPFs is constituted mainly by the following equations.

\begin{enumerate}
\item A constitutive stress rule: Given a dislocation substructure described by $\phi^{\lambda}$ and $\psi^{\lambda}$, the long-range stress field $\vsi$ is determined by solving
\beq \label{eqn_stress_tensor1}
\vsi = 2\mu\left(\text{sym}(\nabla \vu) + \frac{\nu \text{tr}(\nabla\vu)}{1-2\nu}\textbf{I} + \sum_{\lambda}\frac{\phi^{\lambda}}{(b^{\lambda})^2}\text{sym}(\vb^{\lambda} \otimes\nabla\psi^{\lambda})\right)
\eeq
and the equilibrium condition
\beq \label{eqn_stress_tensor2}
\nabla \cdot \vsi = \boldsymbol{0},
\eeq
with boundary conditions given in Eqs.~(\ref{BC_Dirichlet}) and (\ref{BC_Noemann}).

\item A plastic flow rule: The motion of dislocations belonging to the $\lambda$-th slip system is described by an evolution equation for $\phi^{\lambda}$
\beq \label{eqn_evolution_phi}
\pd{\phi^{\lambda}}{t} + v^{\lambda}_{\text{n}} \frac{|\nabla \phi^{\lambda} \times \nabla\psi^{\lambda}|}{|\nabla\psi^{\lambda}|} = g^{\lambda}(\vr)\,(\tau^{\lambda}- \text{sign}(\tau^{\lambda}) \tau^{\lambda}_0(\vr))\,H(\tau-\tau^0(\vr)),
\eeq
where the  dislocation speed $v^{\lambda}_{\text{n}}$ is determined by the mobility law
\beq \label{eqn_mobility_law}
v^{\lambda}_{\text{n}}= m_{\text{g}}b^{\lambda} \tau^{\lambda}_{\text{total}} = m_{\text{g}}b^{\lambda} \left(\tau^{\lambda}_{\text{long}} + \tau^{\lambda}_{\text{self}} \right),
\eeq
 and the shear stress component resolved in the $\lambda$-th slip system $\tau^{\lambda}_{\text{long}}$ is calculated from the long-range stress field $\vsi$ by
 Eq.~(\ref{long_range_stress_res}),
  $\tau^{\lambda}_{\text{self}}$ is the contribution due to the local dislocation line tension effect given in Eq.~(\ref{self_stress_res}), and
$g^{\lambda}(\vr)$ and $\tau^{\lambda}_0(\vr)$ are two functions for the Frank-Read source continuum given by Eqs.~\eqref{effective_source_density} and \eqref{effective_activation_stress}, respectively.

\end{enumerate}

In addition, the Nye dislocation density tensor $\boldsymbol{\alpha}$, the total dislocation density in the system $\rho_{\text{tot}}$, and the total plastic strain rate $\dot{\eps}^{\text{p}}_{\text{tot}}$ are expressed in terms of the DDPFs $\phi$ and $\psi$ by Eqs.~(\ref{Nye_density_tensor_formula}), (\ref{total_number_density}), and (\ref{total_plastic_strain}), respectively.

\subsection{Free energy}

In our DDPF-based continuum model, we can write a total free energy of the system as
\begin{equation}\label{free_energy_exp}
\mathcal{E}= \mathcal{E}_{\text{elastic}} + \mathcal{E}_{\text{self}},
 \end{equation}
 where $\mathcal{E}_{\text{elastic}}$ is the elastic energy
 \beq \label{free_energy_exp_elastic}
 \mathcal{E}_{\text{elastic}}= \frac1{2}\int_{\Omega} \left(\nabla\vu+\sum_{\lambda}\frac{\phi^{\lambda}\vb^{\lambda}}{(b^{\lambda})^2 }\otimes \nabla\psi^{\lambda}\right):\left(\CL:\nabla\vu+\sum_{\lambda}\frac{2\mu\phi^{\lambda} }{(b^{\lambda})^2 } \text{sym}(\vb^{\lambda}\otimes \nabla\psi^{\lambda})\right) \d V - \int_{\Omega_{\text{t}}}\vt\cdot\vu \d S
\eeq
and $\mathcal{E}_{\text{self}}$ is the energy due to  the dislocation line tension effect
 \beq \label{free_energy_exp_self}
\mathcal{E}_{\text{self}}=\sum_{\lambda} \int_{\Omega} \frac{\mu |\nabla\psi^{\lambda}\times \nabla\phi^{\lambda}|}{4\pi}\left(1+\frac{\nu}{1-\nu}\frac{\vb^{\lambda} \cdot\nabla\phi^{\lambda}}{(b^{\lambda}|\vm^{\lambda}\times\nabla\phi^{\lambda}|)^2} \log\frac{b^{\lambda}}{2\pi r_c|\vm\times\nabla\phi^{\lambda}|}\right) \d V.
\eeq
 This total free energy in Eq.~(\ref{free_energy_exp}) in its form agrees with existing theories using other representations of dislocation densities \citep[e.g.][]{Nelson1981,Berdichevsky2006,Le_IJP2014} and our continuum dislocation dynamics model in a single slip plane~\citep{Xiang2009_JMPS}.

 Since the DDPFs $\psi^{\lambda}$ for all ${\lambda}$  stay constant during the deformation process,  we
 can write $\mathcal{E}= \mathcal{E} (\vu, \phi^1, \phi^2, \cdots, \phi^{\lambda},\cdots)$.
  It can be calculated that inside $\Omega$, we have
\beq \label{variation_F_u}
\frac{\delta \mathcal{E}}{\delta \vu} = -\nabla \cdot \vsi,
\eeq
and
\beq \label{variation_F_phi}
\frac{\delta \mathcal{E}}{\delta \phi^{\lambda}} = \frac{|\nabla \psi^{\lambda}|}{b^{\lambda}}\left(\tau_{\text{long}}^{\lambda} + \tau_{\text{self}}^{\lambda}\right),
\eeq
   Recall that the resolved shear stress $\tau^{\lambda}_{\text{long}}$ is from the long-range stress field $\vsi$ given by Eq.~(\ref{long_range_stress_res}), and
  $\tau^{\lambda}_{\text{self}}$ is the contribution due to the local dislocation line tension effect given in Eq.~(\ref{self_stress_res}). Note that the variation of $\mathcal{E}_{\text{self}}$ with respect to $\phi^{\lambda}$ gives the correct leading order contribution in $\tau_{\text{self}}^{\lambda}$.

Using these variations of the free energy, the governing equations in our continuum model can be written as
 \begin{eqnarray}
\mathbf{0} &=&\frac{\delta \mathcal{E}}{\delta \vu},\vspace{1ex}\label{eqn_langevin1}\\
\pd{\phi^{\lambda}}{t}& = &-L^{\lambda}\frac{\delta \mathcal{E}}{\delta \phi^{\lambda}} + s^{\lambda}.
\end{eqnarray}
The first equation gives the equilibrium condition in Eq.~\eqref{eqn_force_balance}, which together with the constitutive relation in Eq.~(\ref{eqn_constitutional_law}) and the boundary conditions in Eqs.~(\ref{BC_Dirichlet}) and (\ref{BC_Noemann}) form the elasticity system. (These boundary conditions can also be obtained from the variation of the elastic energy in Eq.~(\ref{free_energy_exp_elastic}).) The second equation describes the plastic flow rule (dynamics of dislocations), where $f_{\phi^{\lambda}}=-\frac{\delta \mathcal{E}}{\delta \phi^{\lambda}}$ can be understood as the configurational force  and
$L^{\lambda} = \frac{m_{\text{g}}(b^{\lambda})^2|\nabla \phi^{\lambda} \times \nabla\psi^{\lambda}|}{|\nabla\psi^{\lambda}|^2}$ is the kinetic coefficient. These equations are consistent with the governing equations in the DDD models (\citet{Hirth} and those references in the introduction).

When the system evolves to its equilibrium state, $\delta \mathcal{E}/\delta \phi^{\lambda} = 0$, which gives rise to the micro force balance state $
\tau_{\text{long}}^{\lambda} + \tau_{\text{self}}^{\lambda} = 0$. This together with Eq.~(\ref{eqn_langevin1}) agree with the models for equilibrium states of distributions of straight dislocations \citep[e.g.][]{Berdichevsky2006,Le_IJP2014}.

\section{Numerical implementation of the continuum model}\label{sec:implementaion}

In this section, we briefly discuss the numerical implementation of our continuum model as summarized in Sec.~\ref{Sec_summary_constitutive_eqns}. We will focus on
  solving for the long-range stress field $\vsi$ from Eqs.~\eqref{eqn_stress_tensor1} and \eqref{eqn_stress_tensor2} with boundary conditions and the plastic flow rule described by the evolution of DDPF $\phi^{\lambda}$ in Eq.~\eqref{eqn_evolution_phi}.

In the simulations in this paper,  the computational domain is chosen to be a cuboid  $
\Omega = [-D/2,D/2]\times[-D/2,D/2]\times[-L/2,L/2]$. The boundary conditions are imposed as follows. On the bottom surface, no displacement is allowed along the $z$ direction, which is the loading direction, and on the top surface $\left.u_3\right|_{z=L/2} = u_0^{\text{b}}(t)$ is imposed as a result of compression. The shear force is set to be free on these two surfaces. On the other four side surfaces, traction free boundary conditions are imposed. The DDPF $\psi^{\lambda}$ in this paper takes the form in Eq.~(\ref{psi_def_no_cross_slip}) with inter-slip plane distance $d_{\text{sl}}=100b$, and does not change in the plastic deformation process.

\subsection{Finite element formulation for solving for the long-range stress field}

The long-range stress field satisfying  Eqs.~\eqref{eqn_stress_tensor1} and \eqref{eqn_stress_tensor2}
subject to boundary conditions in Eqs.~(\ref{BC_Dirichlet}) and (\ref{BC_Noemann}) yields a weak form that $
\int_{\Omega} \vsi_{\text{long}}:(\nabla \vv) \d V = \int_{\partial \Omega_{\text{t}}} \vt^{\text{b}} \cdot \vv \d S$
for any test vector function $\vv\in\{\vv|\vv=\boldsymbol{0},\text{ on } \partial \Omega_{\text{d}}\}$. Replacing the stress field by the constitutive stress rule given by Eq.~\eqref{eqn_stress_tensor1}, we obtain the weak form for the displacement field $\vu$
\beq \label{weak_form_displacement}
\int_{\Omega} \nabla \vv:\boldsymbol{\CL}:\nabla \vu \d V = \int_{\partial \Omega_{\text{t}}} \vt \cdot \vv\d S - 2\mu \sum_{\lambda}\int_{\Omega} \phi^{\lambda}\text{sym}(\vb^{\lambda}\otimes\nabla\psi^{\lambda}):(\nabla\vv) \d V,
\eeq
where  the forth-order tensor $\boldsymbol{\CL}$ is defined in Eq.~(\ref{L_operator_def}).

In our simulations, $\Omega$ is meshed by C3D8 bricks. We then discretize the weak form in Eq.~\eqref{weak_form_displacement} to get a linear algebraic equation system as
\beq \label{linear_system_u_nodal}
K_{\text{FE}}\vu_{\text{FE}} = \textbf{f}_{\text{FE}},
\eeq
where $\vu_{\text{FE}}$ is a vector of $3N$ dimensions containing all nodal values of $\vu$ with $N$ being the total number of nodes, $K_{\text{FE}}$ is known as the stiffness matrix, and $\textbf{f}_{\text{FE}}$ is assembled by discretizing the right hand side of Eq.~\eqref{weak_form_displacement}.

If the last term is omitted, Eq.~\eqref{weak_form_displacement} is the weak formulation that is widely used in the FE formulation in  linear elasticity. Hence many tools well-developed for solving purely elastic problems, such as meshing, assembling and inversion of $K_{\text{FE}}$, can be inherited by the FE formulation proposed here. The contribution from $\phi^{\lambda}$ and $\psi^{\lambda}$ in Eq.~\eqref{linear_system_u_nodal} can be treated the same way as a body force to the linear system.

\subsection{Finite difference scheme for the evolution of DDPF $\phi^{\lambda}$}

We discretize the evolution equation of DDPF $\phi^{\lambda}$ in Eq.~\eqref{eqn_evolution_phi} using a
 finite difference scheme. Here the grid points of $\phi^{\lambda}$ are chosen coinciding with the vertices of the C3D8 bricks. Combined with the mobility law in Eq.~(\ref{eqn_mobility_law}), Eq.~\eqref{eqn_evolution_phi} can be written as
\beq\label{eqn_phi_convection_term}
\pd{\phi^{\lambda}}{t} +  m_{\text{g}}b^{\lambda} \left(\tau^{\lambda}_{\text{long}} + \tau^{\lambda}_{\text{self}} \right)|\vm^{\lambda}\times\nabla \phi^{\lambda}| = s^{\lambda}.
\eeq
The temporal derivative of $\phi^{\lambda}$  is approximated by  the forward Euler scheme.
For the spatial derivatives of $\phi^{\lambda}$, we follow the methods in \citet{Xiang2003_Acta} that
the first-order upwind scheme is used for those in the term associated with the long-range stress field $\tau_{\text{long}}^{\lambda}|\vm^{\lambda}\times\nabla \phi^{\lambda}|$,
and the
central difference scheme is used to calculate those spatial derivatives in the term $\tau_{\text{self}}^{\lambda}|\vm^{\lambda}\times\nabla \phi^{\lambda}|$ where $\tau_{\text{self}}^{\lambda}$ is given by Eq.~(\ref{self_stress_res}).

Moreover, the regularized $\delta$-function in the source term $s^{\lambda}$ (Eqs.~(\ref{source_term_con_exp})-(\ref{effective_activation_stress})) is given by
\beq\label{delta_regularised}
\delta_{\text{reg}}(\vr) = \frac1{\Delta s_1^2}\cdot\frac{\pi}{\pi^2-4}\left(\cos\frac{\pi |\vm^{\lambda}\times\vr|}{\Delta s_1}+1\right) \cdot \frac1{2\Delta s_2}\left(\cos\frac{\pi(\vm^{\lambda}\cdot\vr)}{\Delta s_2} +1\right)
\eeq
for all $\vr\in\{\vr||\vm^{\lambda}\times\vr|<\Delta s_1,\,|\vm^{\lambda}\cdot\vr|<\Delta s_2\}$,
where $\Delta s_1$ and $\Delta s_2$ are two smoothing parameters.

\section{Numerical examples}

In this section, the derived continuum model is validated through comparisons with DDD simulations. All results presented are obtained by using $10\times10\times20$ C3D8 bricks in the finite element discretization of the cuboid simulation domain described at the beginning of Sec.~\ref{sec:implementaion}.

\subsection{A single Frank-Read source under a constant applied strain}
This example is aimed to provide an illustration of the continuum model.
For the cuboid simulation cell $\Omega$, $L = 24000b$ and $D = 9600b$.
A Frank-Read source of length $l=400b$ with activation stress $7.8\times10^{-4}\mu$ is placed at the center of  $\Omega$, and a $0.3\%$ constant strain is applied by compression on the top surface. All simulations start with a dislocation-free state.

\begin{figure}[!ht]
\centering
\subfigure[$t = \frac{100L}{m_{\text{g}}\mu  b}$]{\includegraphics[width=.24\textwidth]{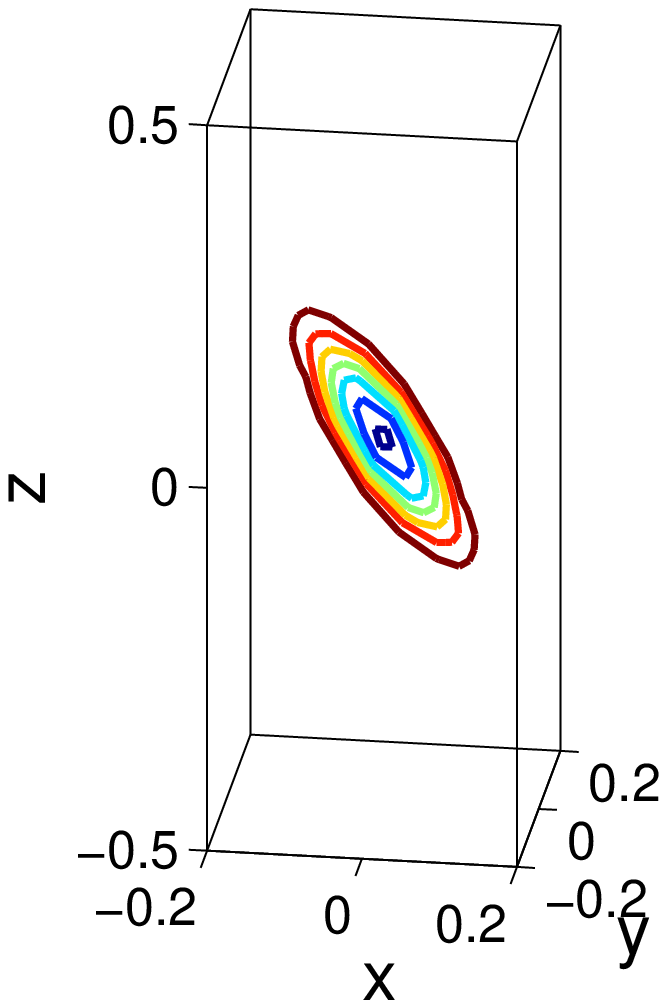}}
\subfigure[$t = \frac{1000L}{m_{\text{g}}\mu  b}$]{\includegraphics[width=.24\textwidth]{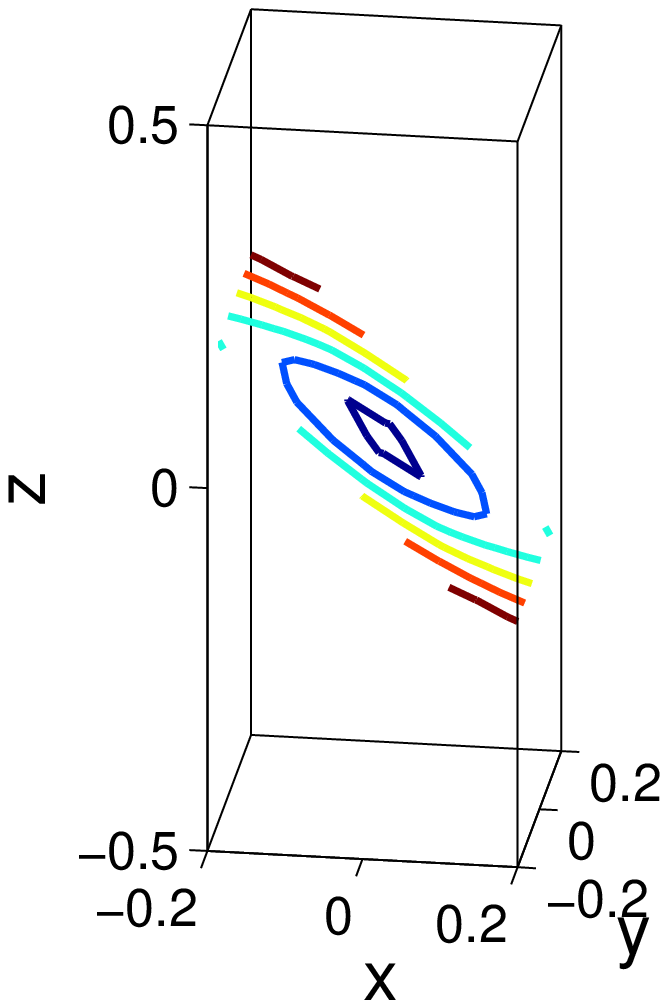}}
\subfigure[$t = \frac{4000L}{m_{\text{g} }\mu b}$]{\includegraphics[width=.24\textwidth]{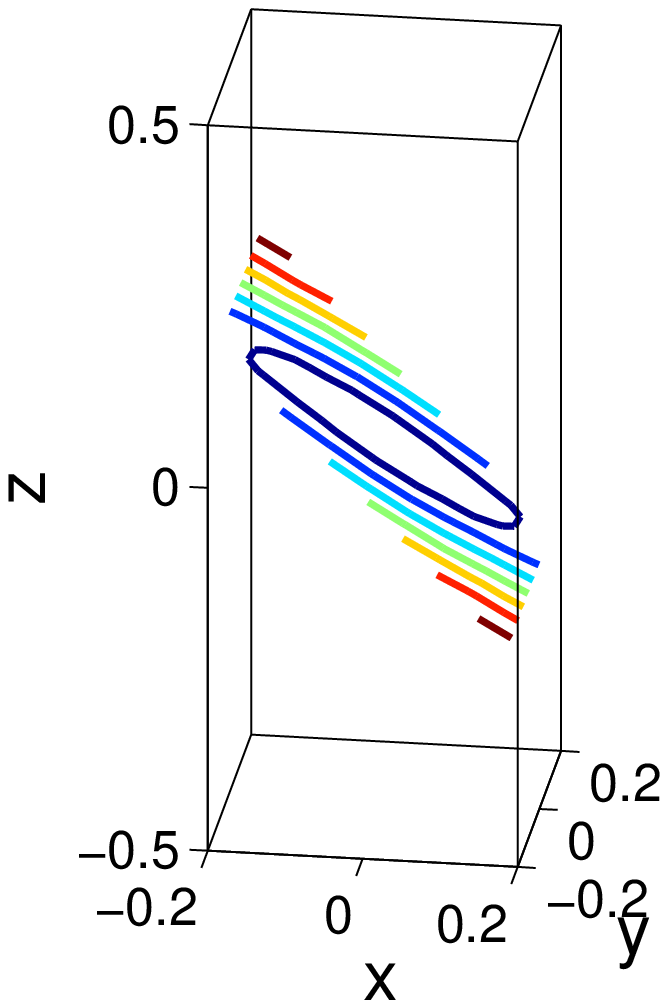}}
\subfigure[$t = \frac{40000L}{m_{\text{g}}\mu  b}$]{\includegraphics[width=.24\textwidth]{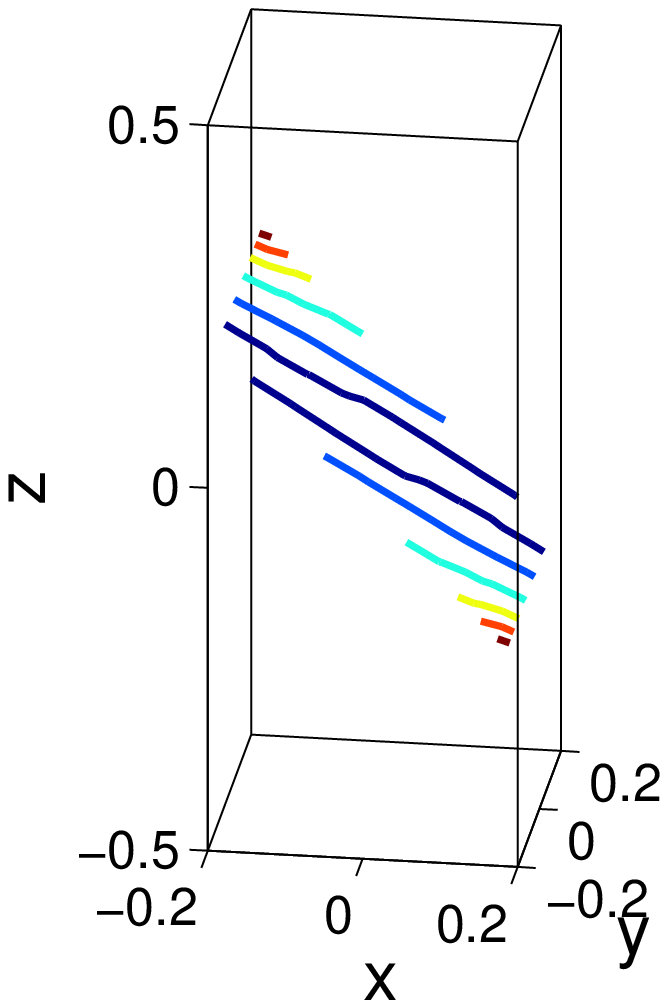}}
\caption{Snap shots of the distribution of dislocation curves generated by an operating Frank-Read source that locates at the center of the simulation domain $\Omega$.   The source  releases dislocation loops in response to a constant strain by compression on the top surface of $\Omega$. The length unit of $\Omega$ is its height $L$. \label{Fig_phi_single_source}}
\end{figure}

\begin{figure}[!ht]
\centering
\includegraphics[width=.5\textwidth]{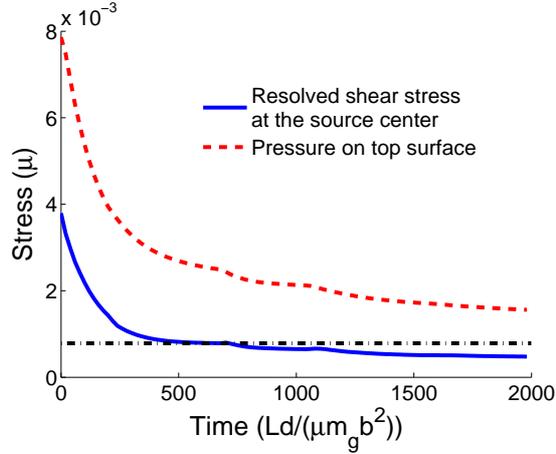}
\caption{As the Frank-Read source keeps releasing dislocation loops, both the resolved shear stress at the source and the pressure on the top surface of the simulation domain drop. When the time $t$ is about $14000L/(m_{\text{g}}\mu b)$, the resolved shear stress falls below the source activation stress (indicated by the dashed-dotted line), and the source is thus deactivated. \label{Fig_traction_resolved_single_source}}
\end{figure}

In Fig.~\ref{Fig_phi_single_source}, we plot the contour lines of the DDPF $\phi$ on one of the slip planes, which give rough locations of the dislocation curves.
It is observed that in response to the applied strain, the source keeps releasing dislocation loops, which exit $\Omega$ from its side surfaces.
As a result, the resolved shear stress drops during this loop-releasing process, and so does the pressure on the top surface as shown in Fig.~\ref{Fig_traction_resolved_single_source}. These findings agree with the common understanding about the role played by a Frank-Read source: It releases dislocation loops so as to soften the materials. When the time $t$ is about $14000L/(m_{\text{g}}\mu b)$, the resolved shear stress finally falls below the source activation stress, the source is then deactivated, see Fig.~\ref{Fig_traction_resolved_single_source}.

\begin{figure}[!ht]
\centering
\subfigure[$u_1$]{\includegraphics[width=.32\textwidth]{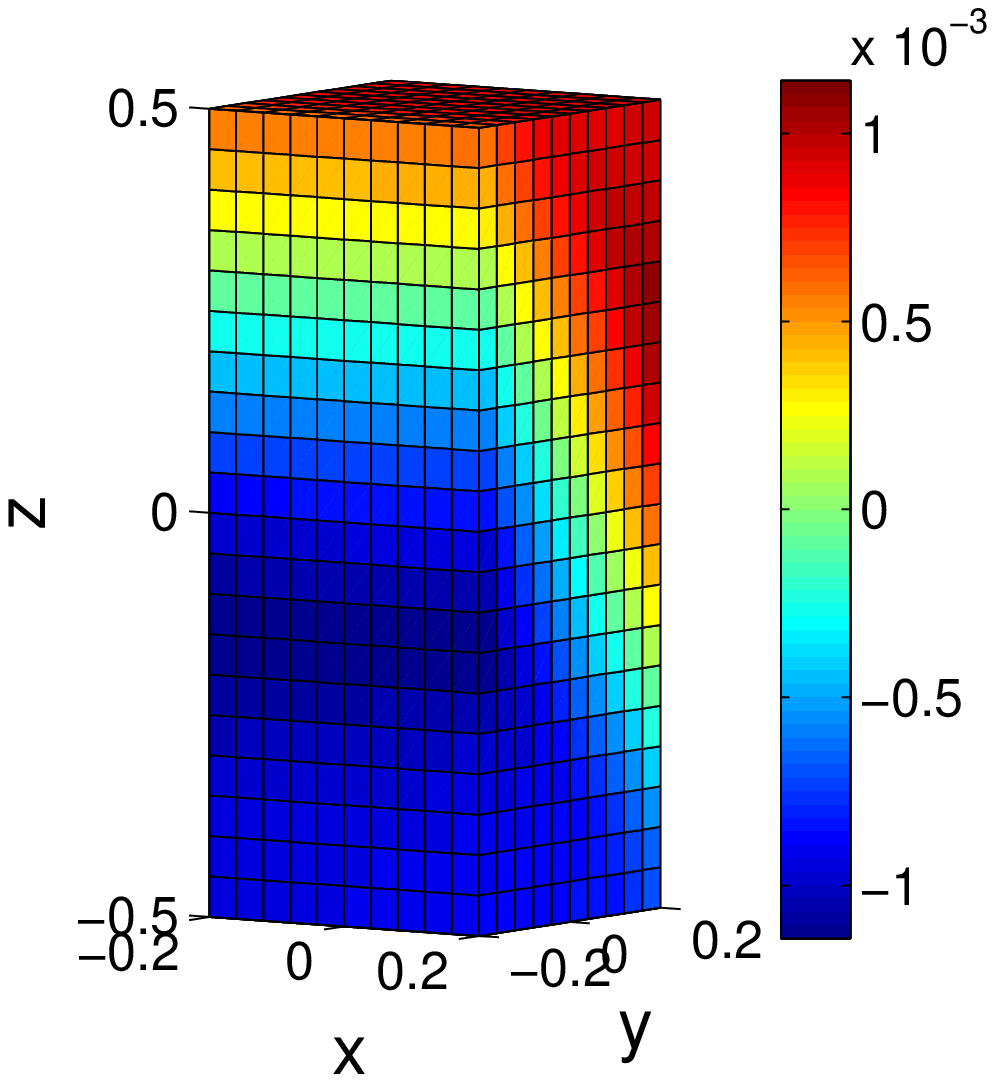}}
\subfigure[$u_2$]{\includegraphics[width=.32\textwidth]{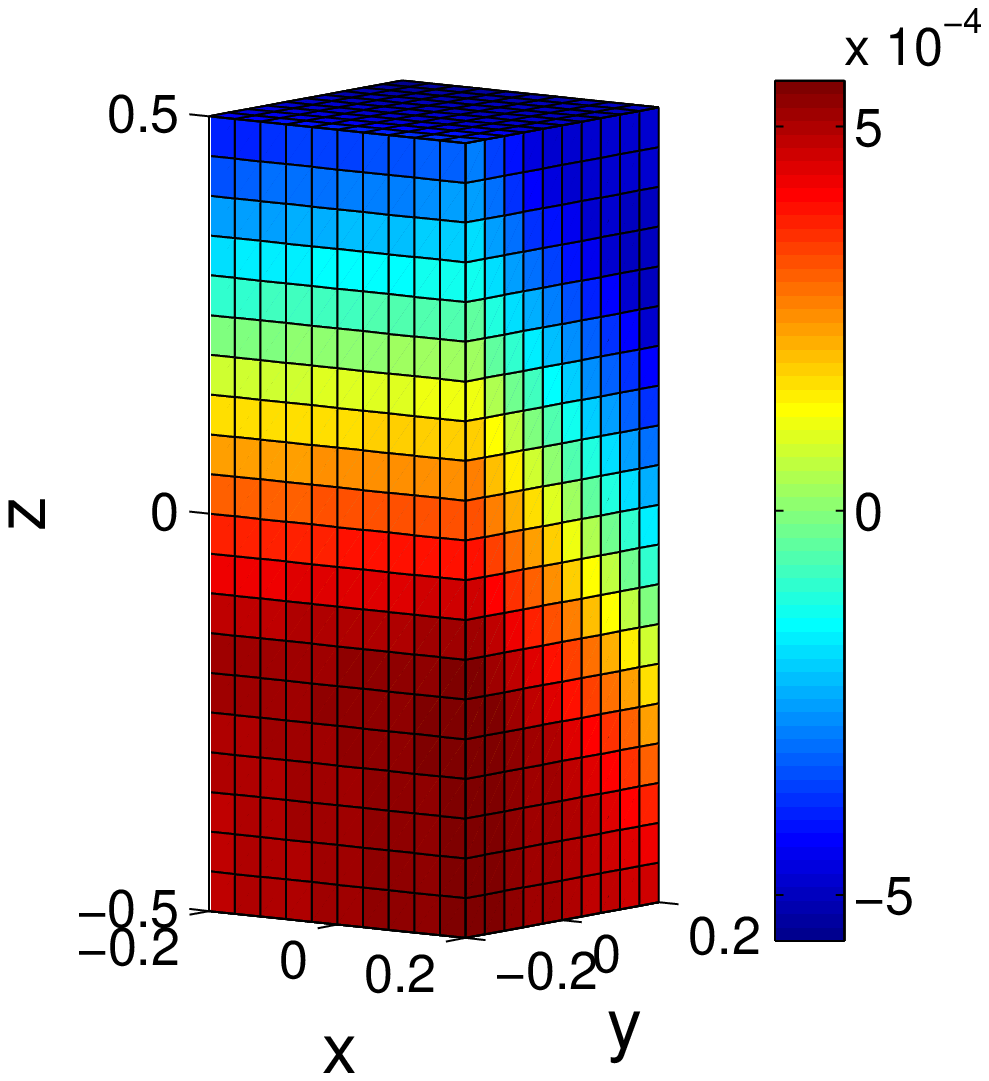}}
\subfigure[$u_3$]{\includegraphics[width=.32\textwidth]{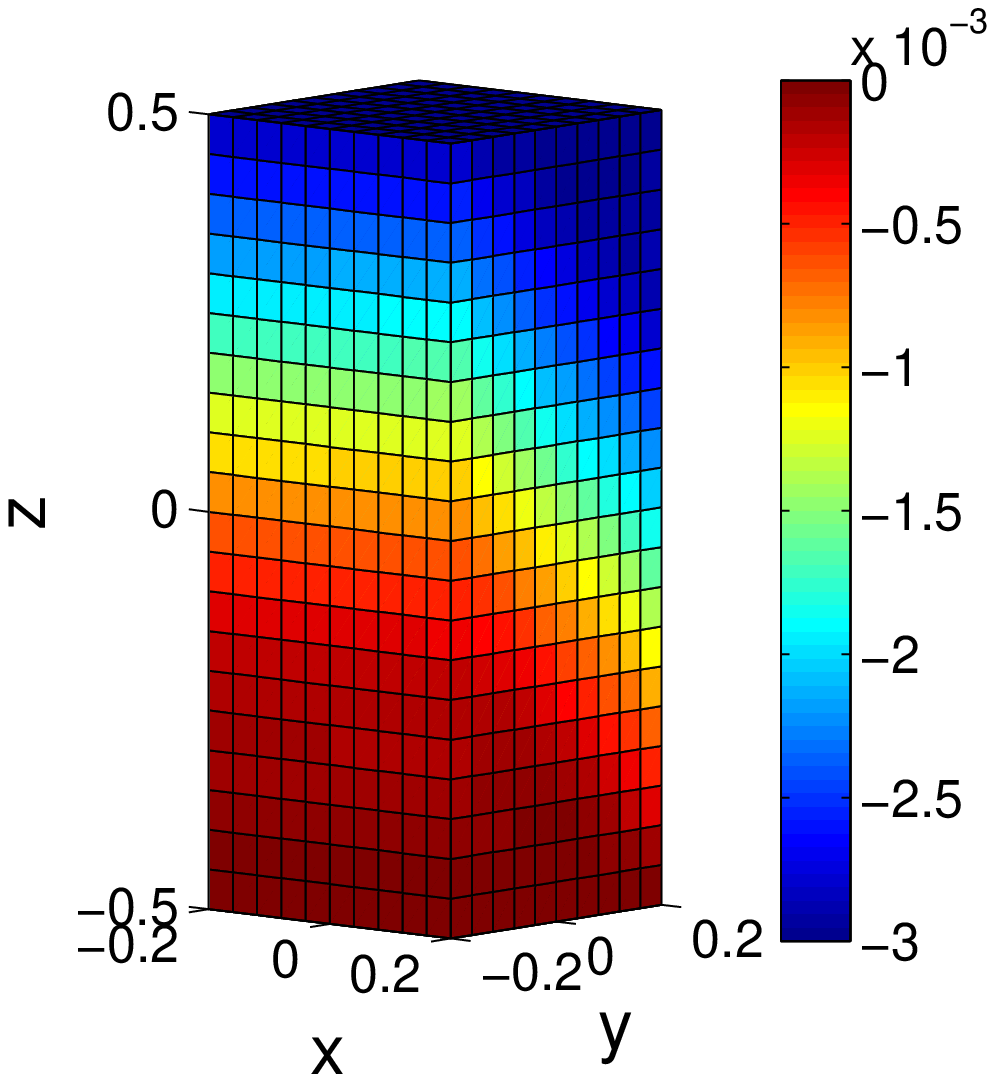}}
\caption{The three components of the displacement vector $\vu$ on the simulation cell surface $\partial \Omega$.  The length unit is $L$, the height of $\Omega$. \label{Fig_displacment_surf}}
\end{figure}

The displacement vector $\vu$ on the domain boundary $\partial \Omega$ is shown in Fig.~\ref{Fig_displacment_surf}. It can be seen that the orientations of the displacement gradients are
in the normal direction of the activated slip plane that contains the Frank-Read source, and they are
localized near this slip plane.
This is because the dislocation loops that have left the specimen form surface steps on the activated slip plane (in a smooth sense).

\subsection{Comparison with DDD simulations \label{Sec_ElAwady}}
To further validate the continuum model, we compare its numerical results with the DDD simulation results obtained by \citet{ElAwady2008}. Following their DDD simulations, we choose a micro-pillar of nickel.
 The loading axis is $<001>$, and a single slip system is activated with slip direction $[0\bar{1}1]$ and slip normal $(111)$. The Schmid's factor $m_{\text{s}}$ is thus 0.4050. The shear modulus is $76$GPa, the Poisson's ratio is $0.31$,  and the length of the Burgers vector is $b=0.25$nm. The strain rate is 200s$^{-1}$. The dislocation glide mobility $m_{\text{g}}$ is unspecified by \citet{ElAwady2008}, we here follow \citet{Weygand2008} to let $m_{\text{g}} = 10^{4}/$(Pa$\cdot$s). In our simulations, the micro-pillars are chosen to be cuboids as described at the beginning of Sec.~\ref{sec:implementaion} for the ease of the finite element implementation. The sample sizes here are defined to be the size of the cuboid base $D$. The (height to base size) aspect ratio of the micro-pillars is $L/D=3$. We choose $D=0.5\mu$m or $1\mu$m.

 \begin{table}[!ht]
\centering
\begin{tabular}{l|lllll}
Sample & Mean source & Standard & Max source & $\tau_{\min}^0$ & Flow stress \\
&length ($\mu$m) & deviation ($\mu$m) &  length ($\mu$m) & (MPa) & (MPa) \\
\hline
1 & 0.1835 & 0.1104 & 0.4832 & 170.7 & 420.0 \\
2 & 0.2403 & 0.0935 & 0.4423 & 132.8 & 331.5 \\
3 & 0.2039 & 0.0992 & 0.3340 & 145.5 & 357.9 \\
4 & 0.1928 & 0.1083 & 0.3711 & 158.9 & 388.8 \\
\end{tabular}
\caption{Statistics of the individual sources whose lengths are obtained randomly following a uniform distribution within $[$20nm, $D]$ for samples of size $D = 0.5\mu$m.  The locations of the sources are also determined randomly following the uniform distribution over the micro-pillar. \label{information_source_05mum_DDD}}
\end{table}
\begin{table}[!ht]
\centering
\begin{tabular}{l|lllll}
Sample & Mean source & Standard & Max source & $\tau_{\min}^0$ & Flow stress \\
&length ($\mu$m) & deviation ($\mu$m) &  length ($\mu$m) & (MPa) & (MPa) \\
\hline
1 & 0.4066 & 0.2252 & 0.9196 & 83.8 & 196.4 \\
2 & 0.4047 & 0.2412 & 0.9995 & 91.0 & 230.2 \\
3 & 0.4596 & 0.2333 & 0.9605 & 80.7 & 204.0 \\
4 & 0.4709 & 0.2275 & 0.9349 & 80.3 & 201.2 \\
\end{tabular}
\caption{Same as Table~\ref{information_source_05mum_DDD} for samples of size $D = 1\mu$m. \label{information_source_1mum_DDD}}
\end{table}

In our simulations, the lengths of the Frank-Read sources are generated randomly following a uniform distribution within [20nm, $D$]. The initial dislocation density $\rho_0$ is also randomly generated within the range $1.6\sim 4\times10^{12}$m$^{-2}$, and these pre-existing dislocation segments of the sources are uniformly assigned to the twelve slip systems of fcc nickel. The statistics of the initial source distributions for various samples are listed in Tables~\ref{information_source_05mum_DDD} and \ref{information_source_1mum_DDD}.
With these isolated Frank-Read sources generated, the corresponding source continuum is developed following Eqs.~\eqref{source_term_con_exp}-\eqref{effective_activation_stress}.
For the source character parameter $C_{\text{s}}$ in the critical stress formula in Eq.~\eqref{activation_stress_single_source}, we choose it to be $1.35$ and $2.02$ for a single edge source and a screw source, respectively, to agree with the simulation set-up in \citet{ElAwady2008}.

In Tables~\ref{information_source_05mum_DDD} and \ref{information_source_1mum_DDD}, we also show the values of $\tau_{\min}^0$, which is defined as the minimum value of the activation stress $\tau_0(\vr)$ of the Frank-Read source continuum given by Eq.~\eqref{effective_activation_stress}, over the points where $g(\vr)\neq0$ inside the pillar. The minimum  activation stress $\tau_{\min}^0$ will be used in the derivation of the scaling law in the next section.

\begin{figure}[!ht]
\centering
\subfigure[$D=0.5\mu$m]{ \includegraphics[width=.48\textwidth]{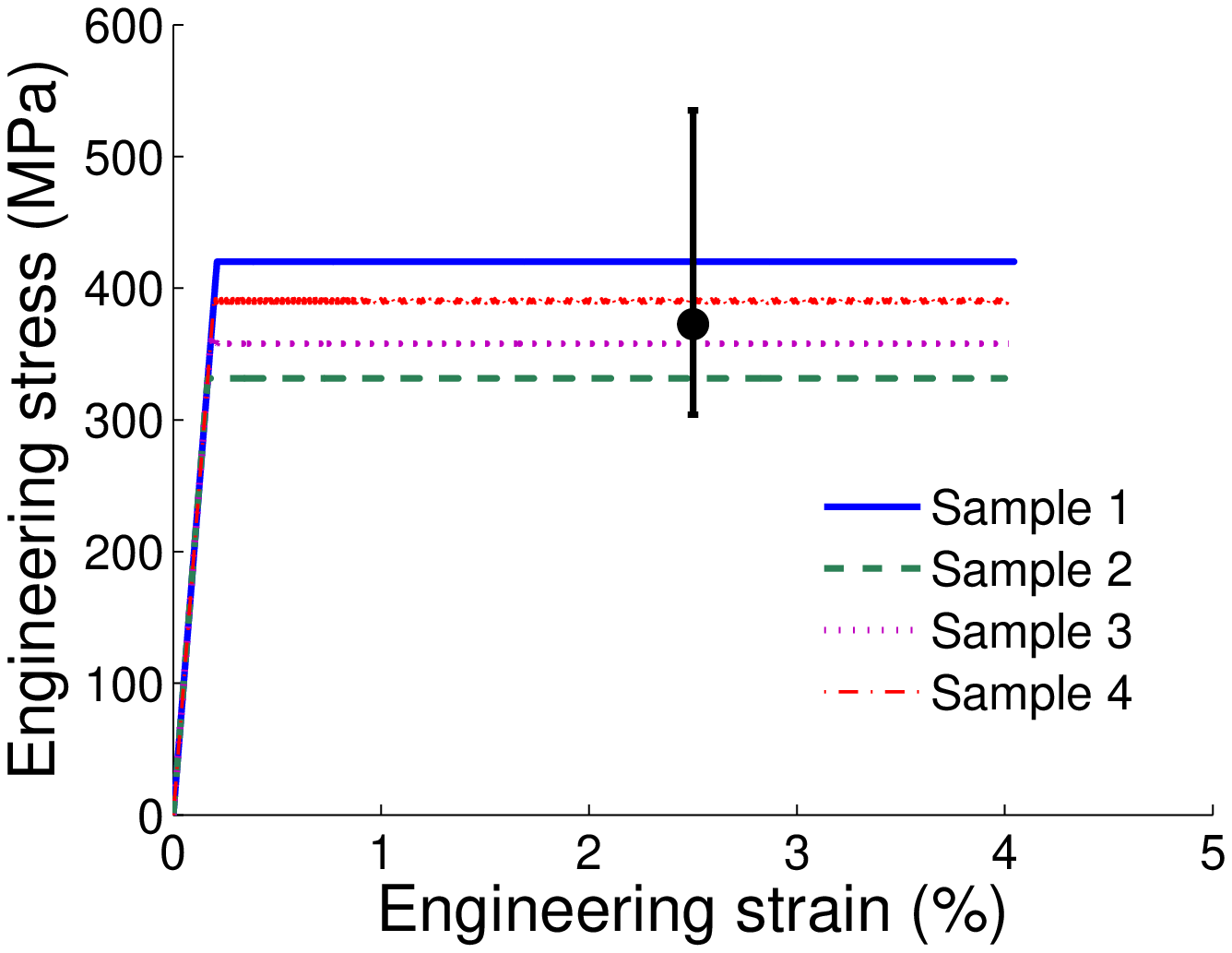}}
\subfigure[$D=1\mu$m]{ \includegraphics[width=.48\textwidth]{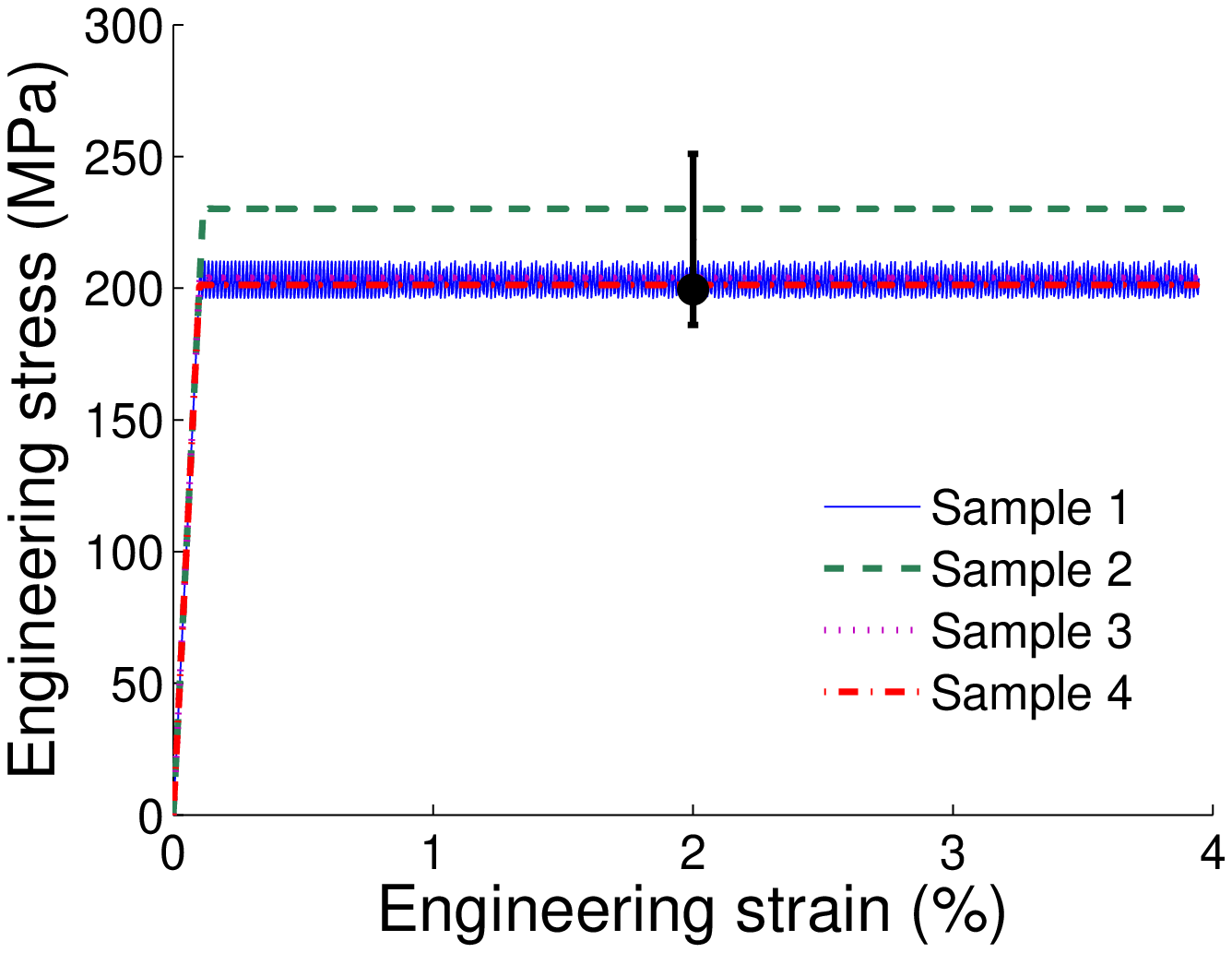}}
\caption{Stress-strain curves obtained by the simulations of our continuum model. The vertical bars denote the ranges of the flow stress predicted by the DDD simulations of \citet{ElAwady2008}, whose averaged values are shown by the black dots. \label{Fig_ElAwardy_ss_curve}}
\end{figure}

The stress-strain curves obtained by our continuum model are shown in Fig.~\ref{Fig_ElAwardy_ss_curve} for samples of sizes $D=0.5\mu$m and  $D=1\mu$m.
Good agreement with the DDD results by \citet{ElAwady2008} are observed. Firstly, both methods predict an initially elastic regime and an almost perfectly plastic regime, where work-hardening effect is barely observed. Secondly, both methods suggest that the engineering stress stays roughly unchanged or oscillate around some constant value in the regime of perfect plastic deformation, and this stress is measured as the flow stress of the micro-pillars.
The ``smaller-being-stronger'' size effect on crystalline strength, which is indicated by the flow stress, is observed in both our simulations and in the DDD simulations of \citet{ElAwady2008}. Moreover,  statistical effects in the flow stress are seen in the simulations of both methods. Such statistical effects have also been examined in other literature \citep[e.g.][]{ElAwady2009, Zbib2014}.
To make quantitative comparisons between the results of our model and those of the DDD simulations of \citet{ElAwady2008}, the respective ranges and averaged values for the flow stress recorded by \citet{ElAwady2008} are drawn in Fig.~\ref{Fig_ElAwardy_ss_curve}. It can be seen that the flow stresses calculated by our continuum model all fall into the respective ranges predicted by their DDD simulations.

These comparisons show that our continuum model is able to provide an excellent summary of the corresponding underlying dynamics of discrete dislocations. In the next section, we will use it to  quantitatively study the size effect on strength observed in the uniaxial compression tests of micro-pillars.

\section{Size effect on strength of single-crystalline micro-pillars}

\subsection{Comparison with the experimental data}

Now we investigate the ``smaller-being-stronger'' size effect in micro-pillars using our continuum model. For the initialization of the Frank-Read sources, we follow \citet{Shishvan2010} that, in analogy to the distribution of grain sizes in polycrystals which has been experimentally measured, the Frank-Read source size $l$ follows a log-normal distribution with the probability density function
\beq \label{log_normal_pdf}
p(l)=\frac1{\sqrt{2\pi}\sigma_{\text{sd}}l} e^{-\frac{(\log l-\log l_{\text{m}})^2}{\sqrt{2}\sigma_{\text{sd}}^2}}
\eeq
with two parameters $l_{\text{m}}$ and $\sigma_{\text{sd}}$ to be determined. The parameter $l_{\text{m}}$ which can be considered as the effective mean source length should decrease with the pillar size $D$. The locations of these source segments follow the uniform random distribution over the micro-pillar. The single-ended sources \citep{Parthasarathy2007} are included, and this happens when part of the source segments are outside the pillar.

\begin{figure}
\centering
\subfigure[$D=1\mu$m]{ \includegraphics[width=.4\textwidth]{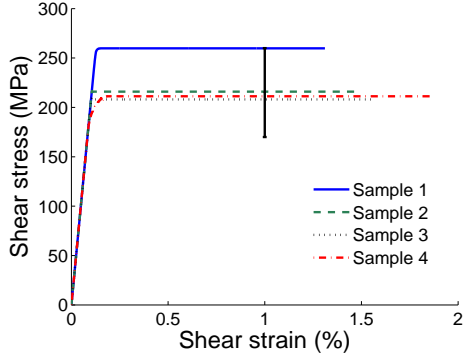}}
\subfigure[$D=2.4\mu$m]{ \includegraphics[width=.4\textwidth]{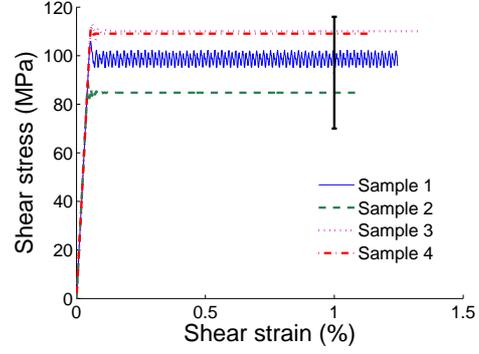}}
\subfigure[$D=5\mu$m]{ \includegraphics[width=.4\textwidth]{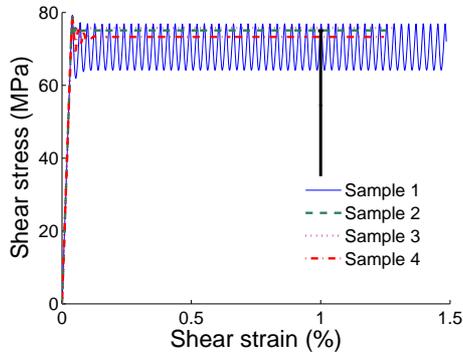}}
\subfigure[$D=10\mu$m]{ \includegraphics[width=.4\textwidth]{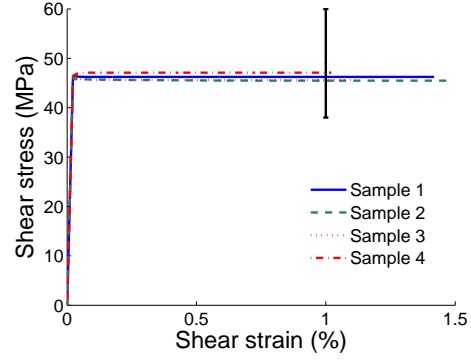}}
\subfigure[$D=20\mu$m]{ \includegraphics[width=.4\textwidth]{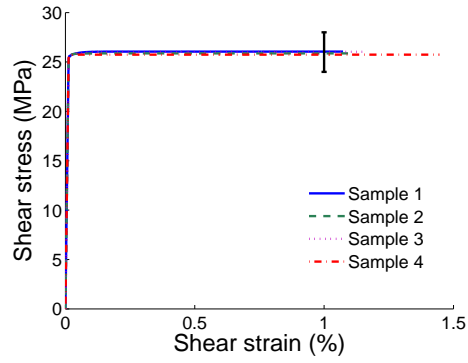}}
\caption{Stress-strain curves obtained by simulations using our continuum model with initial arrangement of Frank-Read sources whose length distribution following the log-normal one given by Eq.~\eqref{log_normal_pdf}. Four simulations are performed for each size of the micro-pillar $D$ that ranges from $1\mu$m to $20\mu$m. The vertical bars identify the ranges of the experimentally measured flow stress  by \citet{Dimiduk2005}. \label{Fig_ss_curve_comp_exp}}
\end{figure}

To validate the initial source distribution in Eq.~(\ref{log_normal_pdf}) and determine the parameters, the numerical results by using our continuum model are compared with the experimental data of nickel micro-pillars by \citet{Dimiduk2005}. In our simulations, most parameters are chosen the same as those used in Sec.~\ref{Sec_ElAwady} with the following exceptions in accordance with \citet{Dimiduk2005}: the loading axis is set along $[269]$ and the singly active slip system is of the slip direction $[101]$ and slip normal $(\bar{1}11)$, the Schmid factor is $m_{\text{s}} = 0.48$, the shear modulus is $78$GPa, and the aspect ratio of the pillar ($L/D$) is chosen randomly between 2 and 3. The total density of all source segments is $3\times10^{-12}$m$^{-2}$ following \citet{Dimiduk2005}.

To fit their experimental results, we choose the effective mean source length
  $l_{\text{m}}=\alpha_m D$, where $\alpha_m=1/15$. The standard deviation $\sigma_{\text{sd}}$ is determined under the assumption that the probability of a source segment that is longer than $D$ is no more than $10^{-7}$,  which gives $\sigma_{\text{sd}}=0.4$.
The computed stress-strain curves with these values of parameters are shown in Fig.~\ref{Fig_ss_curve_comp_exp} for different samples varying in size. It can be seen that the obtained flow stresses are in good agreement with the experimental data of \citet{Dimiduk2005}.
 The size effect on strength is also clearly observed from our simulation results.

\subsection{Scaling law of the size effect on micro-pillar strength\label{Sec_size_effect}}

To describe how the pillar strength depends on its size $D$, we propose a formula
\beq \label{flow_stress_sample_size}
\sigma_{\text{flow}} = \frac{\mu b}{2\pi  m_{\text{s}} \alpha_{\text{eff}} D}  \log\left(\frac{\alpha_{\text{eff}} D}{r_{\text{c}}}\right) + \sigma_{\text{flow}}^0.
\eeq
where $\sigma_{\text{flow}}^0$ is some constant stress, $m_{\text{s}}$ is recalled to be the Schmid factor, and $\alpha_{\text{eff}}$ is a dimensionless constant such that $\alpha_{\text{eff}} D$ measures the length of the weakest source in pillar of size $D$. Eq.~\eqref{flow_stress_sample_size} suggests a scaling law
\beq \label{flow_stress_scale_sample_size}
\sigma_{\text{flow}} \sim \frac{b}{D}\log\left(\frac{D}{b}\right).
\eeq
A comparison of Eq.~\eqref{flow_stress_sample_size} with experimental data (to be shown at the end of this subsection) suggests that Eq.~\eqref{flow_stress_sample_size} pocesses a wider effective zone, that is, it is valid for many (at least three) types of f.c.c. pillars of size ranging from submicrons to tens of microns. We now rationalize Eq.~\eqref{flow_stress_sample_size} by taking the following two steps.

\begin{figure}[!ht]
\centering
\subfigure[]{\includegraphics[width=.48\textwidth]{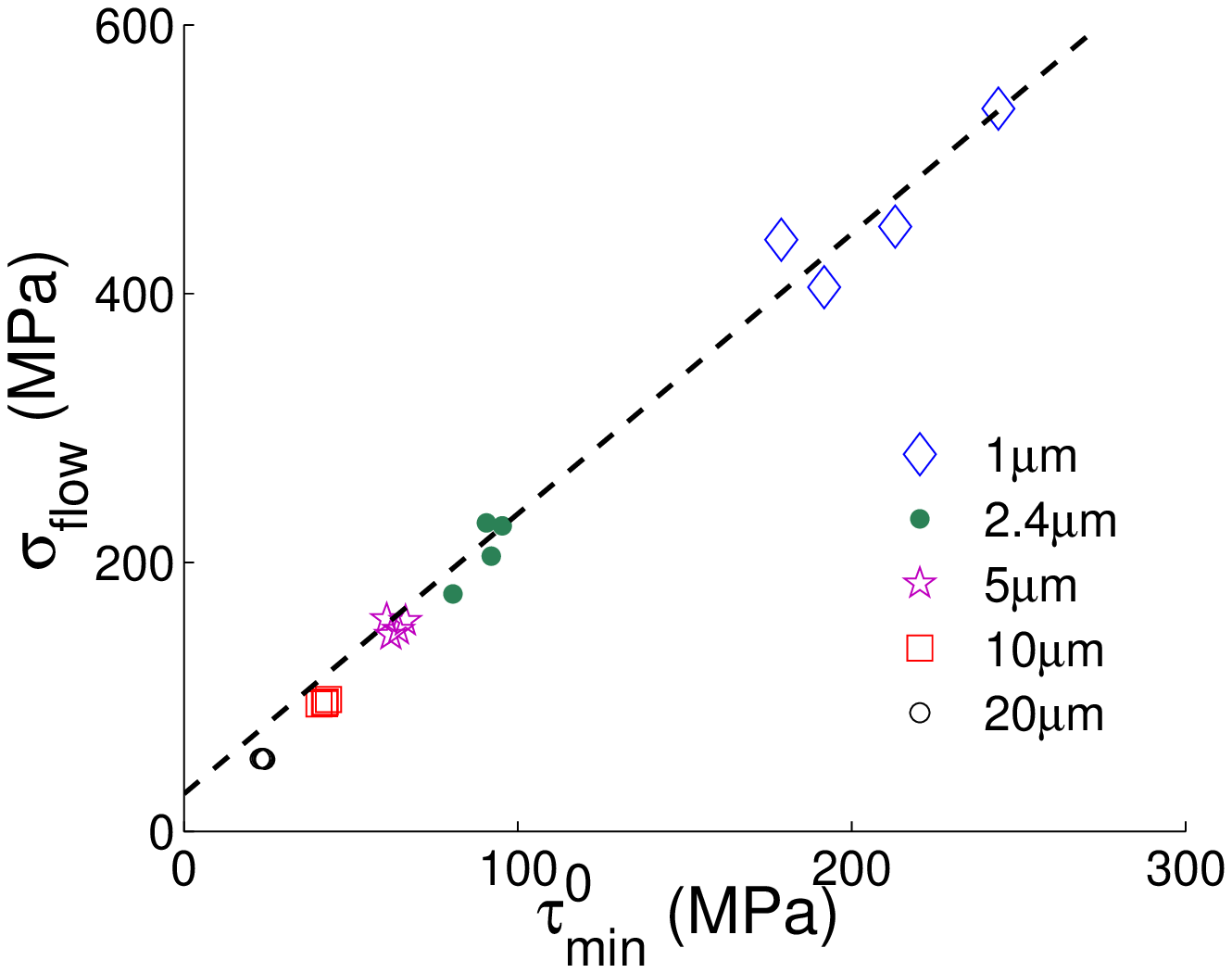}}
\subfigure[]{\includegraphics[width=.48\textwidth]{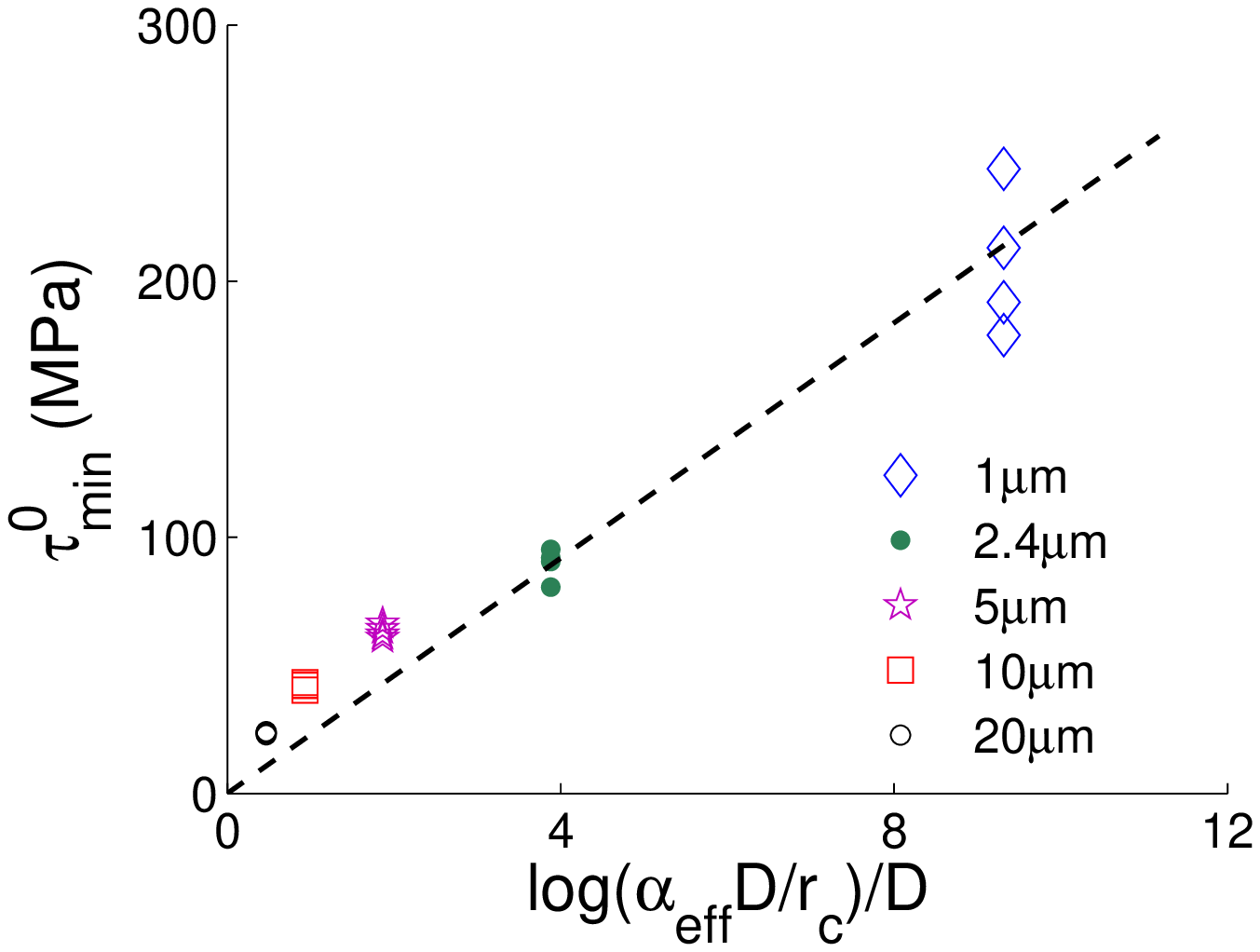}}
\caption{(a) Linear dependence between the flow stress $\sigma_{\text{flow}}$ and the minimum activation stress of the source continuum $\tau^0_{\min}$  given in Eq.~\eqref{flow_tau0_exp} (the dashed line), with  fitted parameter $\sigma_{\text{flow}}^0$. (b) The dependence of the minimum activation stress $\tau_{\min}^0$ on the sample size $D$ given in Eq.~\eqref{tau_min_0_exp} (the dashed line), where
$\alpha_{\text{eff}}=1/6$ and $r_{\text{c}}=0.6b$. These relations are validated by  numerical results obtained using the continuum model for pillars with different sizes (specified in the legend) shown by the dots in (a) and (b).  \label{Fig_sic0_bridge}}
\end{figure}

We first show that the flow stress is related to the minimum  activation stress $\tau_{\min}^0$ by
\beq \label{flow_tau0_exp}
\sigma_{\text{flow}} = \frac{\tau_{\min}^0}{m_{\text{s}}} + \sigma_{\text{flow}}^0,
\eeq
where $\sigma_{\text{flow}}^0$ is some constant stress, and recall that $m_{\text{s}}$ is the Schmid factor.
This means that the flow stress is determined by the weakest Frank-Read source inside the pillar. Comparisons of the results of this linear relation and the computed flow stresses  by using the continuum model for pillars with different sizes are shown in Fig.~\ref{Fig_sic0_bridge}(a), in which the parameter $\sigma_{\text{flow}}^0$ is fitted from the numerical results. The excellent agreement seen in these comparisons demonstrates that Eq.~\eqref{flow_tau0_exp} provides a good quantitative description for the flow stress $\sigma_{\text{flow}}$ in terms of $\tau_{\min}^0$.

The next step is to relate $\tau_{\min}^0$ to the sample size $D$. We follow the formula of the activation stress of a single Frank-Read source to assume
\beq \label{tau_min_0_exp0}
\tau_{\min}^0 = \frac{C_{\text{s}}\mu b}{2\pi l_{\text{eff}}} \log\left(\frac{l_{\text{eff}}}{r_{\text{c}}}\right),
\eeq
where $C_{\text{s}}$ is dependent on the source character chosen to be 1, $l_{\text{eff}}$ can be considered as an effective source length and it is assumed to be a fraction of $D$.  Using simulation results of the continuum model for pillars with different sizes, fitting $\tau_{\min}^0$ against $l_{\text{eff}}$ with reference to Eq.~\eqref{tau_min_0_exp0} gives that $l_{\text{eff}}=\alpha_{\text{eff}} D$, where the parameter $\alpha_{\text{eff}}=1/6$.
Physically, this means that given a micro-pillar of size $D$, the longest source length is most likely $\alpha_{\text{eff}}D$.
Hence $\tau_{\min}^0$ is related to $D$ by
\beq \label{tau_min_0_exp}
\tau_{\min}^0 = \frac {\mu b}{2\pi \alpha_{\text{eff}} D} \log\left(\frac{\alpha_{\text{eff}}D}{ r_{\text{c}}}\right).
\eeq
This relation is validated by  numerical results obtained using the continuum model for pillars with different sizes as shown in Fig.~\ref{Fig_sic0_bridge}(b).

\begin{figure}[!ht]
\centering
\subfigure[Nickel]{\includegraphics[width=.4\textwidth]{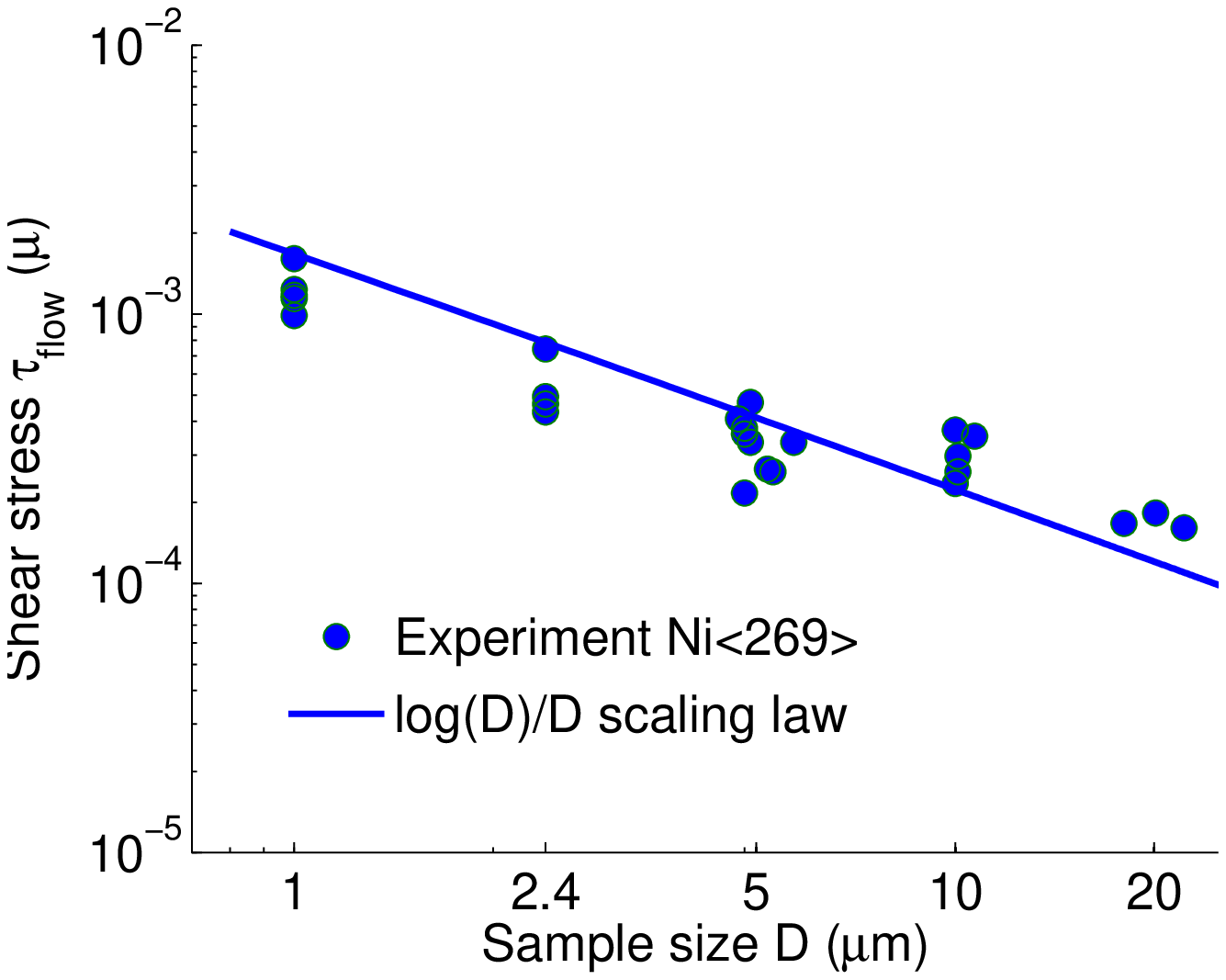}}
\subfigure[Aluminium]{\includegraphics[width=.4\textwidth]{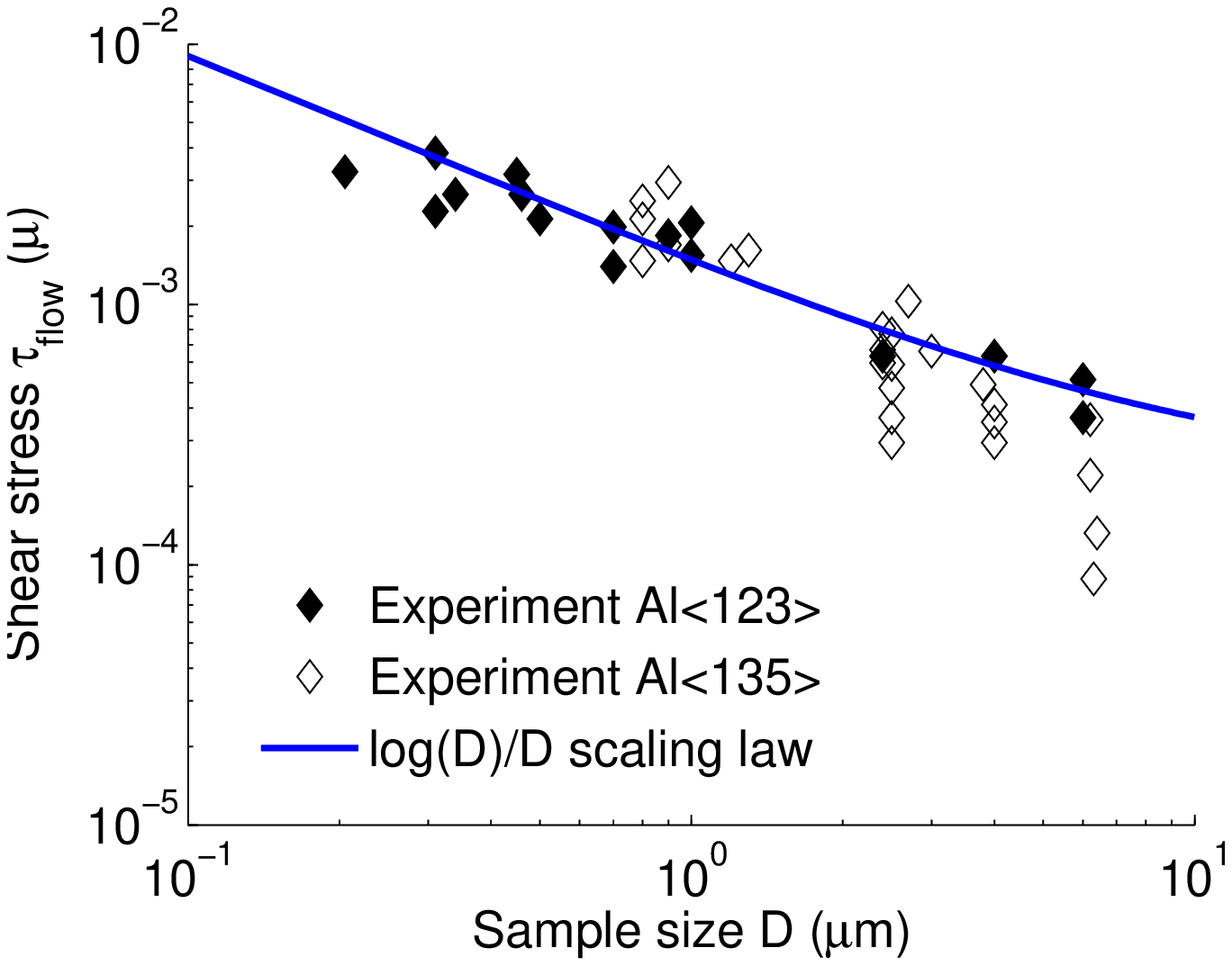}}
\subfigure[Copper $<111>$]{\includegraphics[width=.4\textwidth]{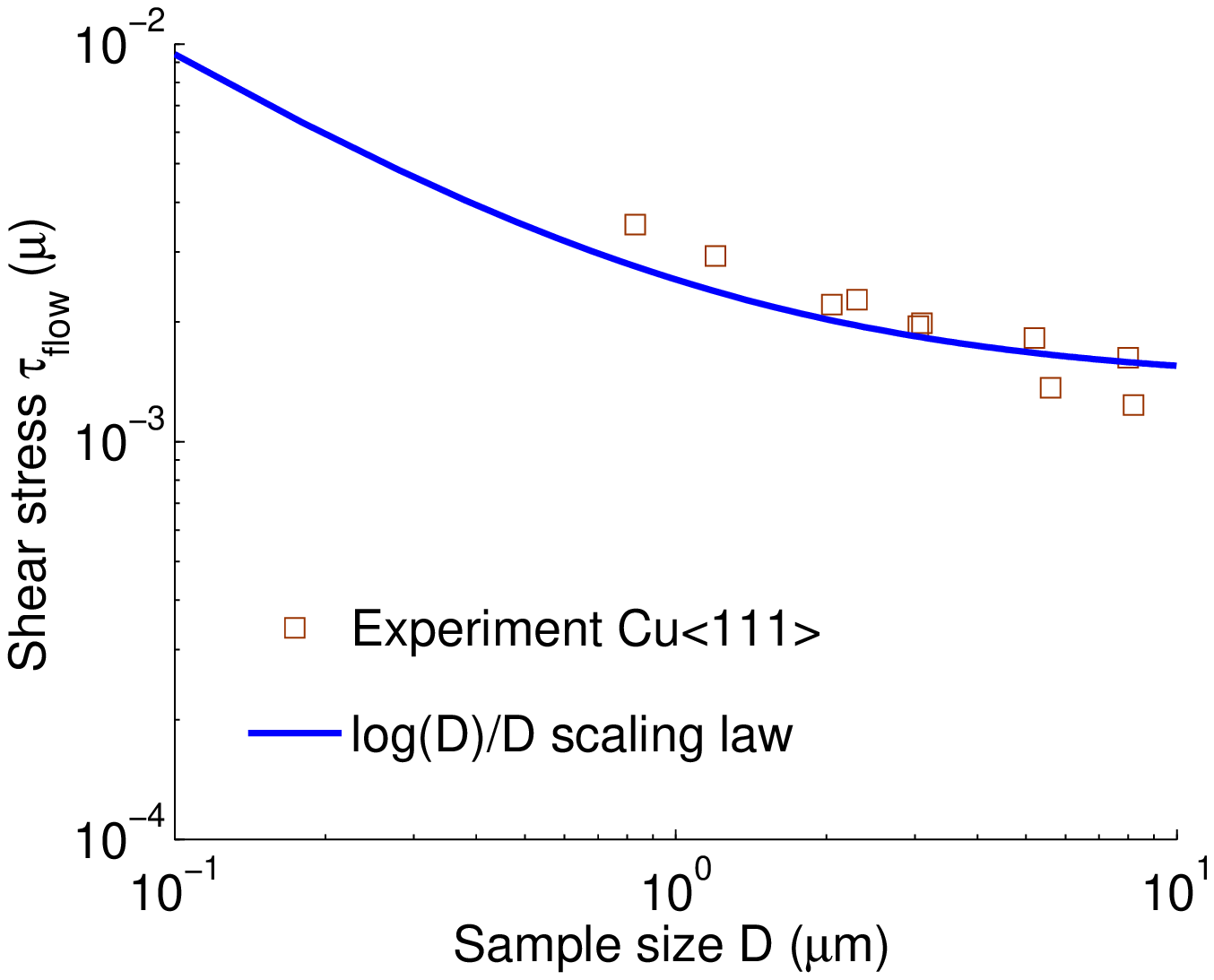}}
\subfigure[Copper $<123>$]{\includegraphics[width=.4\textwidth]{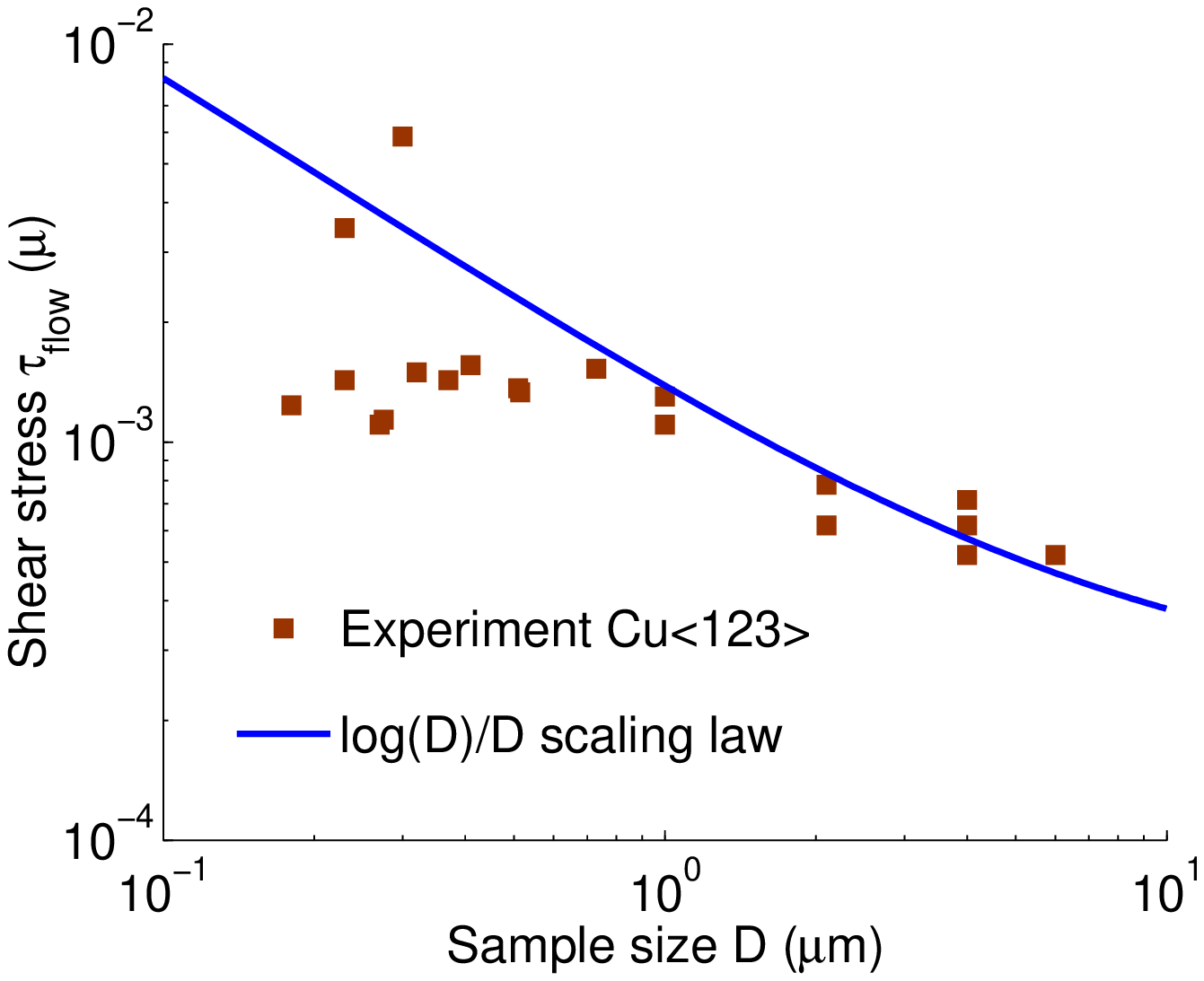}}
\caption{The $\log(D)/D$ scaling law for the size-dependent effect of the flow stress of a micro-pillar given in Eq.~\eqref{flow_stress_sample_size} resolved in the slip planes is examined by the experimental data for single crystal nickel \citep{Dimiduk2005}, aluminium and copper \citep{Uchic_review2009} micro-pillars.   \label{Fig_size_effect_comp}}
\end{figure}

Combining Eqs.~\eqref{flow_tau0_exp} and \eqref{tau_min_0_exp}, we obtain the scaling law in Eq.~(\ref{flow_stress_sample_size}).
Predictions of Eq.~(\ref{flow_stress_sample_size}) are in good agreement with the experimental data as shown in Fig.~\ref{Fig_size_effect_comp}
for single crystal nickel, aluminium and copper pillars. In each comparison in the figure, the resolved flow stress $\tau_{\text{flow}}=m_{\text{s}}\sigma_{\text{flow}}$ is plotted against the pillar size $D$, with $\alpha_{\text{eff}}=1/6$, $r_{\text{c}}=0.6b$ and $\sigma_{\text{flow}}^0$ being a fitted parameter in the scaling law.

It is noted that Eq.~\eqref{tau_min_0_exp} has also been employed to predict the flow stress in polycrystalline thin-film structures \citep{Gumbsch2001, Gruber2008}. In their works, $\alpha_{\text{eff}}$ is suggested to be $1/4$ to $1/3$.

\subsection{Other behaviors of the micro-pillars}

In this subsection, we discuss other plastic behaviors of micro-pillars that can be captured by the continuum model.

\begin{figure}[!ht]
\centering
\subfigure[$D=2.4\mu$m]{\includegraphics[width=.3\textwidth ]{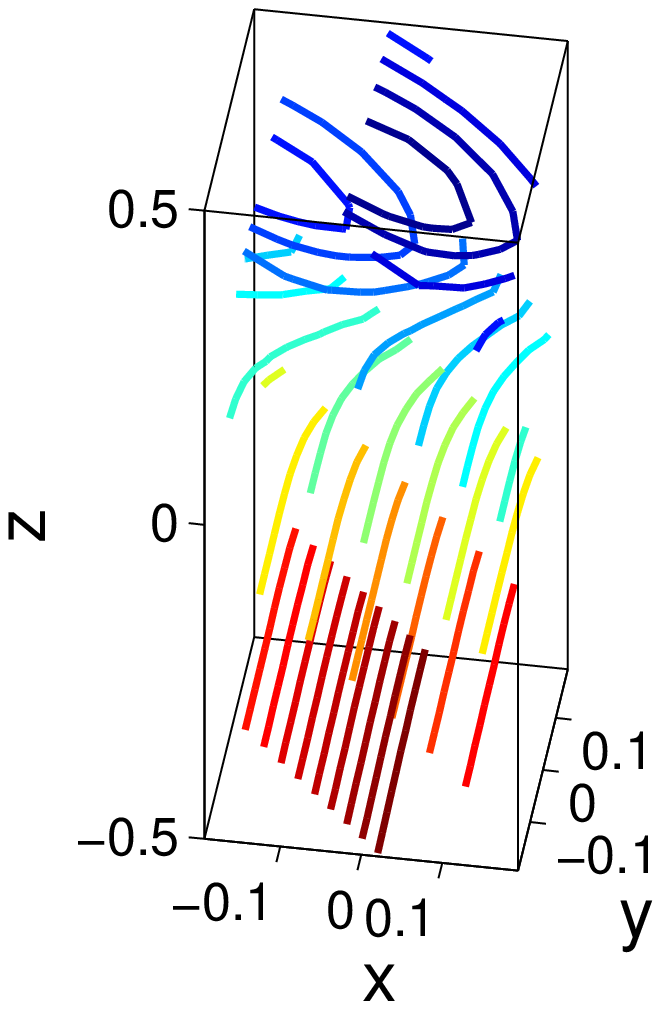}}
\subfigure[$D=5\mu$m]{\includegraphics[width=.3\textwidth ]{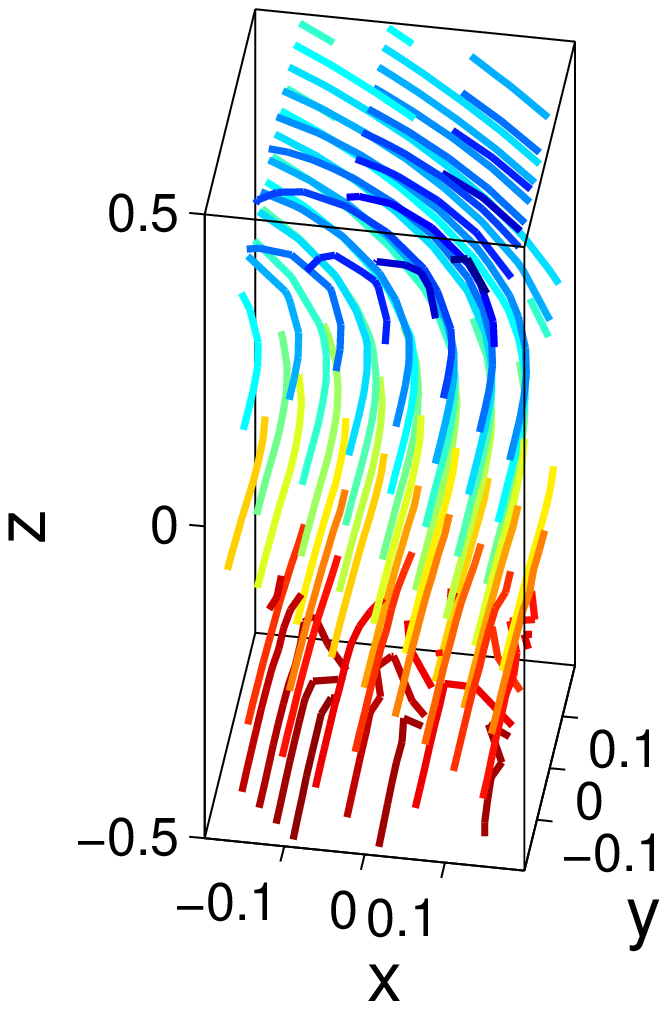}}
\subfigure[$D=10\mu$m]{\includegraphics[width=.3\textwidth ]{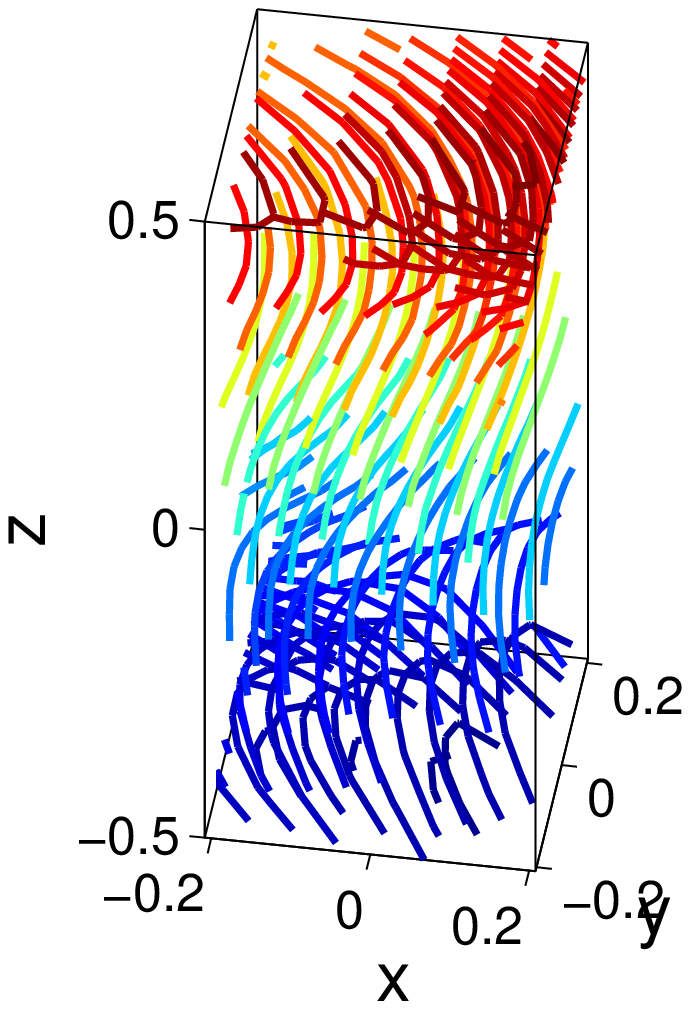}}
\caption{Snapshots of the dislocation substructures (in an average sense as discussed in Sec.~\ref{Sec_DDPFs})
in some simulations using our continuum model  for pillars of size (a) $D = 2.4\mu$m, (b) $5\mu$m, and (c) $10\mu$m. The snapshots are taken from the perfect plastic deformation regime as shown by the stress-strain curves in Figs.~\ref{Fig_ElAwardy_ss_curve} and \ref{Fig_ss_curve_comp_exp}. \label{Fig_dislocation_substructure}}
\end{figure}

Firstly, recall that in our continuum model, the dislocation networks within the specimen (after local homogenization) are described by the DDPFs as defined in Sec.~\ref{Sec_DDPFs} and illustrated by Fig.~\ref{Fig_illu_psi_phi}, that is, the dislocation distribution on  each slip plane  is described by the contour curves of the DDPF $\phi$, and the distribution of the slip plane is represented by another DDPF $\psi$.
Fig.~\ref{Fig_dislocation_substructure} shows some snapshots of the dislocation substructures
in some simulations using the continuum model  for pillars of size $D = 2.4\mu$m, $5\mu$m, and $10\mu$m.
The distributions of dislocation curves in the figure look smoothly varying. This is because the continuum model only resolves the dislocation microstructures in an average sense (see Sec.~\ref{Sec_DDPFs}).

\begin{figure}[!ht]
\centering
\subfigure[]{\includegraphics[width=.48\textwidth]{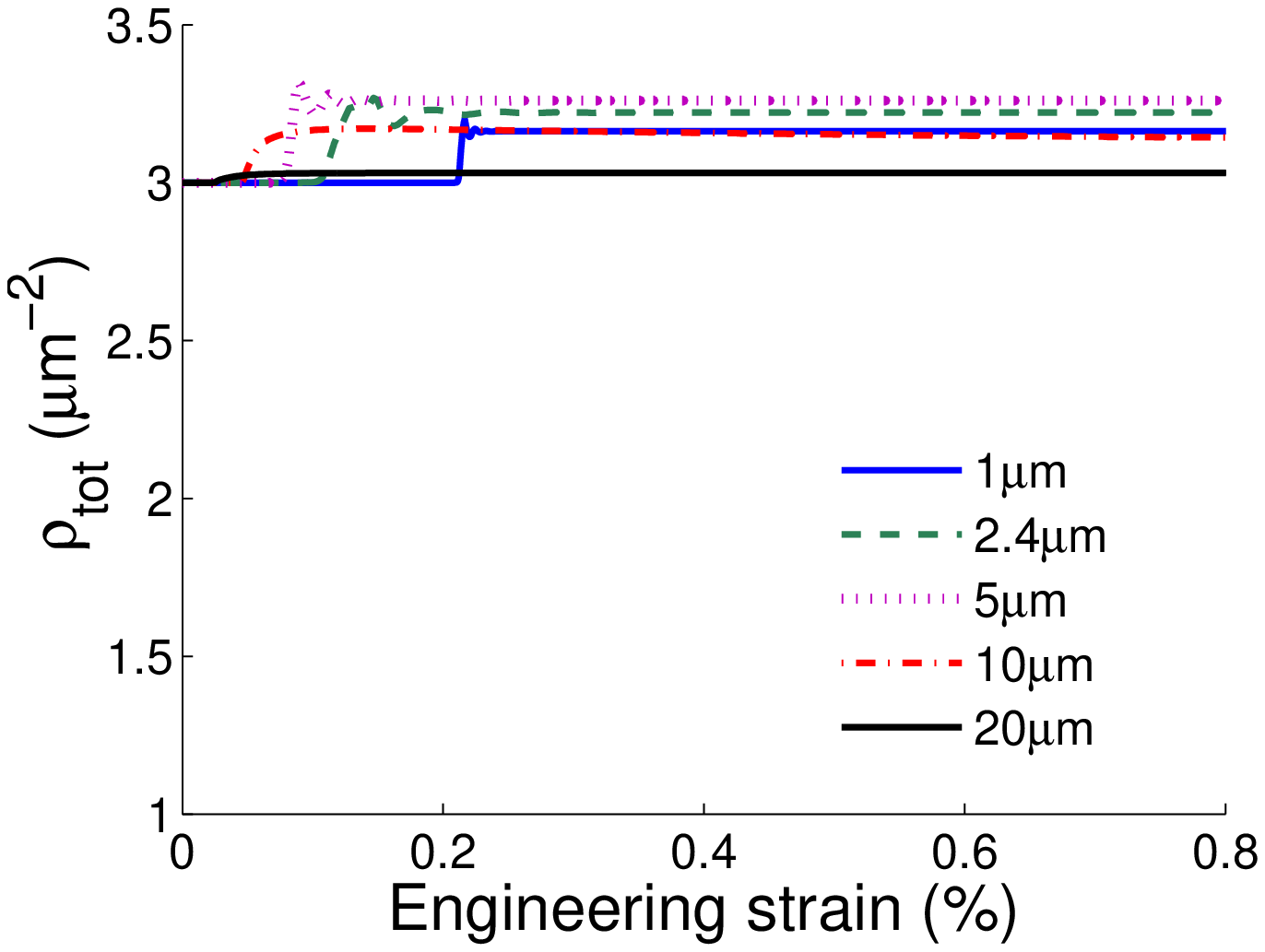}}
\subfigure[]{\includegraphics[width=.48\textwidth]{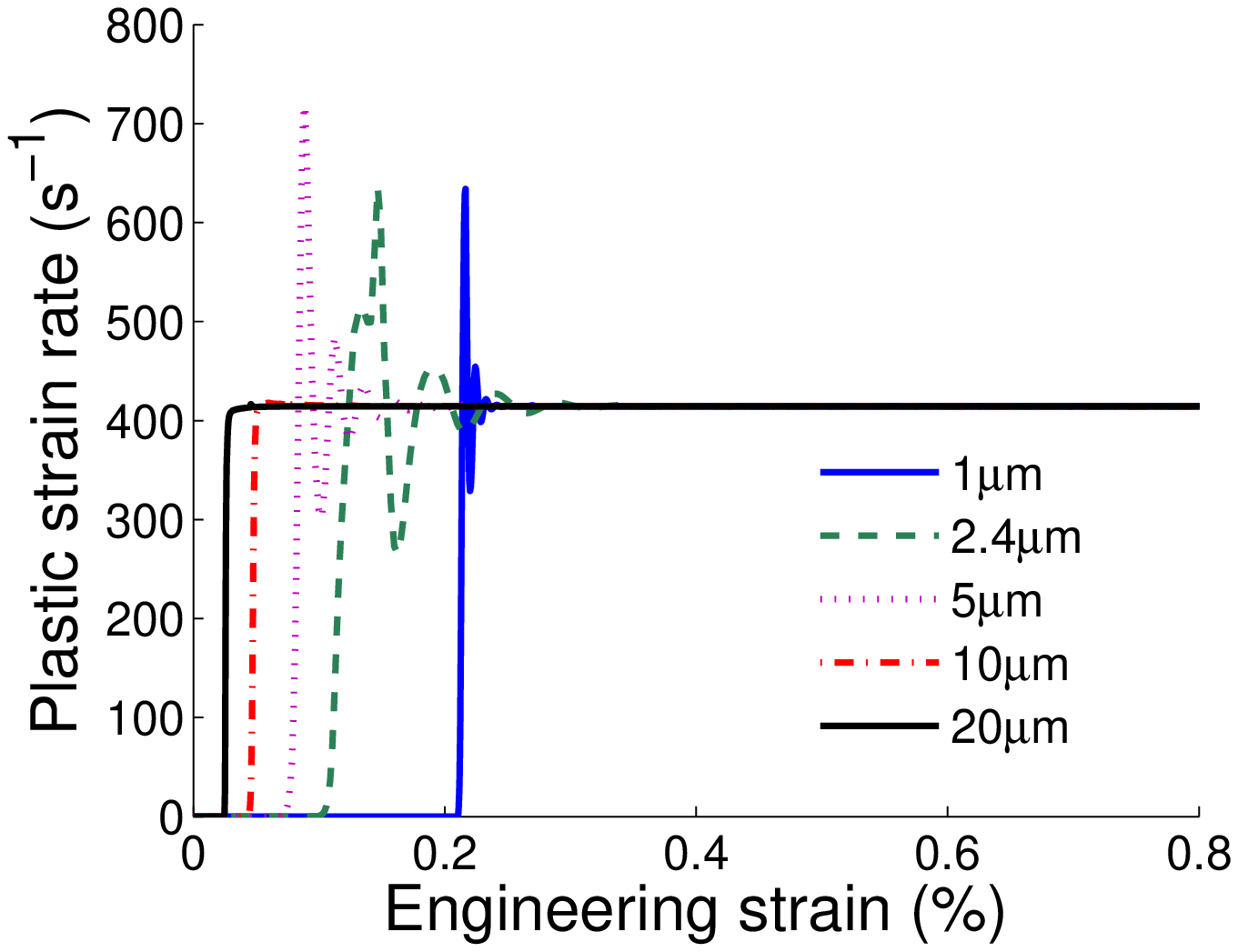}}
\caption{Evolution of (a) the total dislocation density determined by Eq.~\eqref{total_number_density} and (b) the plastic strain rate determined by Eq.~\eqref{total_plastic_strain} in micro-pillars with various sizes, obtained by using the continuum model. \label{Fig_evolution_internal_state}}
\end{figure}

We can also keep track of the two quantities that are commonly used to describe the state of the specimen in a plastic deformation process: the total dislocation density $\rho_{\text{tot}}$ given by Eq.~\eqref{total_number_density} and the plastic strain rate $\dot{\eps}_{\text{tot}}^{\text{p}}$ given by Eq.~\eqref{total_plastic_strain}. Numerical results of $\rho_{\text{tot}}$ and $\dot{\eps}_{\text{tot}}^{\text{p}}$  obtained by simulations using our continuum model are plotted in Fig.~\ref{Fig_evolution_internal_state}. These results along with the stress-strain curves presented in Figs.~\ref{Fig_ElAwardy_ss_curve} and \ref{Fig_ss_curve_comp_exp} suggest that the following evolution process may take place inside the micro-pillars when being compressed under a constant applied strain rate.
When the elastic limit of the samples is reached, dislocation sources start to release dislocation loops, resulting in plastic flows and a rise in the total dislocation density inside the pillars. After a (relatively) short period, the system reaches a steady state corresponding to the perfectly plastic regimes in Figs.~\ref{Fig_ElAwardy_ss_curve} and \ref{Fig_ss_curve_comp_exp}.
At this steady state, the applied strain rate is fully accommodated by the dislocation motion and the resolved shear stress ceases to increase.
Another interesting phenomenon observed from Fig.~\ref{Fig_evolution_internal_state}(b) is that the values of $\dot{\eps}^{\text{p}}_{\text{tot}}$ converge to a same value for samples of various size. This is because this converged values are determined by the applied strain rates which are the same for all samples in the compression tests.

\begin{figure}[!ht]
\centering
\subfigure[2.5\% strain]{\includegraphics[width=.32\textwidth]{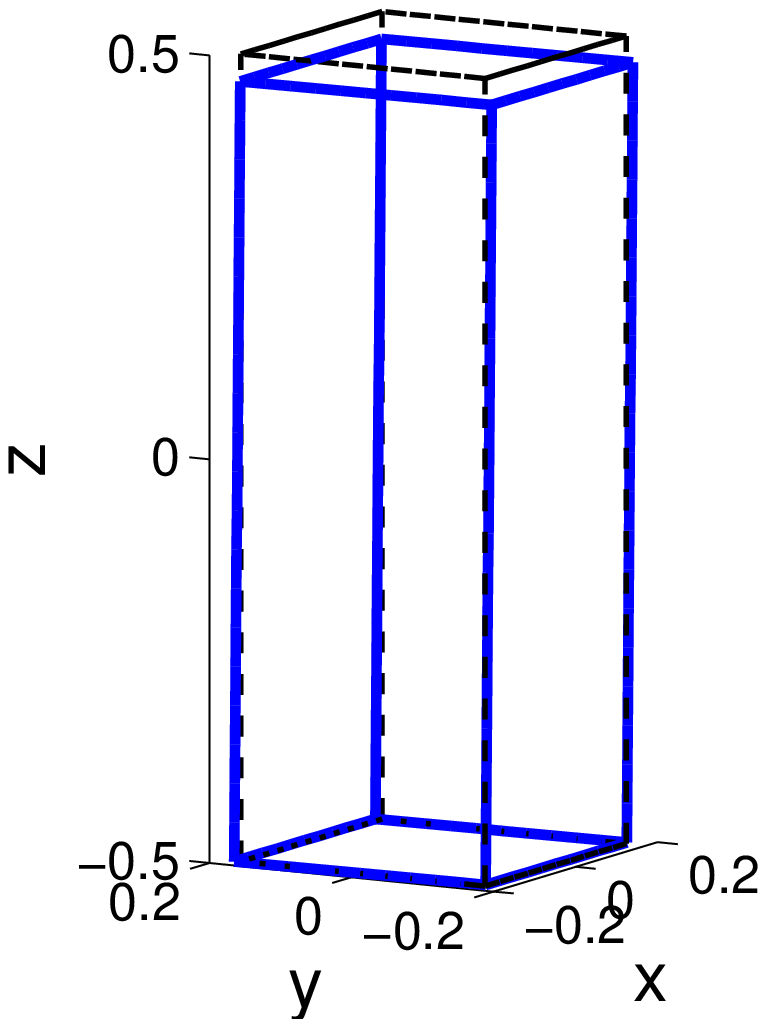}}
\subfigure[9\% strain]{\includegraphics[width=.32\textwidth]{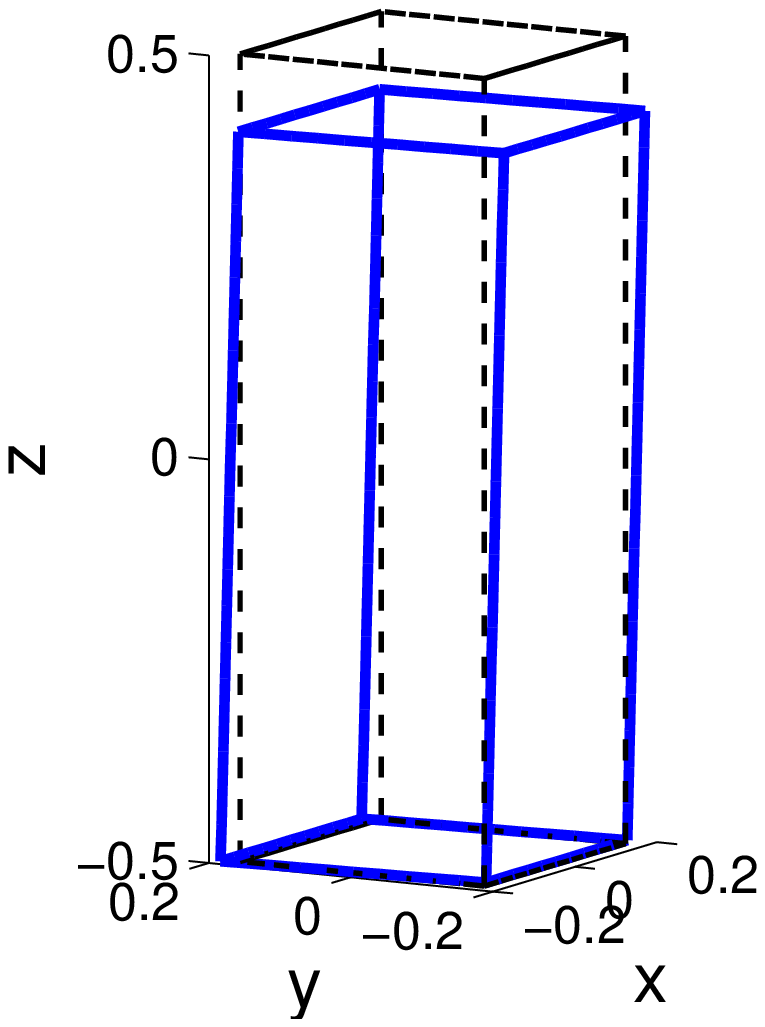}}
\subfigure[12\% strain]{\includegraphics[width=.32\textwidth]{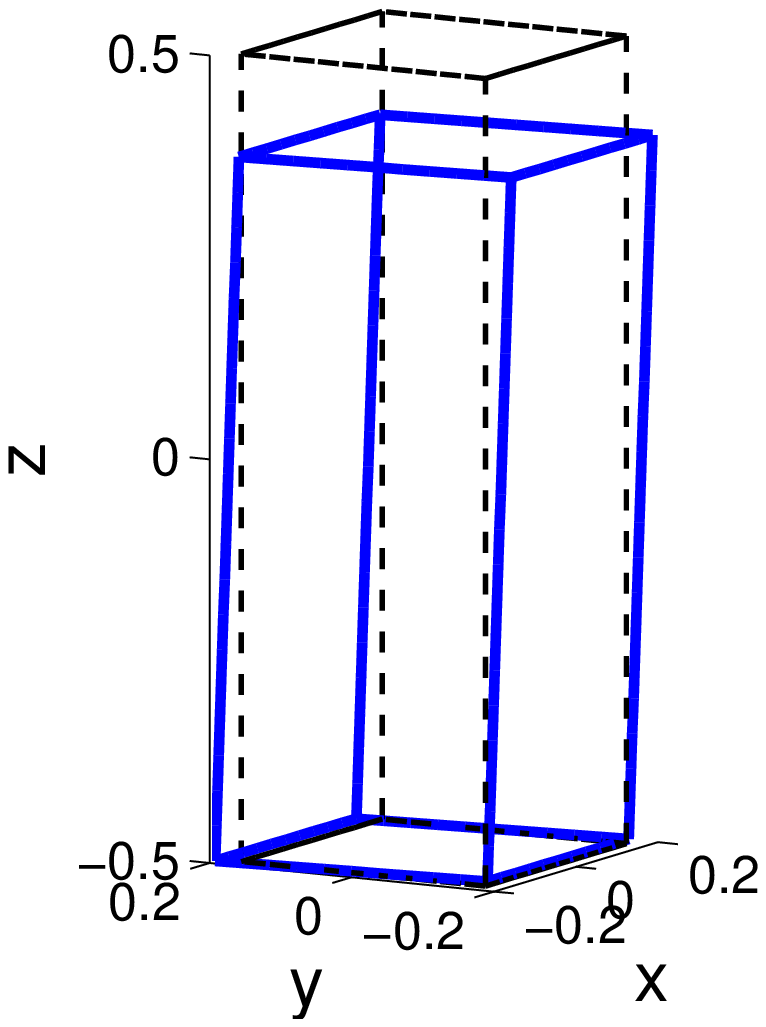}}
\caption{Shapes of a deformed micro-pillar of size $D=5\mu$m under various applied strain: the cuboids formed by the dashed-lines describe the original shapes of the pillar. \label{Fig_deformed_pillars}}
\end{figure}

\begin{figure}[!ht]
\centering
\subfigure[$u_1$]{\includegraphics[width=.32\textwidth]{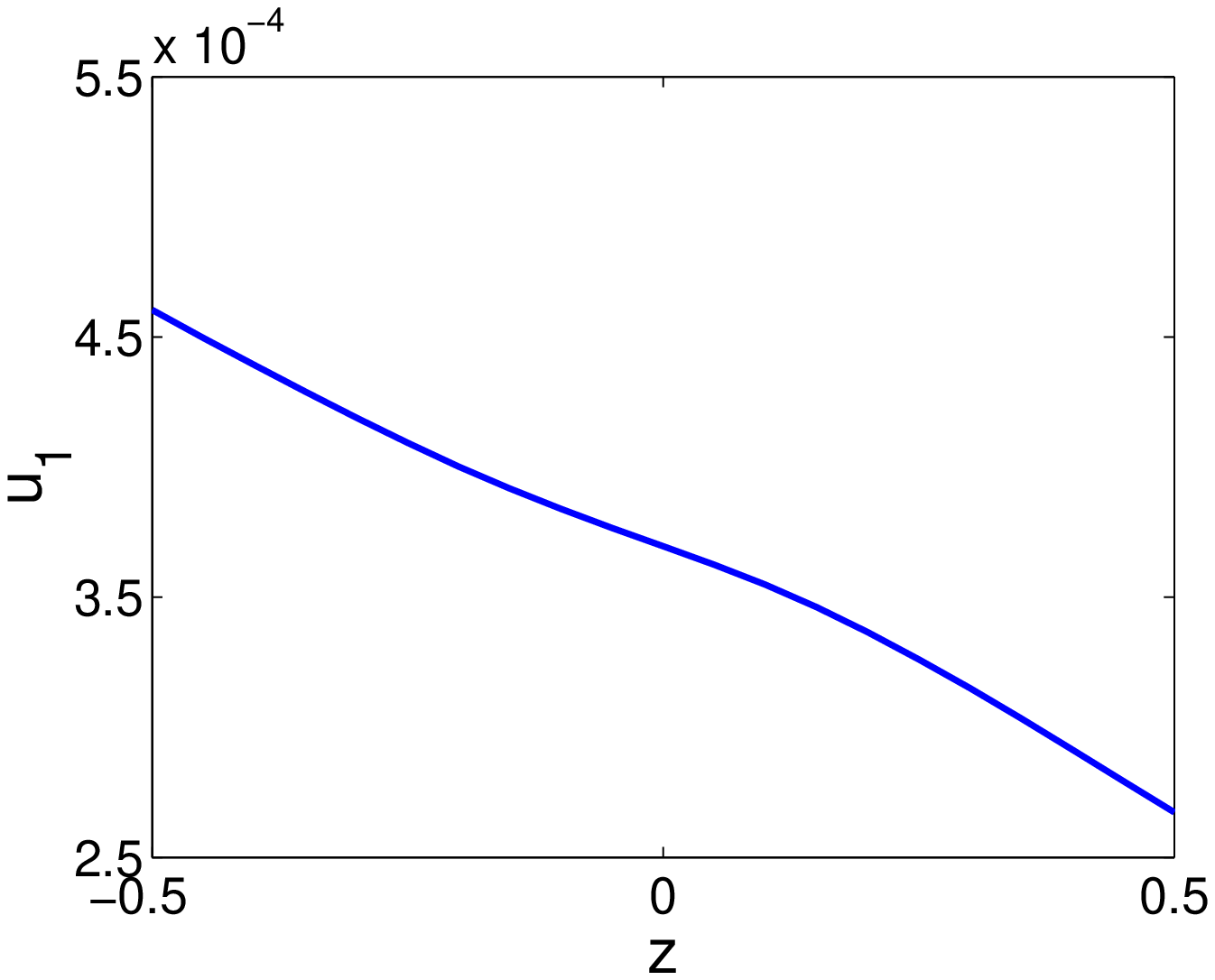}}
\subfigure[$u_2$]{\includegraphics[width=.32\textwidth]{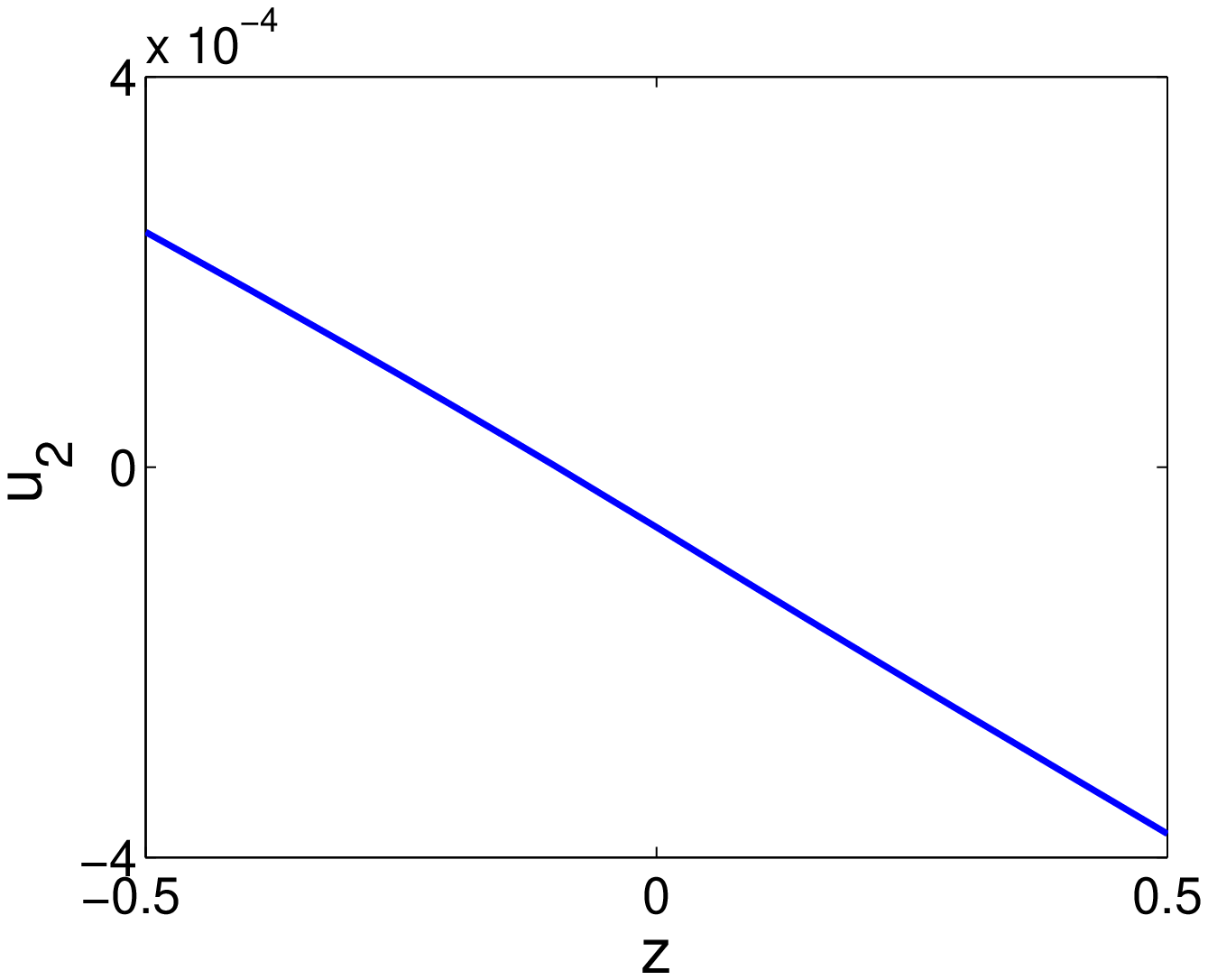}}
\subfigure[$u_3$]{\includegraphics[width=.32\textwidth]{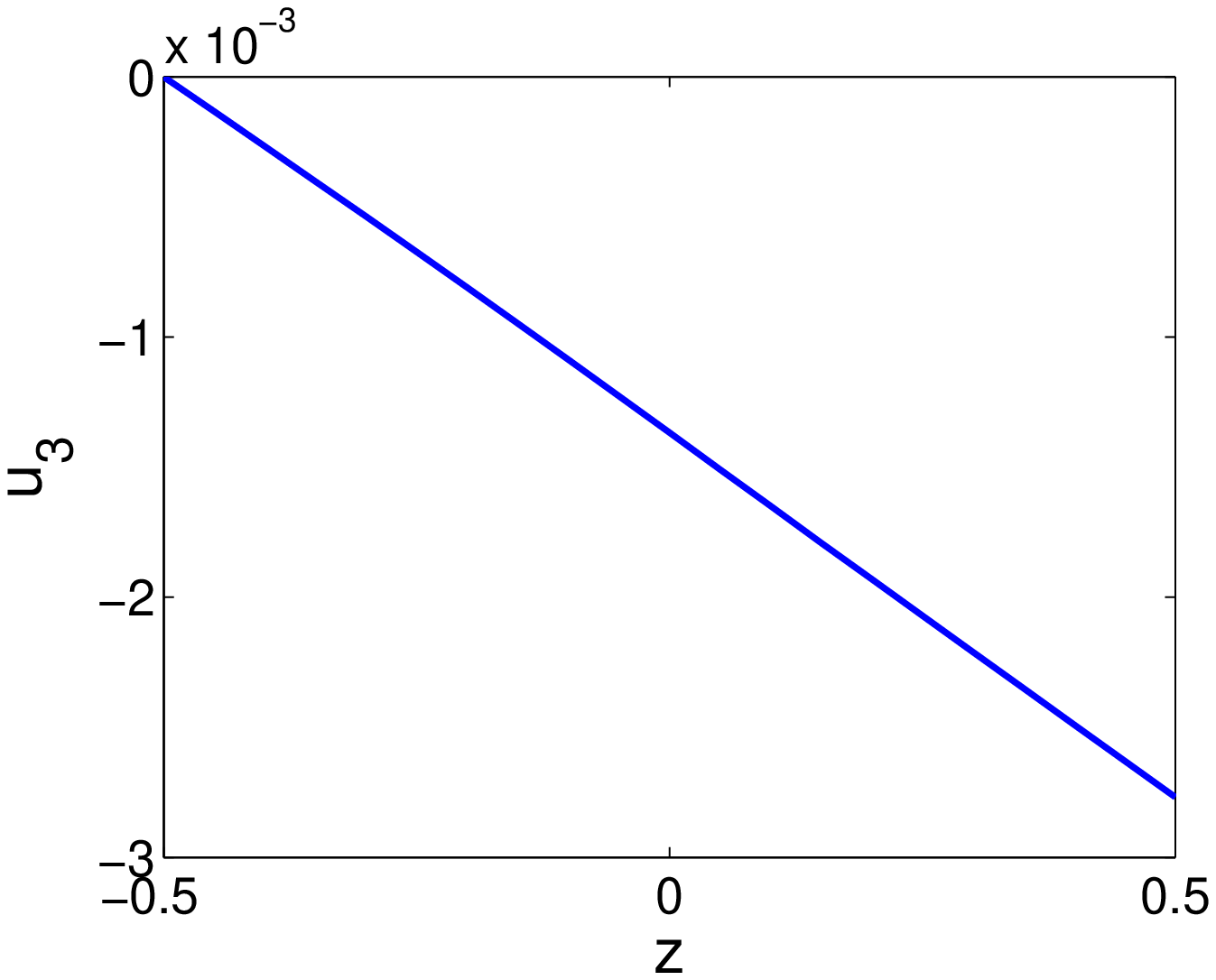}}
\caption{Displacement field $\vu=(u_1,u_2,u_3)$ along a vertical edge of a micro-pillar with size $D=1\mu$m and $8$ Frank-Read sources under $0.3\%$ applied strain. The length unit is $L$.\label{Fig_edge_u}}
\end{figure}

By using the continuum model, we are also able to track the shape changes of the micro-pillars. Given $\vu$ the displacement field, $\vr+\vu$ is the position of a point, whose initial position is at $\vr$. Examples of the profile of a deformed pillar during compression are shown in Fig.~\ref{Fig_deformed_pillars}.
Note that in our simulations, spatial variation in the plastic shear slips induced by a distribution of Frank-Read sources  has been locally averaged over many discrete sources in the source continuum formulation. The pillar profiles in Fig.~\ref{Fig_deformed_pillars} agree in the averaged sense  with the experimental observations (e.g. Fig.~4 in \citet{Dimiduk2005}).  With smaller number of sources, relatively strongly non-uniform displacement fields along $z$ direction can be observed in our simulations, see Fig.~\ref{Fig_edge_u} for an example, which more accurately agree with the slip-band structures observed in the experiments.
For an extreme case, if we check the surface of a micro-pillar containing only one operating Frank-Read source as shown in Fig.~\ref{Fig_displacment_surf}, localized displacement gradient is clearly observed around the activated slip plane. These agree with the conclusion of \citet{Akarapu2010} drawn from simulations using a hybrid elasto-viscoplastic model which couples DDD that such localized deformation, which is strong for small-size pillars, is due to the heterogeneous dislocation distributions that lead to clusters of sources.
More efficient numerical implementation method of our continuum model with fine meshes will help resolve more accurately the pillar profile observed in the experiments, which will be developed in the future work.

\section{Conclusion}

In this paper, we have presented  a dislocation-density-based continuum model to study the plastic behavior of crystals, in which the dislocation substructures are represented by pairs of DDPFs. A discrete dislocation network in three dimensions  is approximated by a dislocation continuum after local homogenization of dislocation ensembles within some representative volume. For the $\lambda$-th slip plane in the dislocation continuum,  a DDPF $\phi^{\lambda}$ is defined such that $\phi^{\lambda}$ restricted on each slip plane describes the plastic slip across the slip plane and identifies the distribution of dislocation curves by its contour lines, and
another  DDPF $\psi^{\lambda}$ is employed to describe the slip plane distribution by its contour surfaces.
The major advantage of this three-dimensional continuum model lies in its simple representation of distributions of curved dislocations. The connectivity condition of dislocations is automatically satisfied with this representation of dislocations.

Based on the DDPFs, the plastic deformation process of crystals can be formulated by a system of equations as given by Eqs.~\eqref{eqn_stress_tensor1}--\eqref{eqn_mobility_law}.
We have shown that the equation system provides an effective summary over the underlying discrete dislocation dynamics, including the long-range elastic interaction of dislocations, dislocation line tension effect, and dislocation multiplication by Frank-Read sources.
Numerically, a finite element formulation is proposed to compute the long-range stress field.

The continuum model is validated by comparisons with DDD simulations and experimental data. As one application of the continuum model characterized by DDPFs, the size effect on the strength of micro-pillars is studied, and the pillar flow stress is found scaling with its (non-dimensionalized) pillar size $D$ by $\log(D)/D$.

Future work may include generalizations of the continuum model characterized by DDPFs to incorporate the out-of-slip-plane dislocation motion (cross-slip or climb) in which the contour surfaces of the DDPF $\psi$ may be curved and evolved in the plastic deformation. We will also take into account in the continuum model other short-range dislocation interactions such as dislocation reactions and junction formation, as well as dislocation interactions with other defects \citep[e.g.][]{Xiang2006,Rickman2010,Xiang_grain_boundary2014}.
Efficient numerical implementation method will also be considered in the future work.

\section*{Acknowledgement}
This work was partially supported by the Hong Kong Research Grants Council General
Research Fund 606313.

\bibliography{mybib}

\end{document}